\documentclass[a4paper,11pt]{article}
\usepackage{jinstpub} % for details on the use of the package, please see the JINST-author-manual
\usepackage{lineno}
\usepackage{graphicx}
\usepackage{subcaption}
\usepackage{multirow} % Required for multirows
\usepackage{lineno}
\usepackage{graphicx}
\usepackage{subcaption}
\usepackage{multirow} % Required for multirows
\usepackage{float} 
\usepackage{overpic}
\usepackage{threeparttable}
\usepackage{textcomp}
\usepackage{setspace}
\usepackage{xcolor}
\title{\boldmath Dark Count Rate Stability of JUNO 20-inch PMTs in Mass Testing}

\author[a,b]{Min Li}
\author[a,b]{Narongkiat Rodphai}
\author[a,b]{Caimei Liu}
\author[a,b,1]{Zhimin Wang\note{Corresponding author.}}
\author[a,b]{Zhaoyuan Peng}
\author[e]{Jun Wang}
\author[f]{Nikolay Anfimov}
\author[f]{Denis Korablev}
\author[c]{Tobias Lachenmaier}
\author[f]{Alexander G. Olshevskiy}
\author[a]{Zhonghua Qin}
\author[c]{Tobias Sterr}
\author[c]{Alexander Felix Tietzsch}
\author[e]{Rong Zhao}
\author[e]{Wei Wang}
\author[a]{Kaile Wen}
\author[d]{Bjoern Soenke Wonsak}
\author[a]{Wan Xie}
\author[a]{Meihang Xu}
\author[a,b]{Yu Zhang}
\affiliation[a]{Institute of High Energy Physics, Beijing 100049, China}
\affiliation[b]{University of Chinese Academy of Sciences, Beijing 100049, China}
\affiliation[c]{Eberhard Karls Universität Tübingen, Physikalisches Institut, Tübingen, Germany}
\affiliation[d]{Institute of Experimental Physics, University of Hamburg, Hamburg, Germany}
\affiliation[e]{Sun Yat-Sen University, Guangzhou, China}
\affiliation[f]{Joint Institute for Nuclear Research, Dubna, Russia}
\emailAdd{wangzhm@ihep.ac.cn}
\abstract{The Jiangmen Underground Neutrino Observatory (JUNO) is an ambitious multipurpose neutrino experiment designed to determine the neutrino mass ordering, with an impressive energy resolution goal of at least 3\% at 1\,MeV. To achieve a photon detection coverage of approximately 75\%, JUNO will utilize two types of 20-inch photomultiplier tubes (PMTs): the large PMT (LPMT) and the microchannel plate PMT (MCP-PMT).
A significant concern in high-precision neutrino measurements is the dark count rate (DCR) of PMTs, which introduces noise that can adversely affect energy measurement accuracy. During the mass testing phase of the JUNO 20-inch PMTs, comprehensive measurements of the DCR were undertaken. These measurements not only captured the DCR values of individual PMTs but also examined the stability and temperature dependence of the DCR at an operating gain of ($1 \times 10^7$).
This paper presents a detailed characterization of the DCR of the JUNO 20-inch PMTs, investigating factors such as cooling time, temperature variations, and long-term stability using the JUNO Pan-Asia PMT testing facilities. The results reveal distinct DCR characteristics between the two types of PMTs, providing valuable insights into the nature of DCR and its implications for JUNO's scientific objectives.
In addition to performance characterization, we implemented a monitoring system to track DCR stability over time. Notably, several spikes in DCR were identified, prompting a preliminary investigation into their causes. Potential factors contributing to these spikes, such as flasher events, were explored using coincidence rate analysis and complementary imaging techniques.
The findings from this study are crucial for optimizing the performance of PMTs in JUNO, ultimately aiding the experiment in achieving its goals related to neutrino physics.}


\begin{document}
\maketitle
\flushbottom

 \section{Introduction}
 \label{sec:intro}
 The Jiangmen Underground Neutrino Observatory (JUNO) \cite{JUNO-yellow-osti-1354633,2022103927} is a multi-purpose neutrino experiment with the primary objective of precisely measuring the neutrino mass ordering at a significance level of 3-4\,$\sigma$ within six years. The JUNO detector is strategically located approximately 53\,km from both the Yangjiang and Taishan nuclear power plants.
The detector comprises several sub-detectors, including the center detector (CD)\cite{JUNO-CD}, a veto detector featuring a water pool Cherenkov detector (WP), and a muon Top Tracker (TT). The CD is a liquid scintillator detector filled with 20\,kton of liquid scintillator, which serves as the target. It will deploy a total of 17,612 20-inch photomultiplier tubes (LPMTs) \cite{PMT:JUNO:2022hlz} and 25,600 3-inch small photomultiplier tubes (SPMTs) \cite{JUNO:3inch:CAO2021165347} in pure water. This configuration aims to achieve an energy resolution of $3\%/\sqrt{E(MeV)}$.
In addition to the CD, another 2,400 LPMTs are employed in the WP to detect Cherenkov light produced by cosmic muons. The PMT coverage is designed to exceed 78\% (75\% with LPMTs alone), aided by an average photon detection efficiency (PDE) exceeding 27\%, which is essential for reaching the specified energy resolution. In total, JUNO will utilize 20,012 LPMTs, including 5,000 R12860-50 dynode-PMTs by Hamamatsu Photonics K.K. (HPK) and over 15,000 microchannel plate (MCP) PMTs from North Night Vision Technology Co., Ltd. (NNVT).
These large-area PMTs have been successfully implemented in numerous large-scale particle physics experiments, including KamLand \cite{suekane2006kamland}, Borexino \cite{alimonti2009borexino}, SNO \cite{boger2000sudbury}, Super-Kamiokande \cite{fukuda2003super}, and Daya Bay \cite{an2016detector}, among others.

 The large photomultiplier pubes (LPMTs) of the JUNO experiment will operate with a gain of ($1 \times 10^7$) \cite{PMT:gain:Zhang2021GainAC}. Extensive studies have been conducted on various parameters of these PMTs, including their timing characteristics, gain, and single-photon response.
 
Among the PMTs utilized, the 20-inch dynode PMT, model R12860-50 from Hamamatsu Photonics (HPK), has a dark count rate (DCR) that can be reduced to lower than 20\,kHz \cite{maekawa2021performance}. Additionally, the newly developed 20-inch microchannel plate PMTs (MCP-PMT) from NNVT are anticipated to exhibit superior performance characteristics \cite{kim2022characterization,MCPPMT:CHANG2016143}.

To ensure thorough testing, all LPMTs for JUNO are evaluated using a specially designed mass testing system capable of assessing most relevant parameters \cite{wonsak2021container,Anfimov2017LargeP2,JUNO:PMT:database}. This facility was located in Zhongshan City, Guangdong Province, China. The key characteristics of the LPMTs, including DCR, photon detection efficiency, and charge resolution, have been measured in line with JUNO's requirements \cite{PMT:JUNO:2022hlz,Zhao:2022gks,Liu:2022nhe}.

Given JUNO's stringent energy resolution requirements, understanding and addressing the DCR of PMTs is vital for accurate energy measurement and reconstruction. The DCR is influenced by various sources, including thermionic emission from the photocathode and dynodes, leakage current, scintillation-induced photocurrents from the glass envelope or electrode supports, and ionization currents originating from residual gases. Moreover, gaining a comprehensive understanding of the DCR, alongside factors such as cooling time, temperature effects, and long-term stability, is essential for both the LPMTs and the JUNO experiment, beyond the traditional parameters evaluated in mass testing \cite{PMT:JUNO:2022hlz}.
Flasher effects related to PMTs, which can lead to heightened DCR or substantial pulses in the PMTs, pose significant challenges for detectors designed for rare event detection \cite{CAO201662,Yang_2020,Qian_2020}. Furthermore, large pulses generated from the PMT glass due to muon interactions and other radioactivities have been investigated in \cite{Zhang_2022,Zhang_2024}. Consequently, further exploration of potential flasher PMT candidates is valuable.

The layout of this paper is organized as follows: In Section \ref{sec:setup}, we will provide an overview of the LPMTs and the mass testing system. Section \ref{sec:resanddis} will present and analyze the results, focusing on the DCR of the 20-inch PMTs with respect to cooling time and temperature effects. Section \ref{sec:flasher} will examine characteristics of potential flasher PMTs. Finally, a brief summary will be provided in Section \ref{sec:conclu}.

 \section{System Setup}
\label{sec:setup}

 \subsection{20-inch Photomultiplier Tubes (PMTs)}
 \label{sec:pmt}
The JUNO experiment employs two types of 20-inch PMTs: approximately 5,000 box and linear-focused dynode PMTs (R12860-50 HQE or R12860)~\cite{HPK-R12860} from Hamamatsu Photonics (HPK), and around 15,000 microchannel plate PMTs (MCP PMTs) (GDB-6201 or N6201)~\cite{NNVT-GDB6201-note,NNVT-GDB6201-improvement}, from North Night Vision Technology Co., Ltd. (NNVT). The selection of these PMTs was influenced by various factors, including performance parameters, costs, associated risks, and the overarching physics objectives of the JUNO project~\cite{JUNOPMTliangjian}.

Designed to operate with a gain of ($1 \times 10^7$), the 20-inch PMTs in JUNO are subjected to a positive high voltage (HV) that takes into account considerations of cost, assembly, and noise. The detector is also designed to function within a temperature range of (21 $\pm$ 1 $^\circ$C)~\cite{zhang2021method}.

Both the DCR and PDE of the PMTs play critical roles in event reconstruction and have significant ramifications for physics measurements. The theoretical implications of the DCR for large liquid scintillator (LS) or water-based neutrino experiments have been discussed in the literature~\cite{JUNOPMTliangjian}. Extensive previous research has addressed various characteristics of DCR, as seen in studies such as~\cite{YU2021165433,ICARUS-2018,PMT-2016,MCPPMT-DCR2021}. For the PMTs accepted for JUNO, the mean DCR values are approximately 15.3\,kHz for HPK PMTs and 49.3\,kHz for NNVT PMTs, with a requisite cooling period of no less than 12 hours~\cite{PMT:JUNO:2022hlz}.

 \subsection{Container System}
 \label{sec:consys}

 During the acceptance testing of JUNO, all 20-inch PMTs have been evaluated under several configurations. This includes testing bare PMTs with a plug-able high voltage (HV) divider for initial acceptance checks and water-proof potted PMTs paired with JUNO electronics.

\begin{figure*}[!ht]
     \subfloat[Container A and B]{\label{fig:containerab}\includegraphics[width=0.495\textwidth,height=0.3\textwidth]{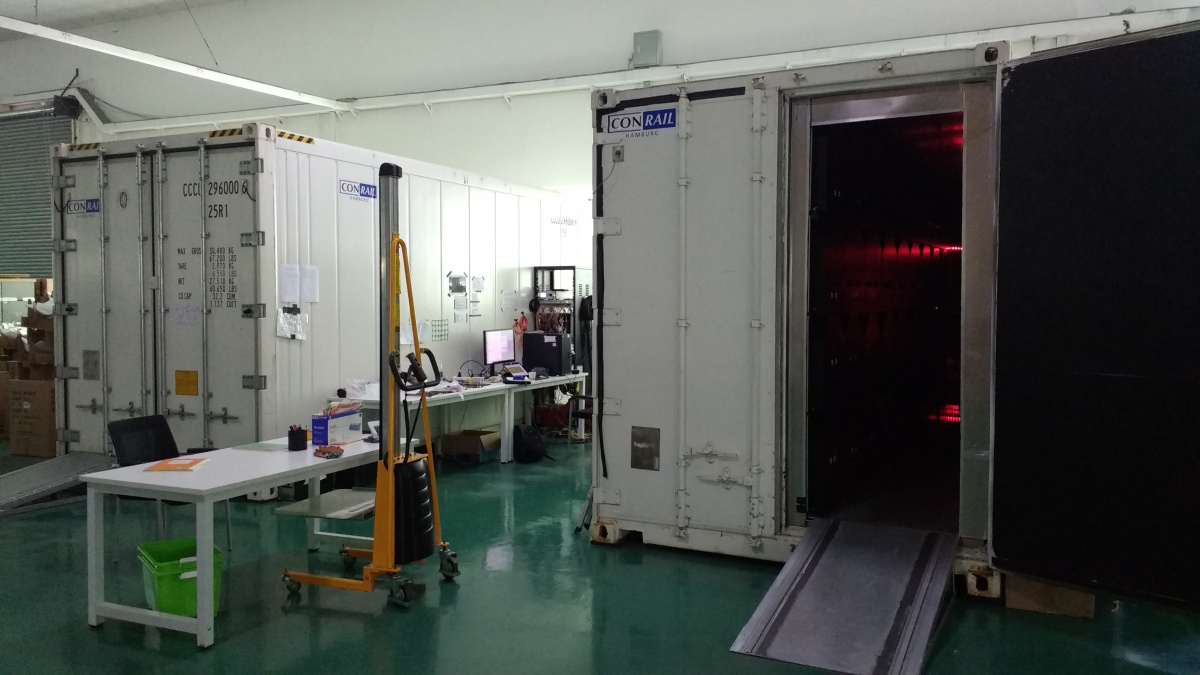}}
    \hfill 	
     \subfloat[Container D]{\label{fig:containerd}\includegraphics[width=0.495\textwidth,height=0.3\textwidth]{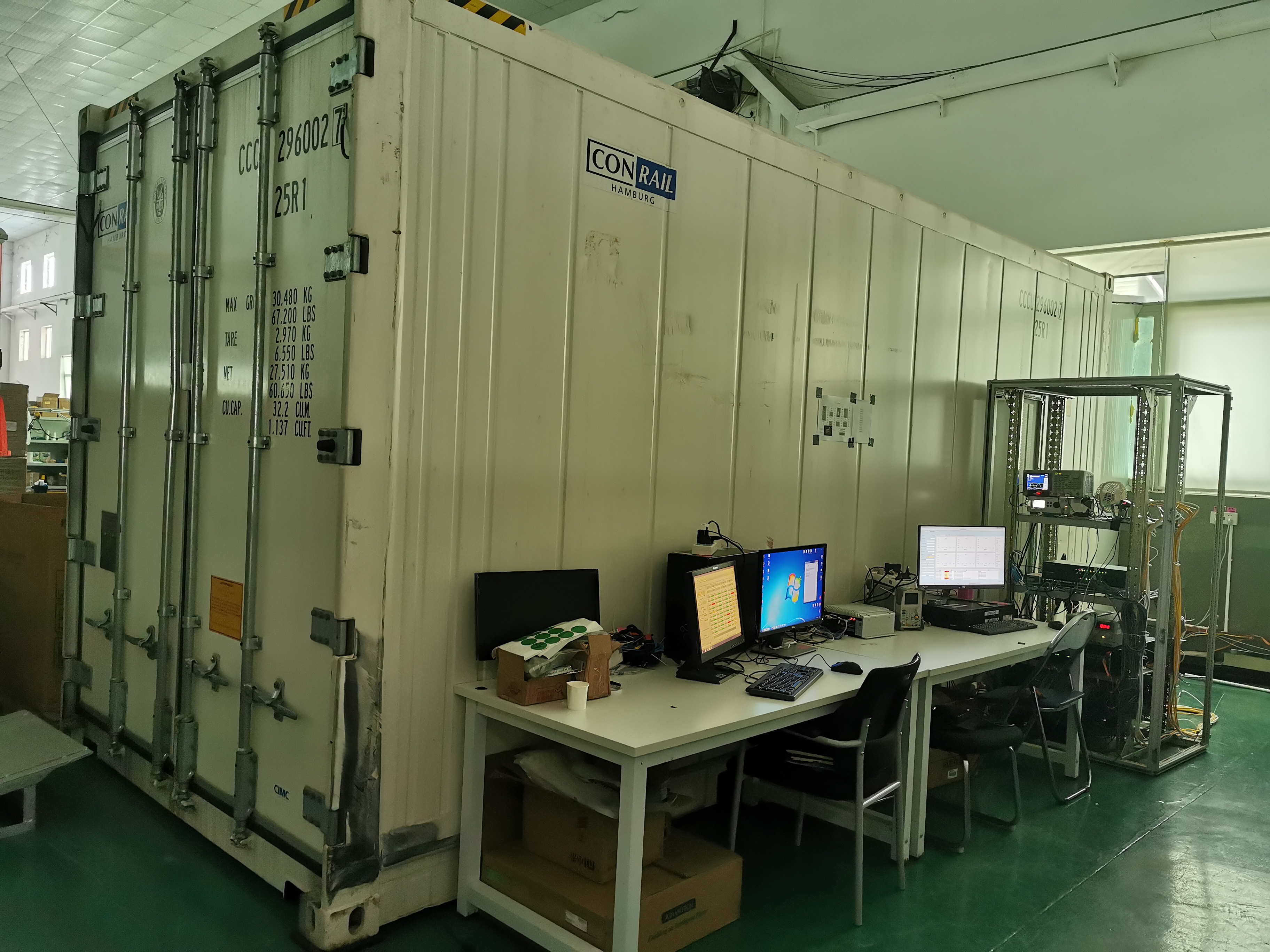}}
     \newline	
       \subfloat[Drawer box]{\label{fig:drawerbox}\includegraphics[width=0.495\textwidth,height=0.3\textwidth]{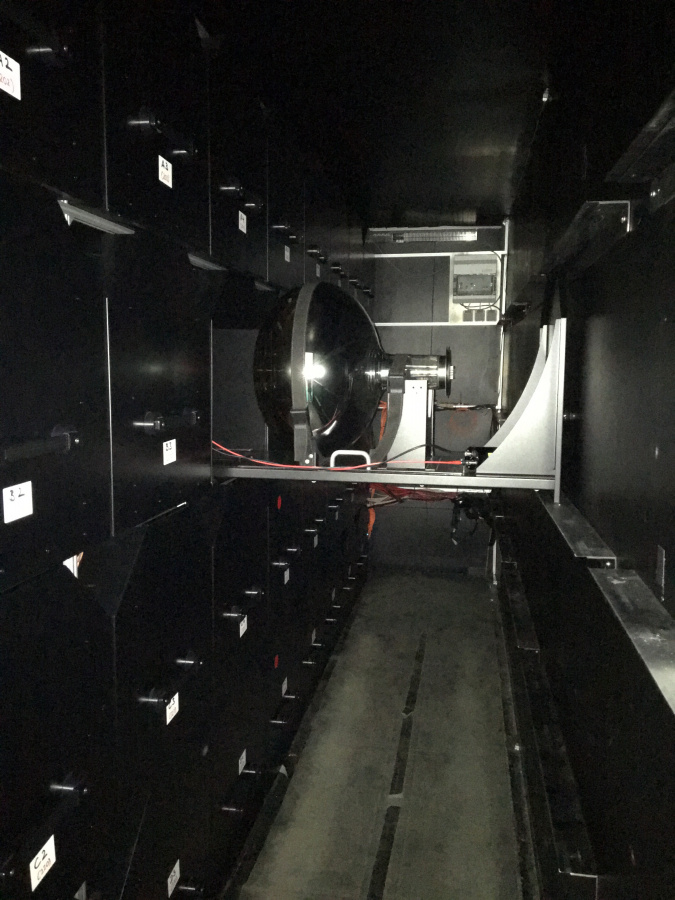}}
      \hfill 	
       \subfloat[Part of the 1F3 Electronics]{\label{fig:1f3elec}\includegraphics[width=0.495\textwidth,height=0.3\textwidth]{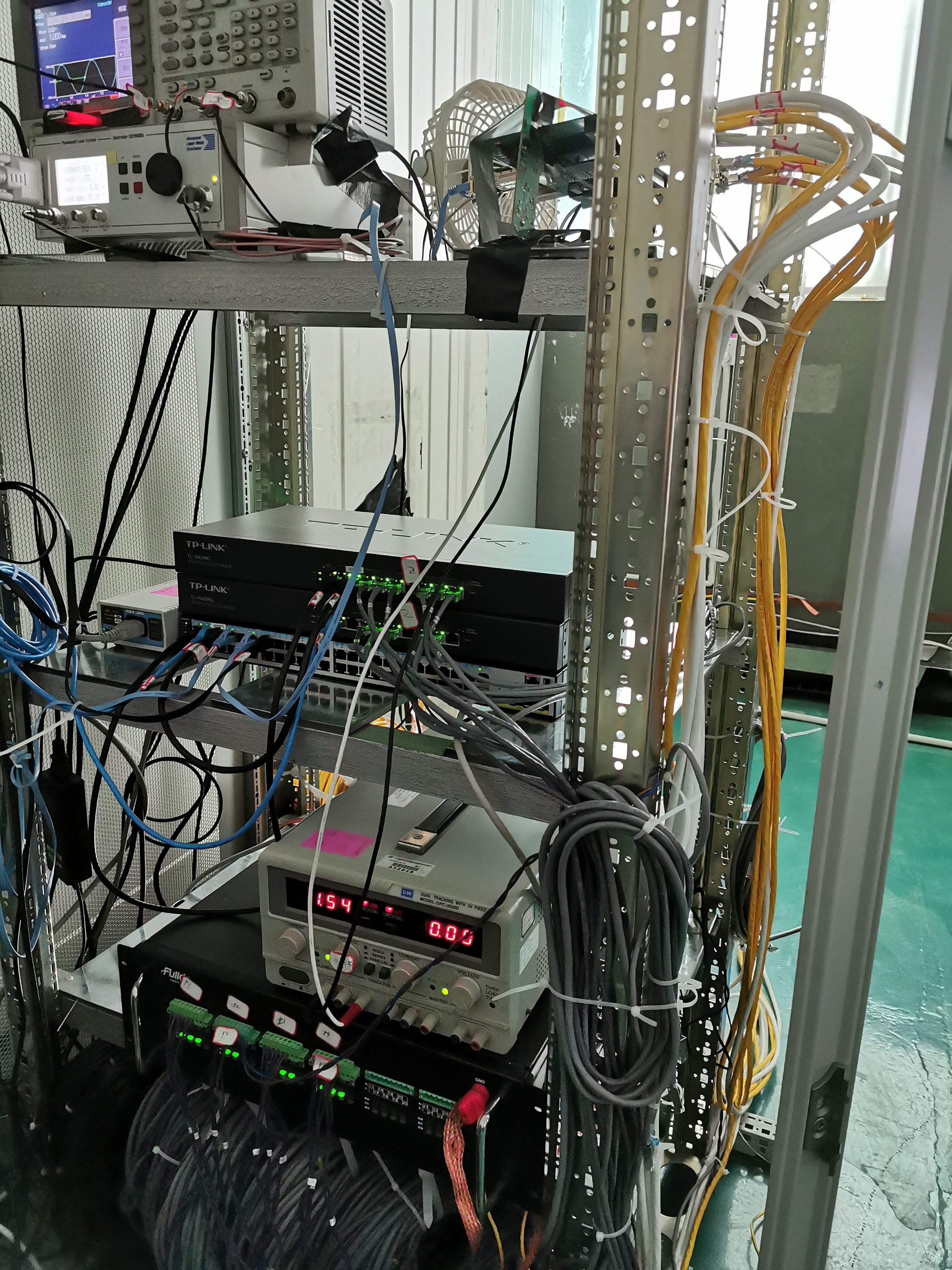}}
   \caption{Container system}
   \label{fig:cs}
 \end{figure*}

The mass testing of these 20-inch PMTs was conducted using a container system (see Figure \,\ref{fig:cs}) \cite{wonsak2021container} at the JUNO Pan-Asia testing and potting station in Zhongshan, China. Containers A and B are equipped with commercial electronics made by CAEN, located outside the containers, containing 36 individual channels (or drawers) each. Conversely, container D, which has 32 channels/drawer, is fitted with JUNO-designed electronics housed within, known as the 1F3 electronics (one electronics box servicing three PMT channels).

For DCR measurements, containers A and B employ a discriminator and scalar to count PMT pulses at a threshold of 3 $\pm$ 1\,mV (around 0.3\,p.e.). In contrast, container D utilizes the 1F3 electronics to measure DCR with a more precise threshold of 2 $\pm$ 0.1\,mV (around 0.25\,p.e.) \cite{Liu:2022nhe}. The DCR is calculated by counting the PMT pulses over a 30-second interval, with results automatically retrieved by the data acquisition system based on the specified configuration.

To maintain stable testing conditions, the containers are equipped with a high-power HVAC (heating, ventilation, and air conditioning) system, which regulates the internal temperature to within 1  $^\circ$C~ across a range of -20 $^\circ$C~to +45 $^\circ$C. Temperature sensors monitor all channels, with periodic readings taken every 60 seconds, achieving an accuracy of 0.25 $^\circ$C. During routine PMT tests, only container D operates under controlled conditions at 22 $^\circ$C\, to ensure that the 1F3 electronics function in an environment similar to that of the JUNO detector, as they generate approximately 10\,W of heat per channel. Containers A and B, however, are mostly operated without the HVAC and are subject to the ambient conditions of the warehouse, typically averaging 25 $^\circ$C~(with fluctuations between 22 $^\circ$C to 26 $^\circ$C).

 \section{Results and Discussions}
 \label{sec:resanddis}

% %\subsection{test plan}

It is well established that the measured DCR is influenced by several factors, including cooling time, environmental conditions, and system gain. The long-term stability of the DCR is crucial for ensuring stable operation. This study aims to comprehensively address the various effects of DCR within the context of the container system.

 \subsection{DCR and Cooling Time}
 \label{sec:time}
 
 \iffalse
 \fi

According to the standard procedure for JUNO PMT testing, it is inevitable that the PMT is exposed to light, even in dim conditions, during the loading and unloading processes. This exposure temporarily increases the measured DCR of the PMT, leading to the necessity of a cooling period to ensure accurate DCR measurement.

 \begin{figure}[!ht]
     \subfloat[DCR over time of NNVT PMTs]{\label{fig:NNVT time effect}\includegraphics[width=0.495\textwidth,height=0.4\textwidth]{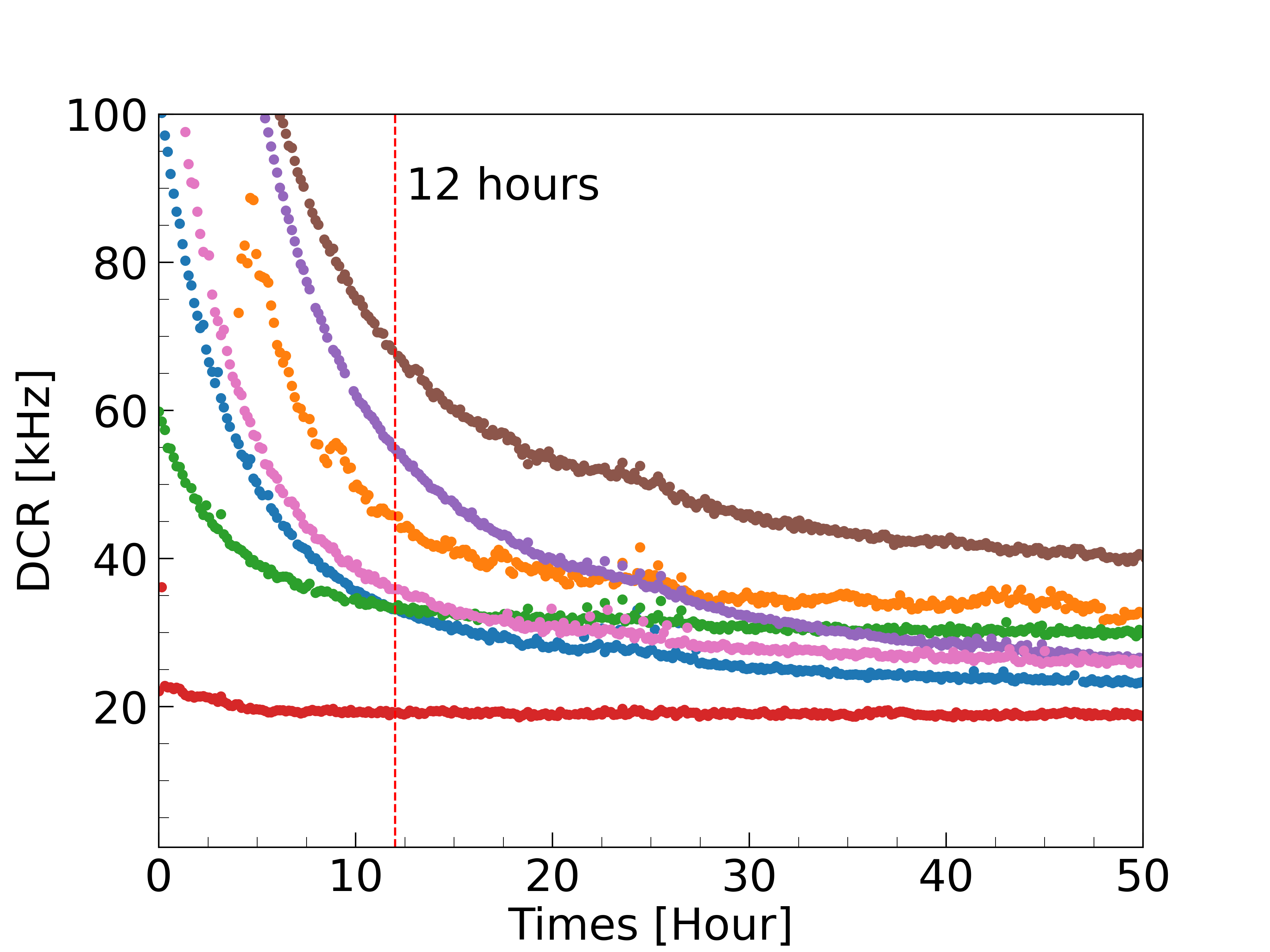}}
    \hfill 	
     \subfloat[DCR over time of HPK PMTs]{\label{fig:HPK time effect}\includegraphics[width=0.495\textwidth,height=0.4\textwidth]{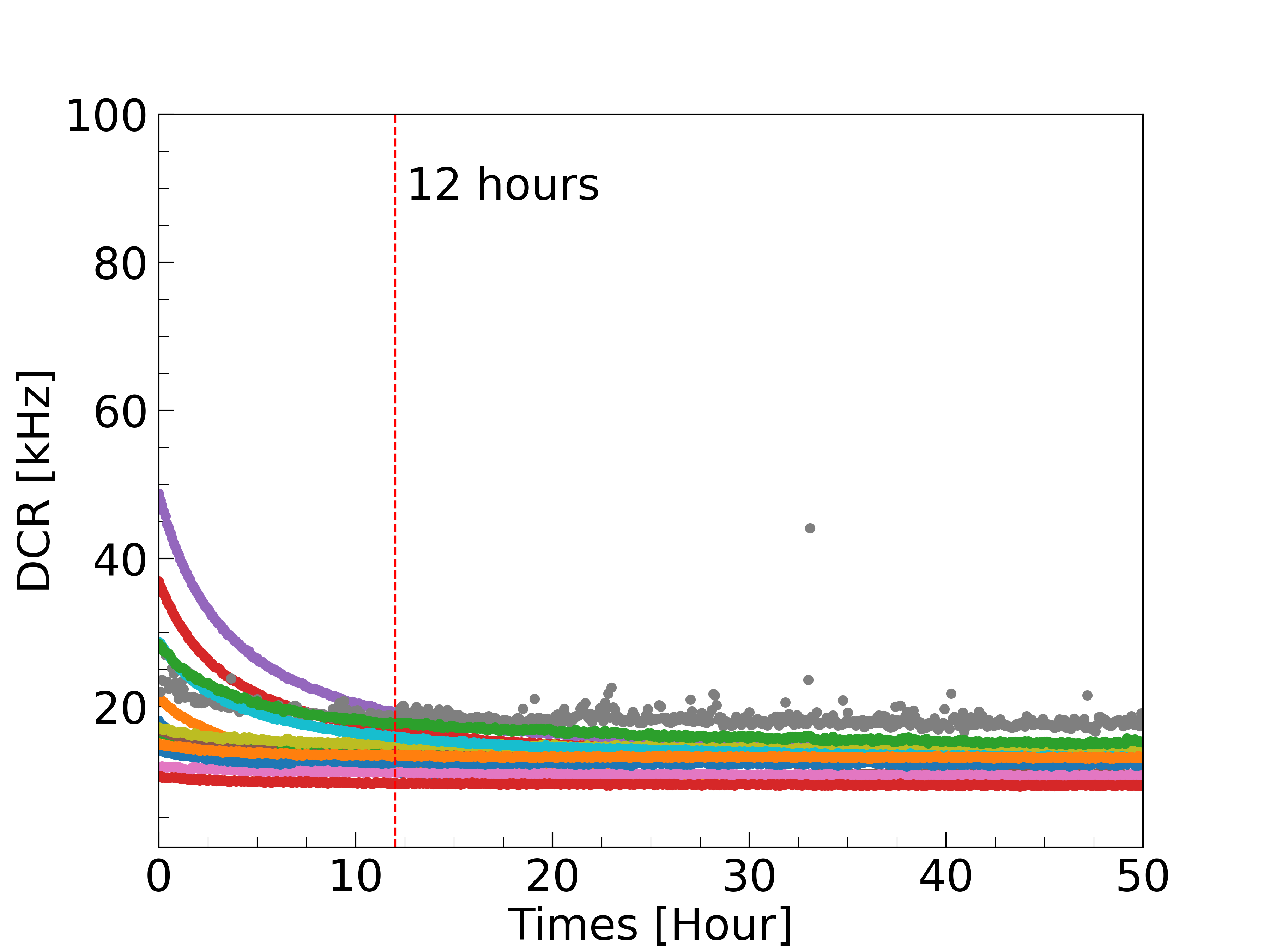}}
   \caption{DCR and\,cooling time}
   \label{fig:timeeffect}
 \end{figure}

The relationship between the DCR and the cooling time of potted 20-inch PMTs is illustrated in Figure \ref{fig:timeeffect}. The monitoring period for the DCR spans approximately 50 hours after the PMTs have been loaded and the high voltage (HV) has been applied.  During this time, all PMTs are subjected to a comparable light intensity. The monitored DCR for each PMT is represented in distinct colors in Figure \ref{fig:NNVT time effect} (NNVT) and Figure \ref{fig:HPK time effect} (HPK). Notably, the initial DCR of most NNVT PMTs is higher than HPK PMTs, which is likely due to the difference in the internal characteristics of the PMTs. The DCR of HPK PMTs stabilizes rapidly, whereas the DCR values among NNVT PMTs exhibit substantial variability. A minimum cooling period of 12 hours is implemented before DCR measurement to facilitate an accurate assessment, while also considering preliminary monitoring and operational time constraints. Furthermore, gaining additional insights into the cooling characteristics of each PMT type could enhance predictions of the DCR over time.\\

% %\textcolor{blue}{The cooling time factor??}

\begin{figure}[!ht]
     \subfloat[Normalized DCR and\,time]{\label{fig:dcrratio}\includegraphics[width=0.495\textwidth,height=0.4\textwidth]{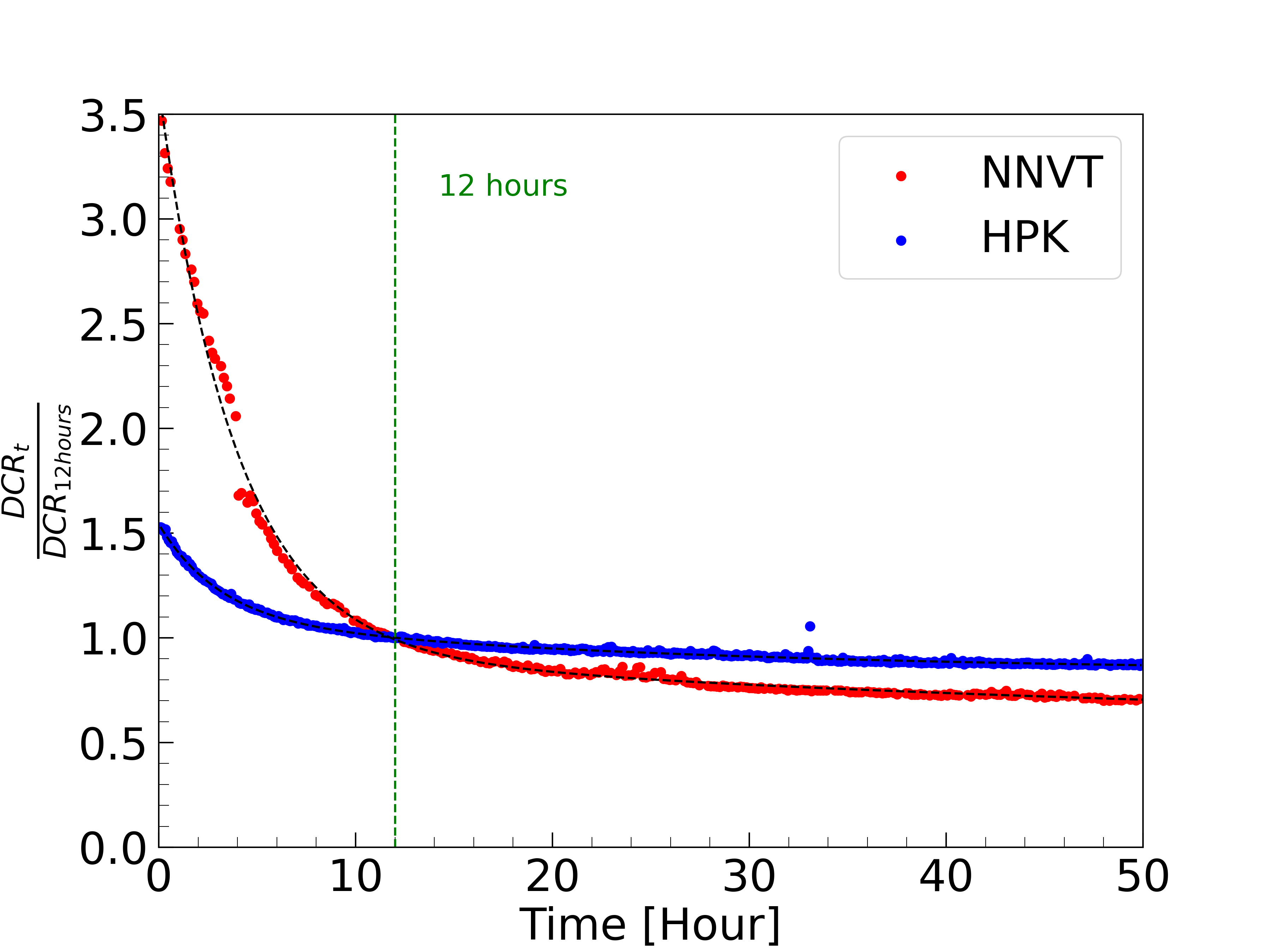}}
    \hfill 	
     \subfloat[DCR of 20-inch PMTs at 12 hours]{\label{fig:bareresult}\includegraphics[width=0.495\textwidth,height=0.4\textwidth]{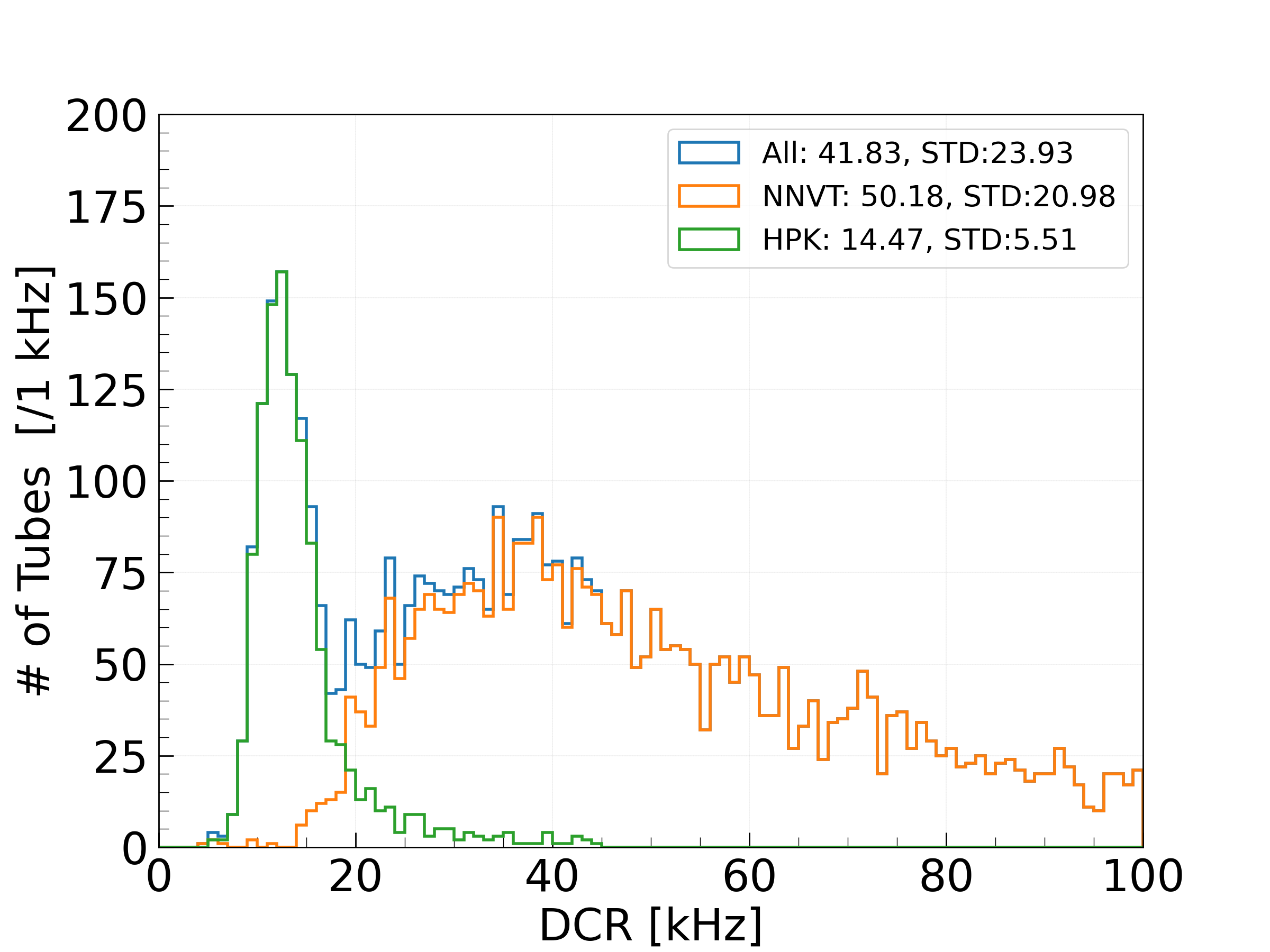}}
   \caption{DCR test results in Pan-Asia}
   \label{fig:dcrresults}
 \end{figure}

The DCR variation ratio is defined based on the tested samples of each PMT type, as illustrated in Figure \ref{fig:dcrratio}. This ratio is normalized to the measurement taken at 12th hour. Both types of PMTs exhibited a rapid decline during the initial 12 hours; however, the rate of decrease slowed after this period.

To analyze the ratio plot shown in Figure \ref{fig:dcrratio}, we applied a fitting function described by:

\begin{equation}
\label{eq:DCR_variation_ratio}
y = \frac{a_{1}}{\tau_1} \cdot e^{-\frac{x}{\tau_1}} + \frac{a_{2}}{\tau_2} \cdot e^{-\frac{x}{\tau_2}} + c
\end{equation}

The fitting function \ref{eq:DCR_variation_ratio} contains two components which represent the cooling factors, light exposure, and thermal effects. The results of this analysis are summarized in Table \ref{table:timeeffectpara}. The parameters $ a_{1} $ and $ a_{2} $ indicate the contributions of the respective components associated with $ \tau_1 $ and $ \tau_2 $. Notably, the component associated with $ \tau_1 $ has a more significant impact than that of $ \tau_2 $. 
%Tobi_S: I'm not entirely sure, what you want to say with the following sentecne. I guess you want to point out, that c is the expected DCR after a time that is long enough to exclude any contribution of external illumination? If so, this sentence needs to be change to point this out. If not, never mind...
The parameter $ c $ denotes the expected DCR when the cooling time is sufficiently extended, though its value is influenced by the specific test sample and the statistics recorded in Table \ref{table:timeeffectpara}.

After 12 hours, the equation \ref{eq:DCR_variation_ratio} related to the rapid factor $ \tau_1 $ (approximately 3-4 hours) has nearly diminished. Additionally, it can be observed that the DCR of the potted NNVT PMT appears to require a longer time (52 hours) to attain a stable status compared to the potted HPK PMTs, which reach stability in 25 hours.

%The ideal DCR of the measured PMTs corresponds to a fraction of the DCR measured after 12 hours of cooling time. These fractions are 56\% in the case of NNVT tubes and 83\% in the case of HPK tubes, respectively.

The ideal DCR of the measured PMTs is determined as a fraction of the DCR measured after 12 hours of cooling time. For NNVT tubes, the ideal DCR corresponds to 56\% of the measured DCR after cooling, while for HPK tubes, it corresponds to 83\% of the measured DCR after the same cooling period.
%For the potted NNVT PMTs, the ideal DCR corresponds to approximately 56\% of the value measured at 12 hours, while for the potted HPK PMTs, it is around 83\%.

%\begin{figure}[!ht]
%     \centering
%      \includegraphics[width=0.6\textwidth,height=0.5\textwidth]{figure/DCR_change_compare.png}
%   \caption{Normalized DCR and\,time}
%   \label{fig:timeeffect:ratio}
% \end{figure}

 \begin{table}[]
 \centering
 \caption{Parameter distribution of time effect of potted PMTs}
 \label{table:timeeffectpara}
 \begin{tabular}{cccccc}
 \hline
 func. & $a_1$    & $\tau_1$  & $a_2$   & $\tau_2$ & $c$    \\
 \hline
 HPK  & 1.42 & 3.11  & 6.30 & 24.99 & 0.83 \\
 NNVT &10.52  &3.93  & 20.19 &51.74& 0.56 \\
 \hline
 \end{tabular}
 \end{table}

\begin{figure}[!ht]
     \subfloat[Potted NNVT]{\label{fig:potteddcr_compare_nnvt}\includegraphics[width=0.495\textwidth,height=0.4\textwidth]{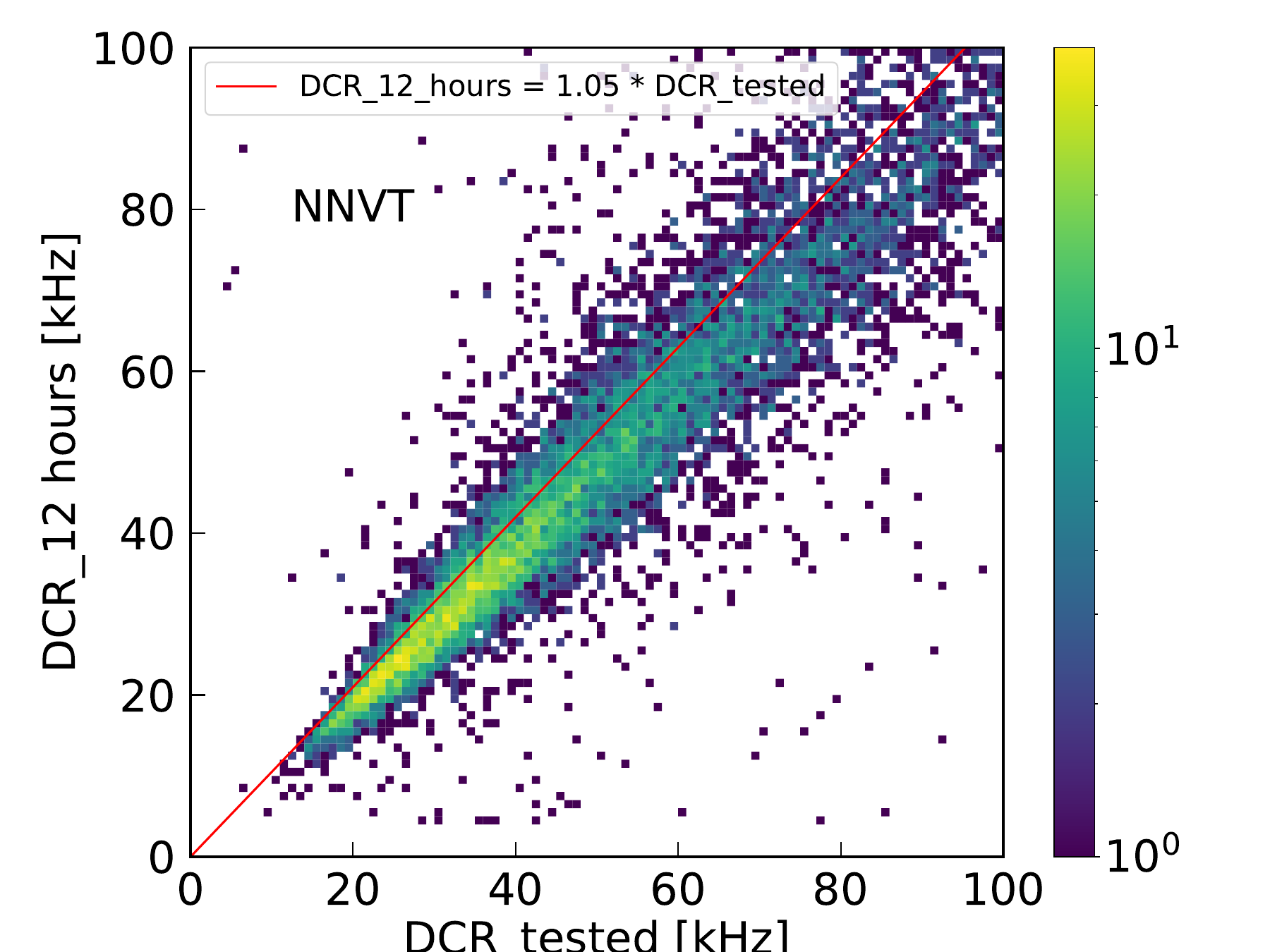}}
    \hfill 	
    \subfloat[Potted HPK]{\label{fig:potteddcr_compare_hpk}\includegraphics[width=0.495\textwidth,height=0.4\textwidth]{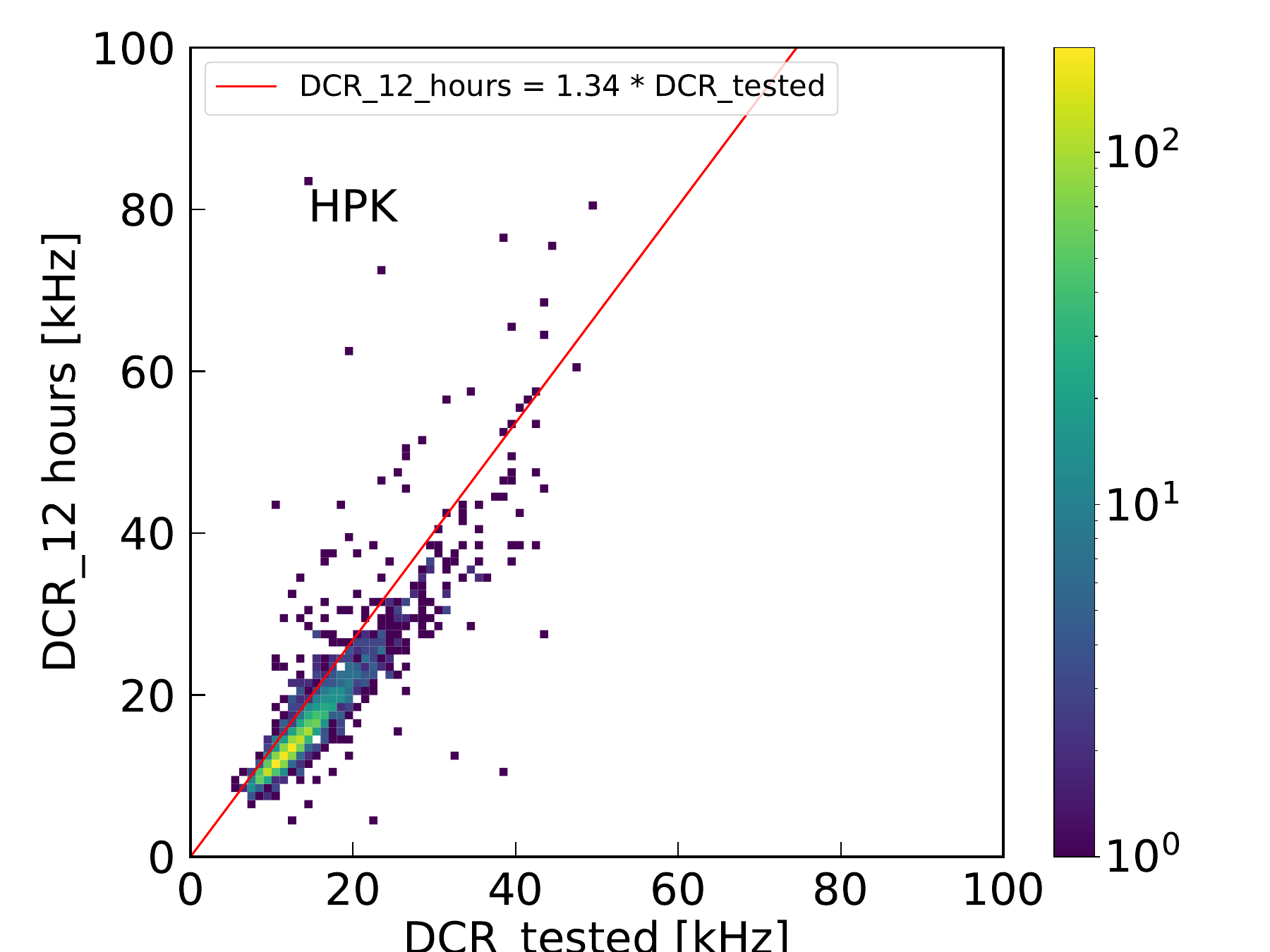}}
  \caption{DCR comparison between potted NNVT and HPK PMTs, which the rates at 12h are in general higher than the rates at "tested"} %\textcolor{red}{The rates at 12h are in general higher than the rates at "tested". The reason is from the tests with light between 12h and "tested", which will increase the measured DCR as discussed in the content.}}
   \label{fig:dcrcomare}
 \end{figure}

%In addition to the DCR measurement after 12 hours, another measurement after 16 hours was performed. In between the two measurements, other parameters, e.g., the PDE have been measured. 
In the additional to the 12 hours DCR measurement, the DCR also was measured after 16 hours, which other parameters and the PDE also were analyzed during this measurement as well. However, since the measurement of these parameters included the use of a light source operated in the single photon regime, an increase of the in DCR in the second measurement was observed. Specifically, the increase in DCR for the NNVT is approximately 5 $\pm$ 2\%, while for the HPK, it is about 34 $\pm$ 3\%. A comparison between the two measurements at 12 hours can be seen in Figure\,\ref{fig:dcrcomare}.\\
%The DCR of the same PMT was measured after cooling periods of 12 hours and 16 hours. During the PMT characterization, the device was exposed to LED flashing at a Single Photoelectron (SPE) level for approximately 3 hours. The results of the DCR measurements are presented in Figure \,\ref{fig:dcrcomare}.

%It was observed that the DCR measured at 16 hours is higher than that at 12 hours, which can be attributed to the exposure to the LED. Specifically, the increase in DCR for the NNVT is approximately 5\%, while for the HPK, it is about 34\%.
 %

\begin{figure}[th]
     \subfloat[NNVT PMTs]{\label{fig:dcrcomparennvt}\includegraphics[width=0.495\textwidth,height=0.4\textwidth]{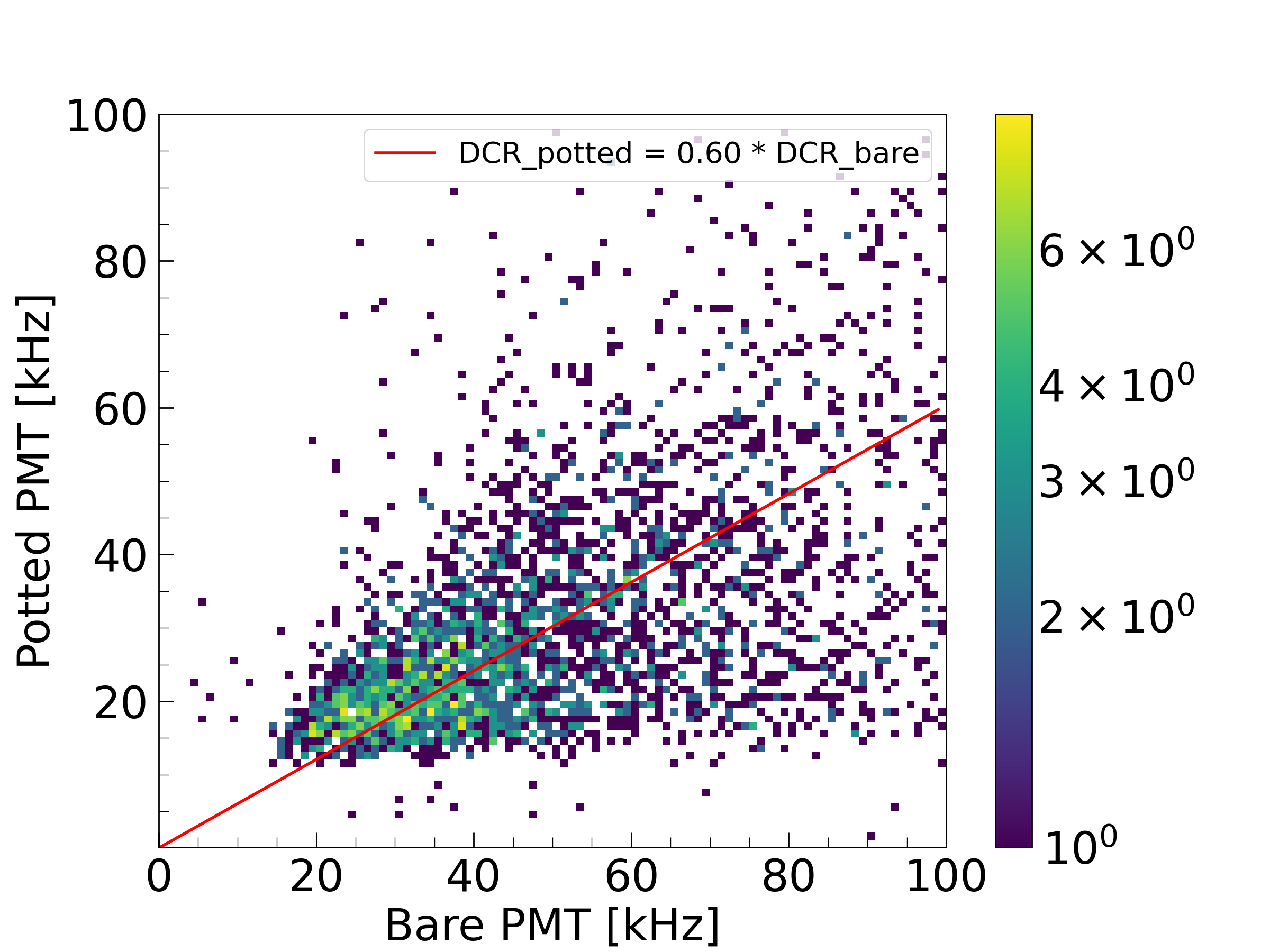}}
     \hfill
     \subfloat[HPK PMTs]{\label{fig:dcrcomparehpk}\includegraphics[width=0.495\textwidth,height=0.4\textwidth]{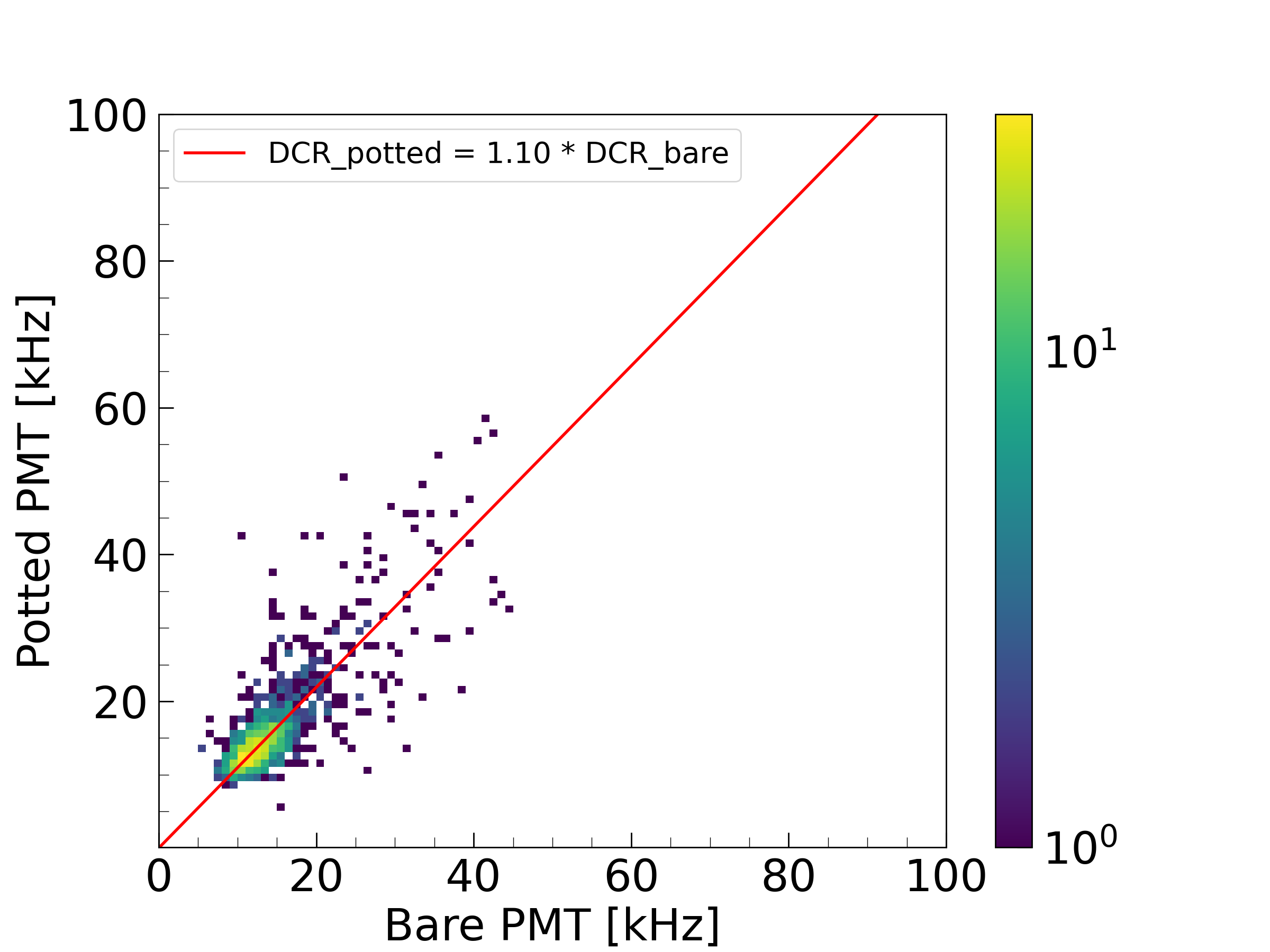}}
   \caption{DCR comparison between bare and potted PMTs}
   \label{fig:dcrxompareresults}
 \end{figure}

% %\textcolor{blue}{(do we keep it or not?)} 
Moreover, the DCR between bare and potted PMTs also was measured and analyzed, which observed a DCR decreasing after PMT's waterproof potting. This PMT's waterproof process indicates a significant difference between the behavior of NNVT and HPK PMTs under similar conditions. For the NNVT PMTs, the average DCR was reduced at least 60\% to 31 $\pm$ 2\,kHz, by tested PMTs after potting with containers A and B. This decrease suggests that the waterproof potting may be effectively mitigating certain noise contributions, associated with the bare NNVT PMTs. In contrast, the HPK PMTs exhibited stability in their mean DCR value, indicating that their design or materials might be inherently less susceptible to the noise factors which affected the NNVT PMTs. The general result of this comparison can be found in Figure \ref{fig:dcrxompareresults}.
%The analysis accounted for various external factors such as noise levels and temperature variations, thus providing a clearer insight into the impact of the waterproof potting process itself. The hypothesis regarding the differences

 \subsection{DCR and~Temperature}
  \label{sec:temperature}

  \iffalse
  \fi

The temperature effect on the DCR was investigated after accounting for the cooling time effect. The study included three measurements including:
\begin{itemize}
    \item Bare PMTs in Container B: Tagged as commercial electronics, this setup included two separate measurements—one with NNVT PMTs and another with HPK PMTs.

    \item Potted PMTs in Container D: Tagged as 1F3 electronics, measurements were conducted simultaneously using different types of PMTs.

    \item Room Temperature Variations in Container A: This setup, also tagged as commercial electronics, investigated the relationship between DCR and variations in room temperature.
\end{itemize}
Note that all measurements incorporated both NNVT and HPK PMTs.
\par During the experiment, the HVAC system of the container was used to manage the temperature inside the containers. The system was designed to maintain a temperature gradient of less than 2~$^\circ$C~per hour within a thermal range of 14~$^\circ$C~to 28~$^\circ$C. Although the monitored temperature of each drawer adjusted rapidly in response to the set points, it was ensured that the temperature remained stable for an additional 2-3 hours at each set point, allowing the PMTs to achieve thermal equilibrium.

For the analysis of the relation between DCR and room temperature, we conducted a detailed review of long-term monitoring data. This was compared against measurements taken during short-term temperature surveys to assess the consistent effects of temperature variation of the DCR.

 \subsubsection{Bare PMTs with Commercial Electronics}
 \label{sec:temperaturebare}

 \iffalse
 \fi
The bare NNVT and HPK tubes were tested at different temperatures in two separate batches. This section will discuss each aspect individually.

% The bare NNVT and HPK PMTs for temperature survey with heating and cooling are tested in two loadings separately, and the results will be also discussed individually.

 %\subsubsection{NNVT}
 \begin{itemize}
     \item{Heating and Cooling of NNVT PMTs}                                             
\par The results of the bare NNVT PMTs are presented in Figure \ref{fig:barennvt}, illustrating the heating and cooling processes. The monitored temperature, as depicted in Figure \ref{fig:barennvt:T}, indicates that a waiting time of 2-3 hours is required for thermal equilibrium at each set point. The DCR observed during the heating and cooling phases is shown separately in Figure \ref{fig:dcrdistributionnnvtheating} and Figure \ref{fig:dcrdistributionnnvtcooling}.

To facilitate comparison, the DCR curve for each PMT was normalized to its DCR value at 21 $^\circ$C and averaged across all samples, as represented in Figure \ref{fig:ratiobarennvt}. From this analysis, it can be concluded that the PMTs exhibit a similar trend during both heating and cooling processes. The maximum and minimum values of DCR variation are marked for each process.

For the bare NNVT PMTs, the average DCR variation is approximately 12\%/$^\circ$C within the measured temperature range, equating to an absolute change of about 6\,kHz/$^\circ$C (as summarized in Table \ref{table:tempeffectparabare}). Notably, the variation in DCR during the heating process is smoother compared to the cooling process. This suggests that the PMTs require more time to achieve thermal equilibrium while cooling, particularly at the highest temperature ( $\sim$27$^\circ$C) observed.

 \begin{figure}[!ht]
    \subfloat[Monitored temperature]{\label{fig:barennvt:T}\includegraphics[width=0.495\textwidth,height=0.4\textwidth]{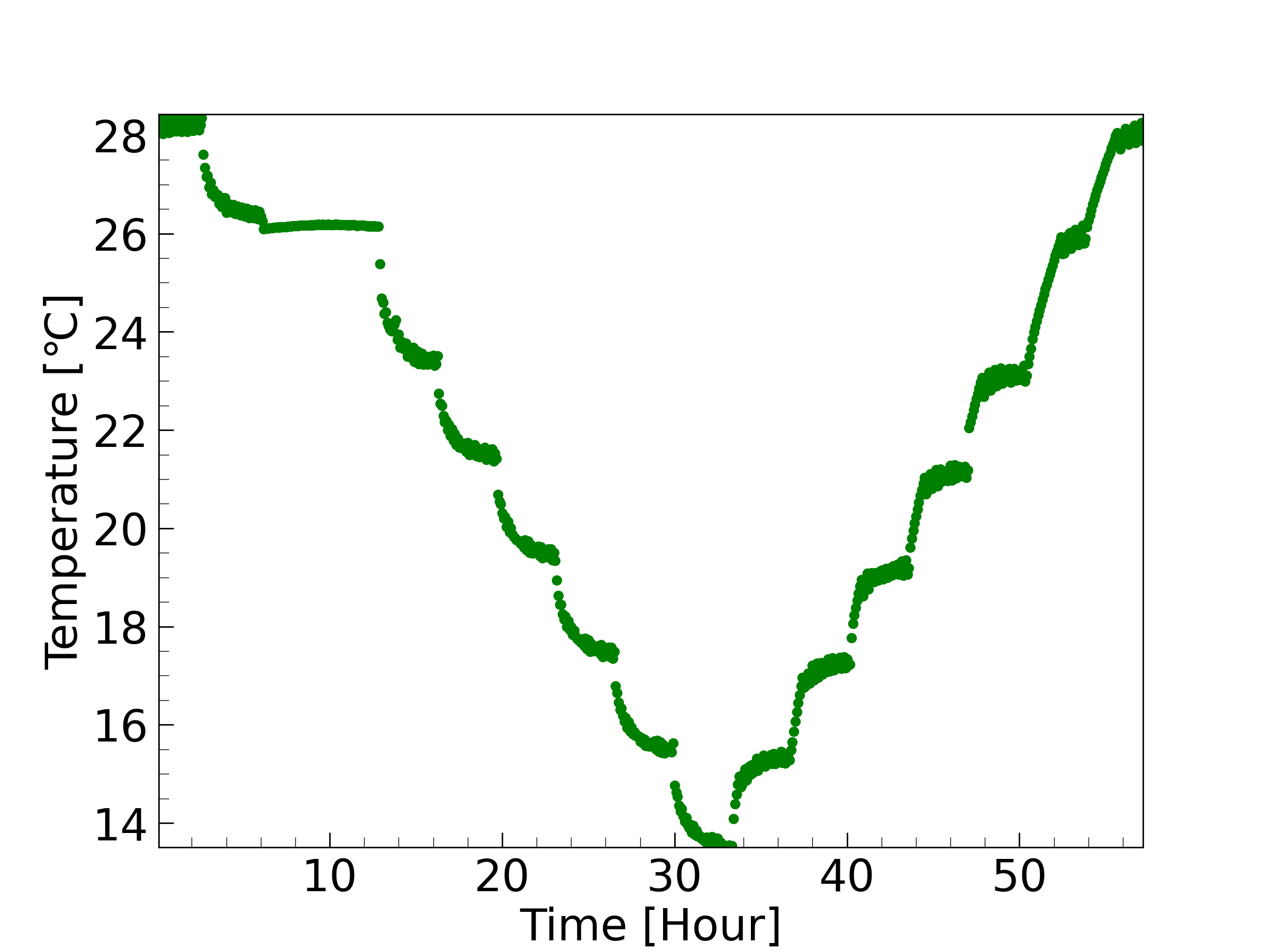}}
   \hfill 	
    \subfloat[DCR and heating]{\label{fig:dcrdistributionnnvtheating}\includegraphics[width=0.495\textwidth,height=0.4\textwidth]{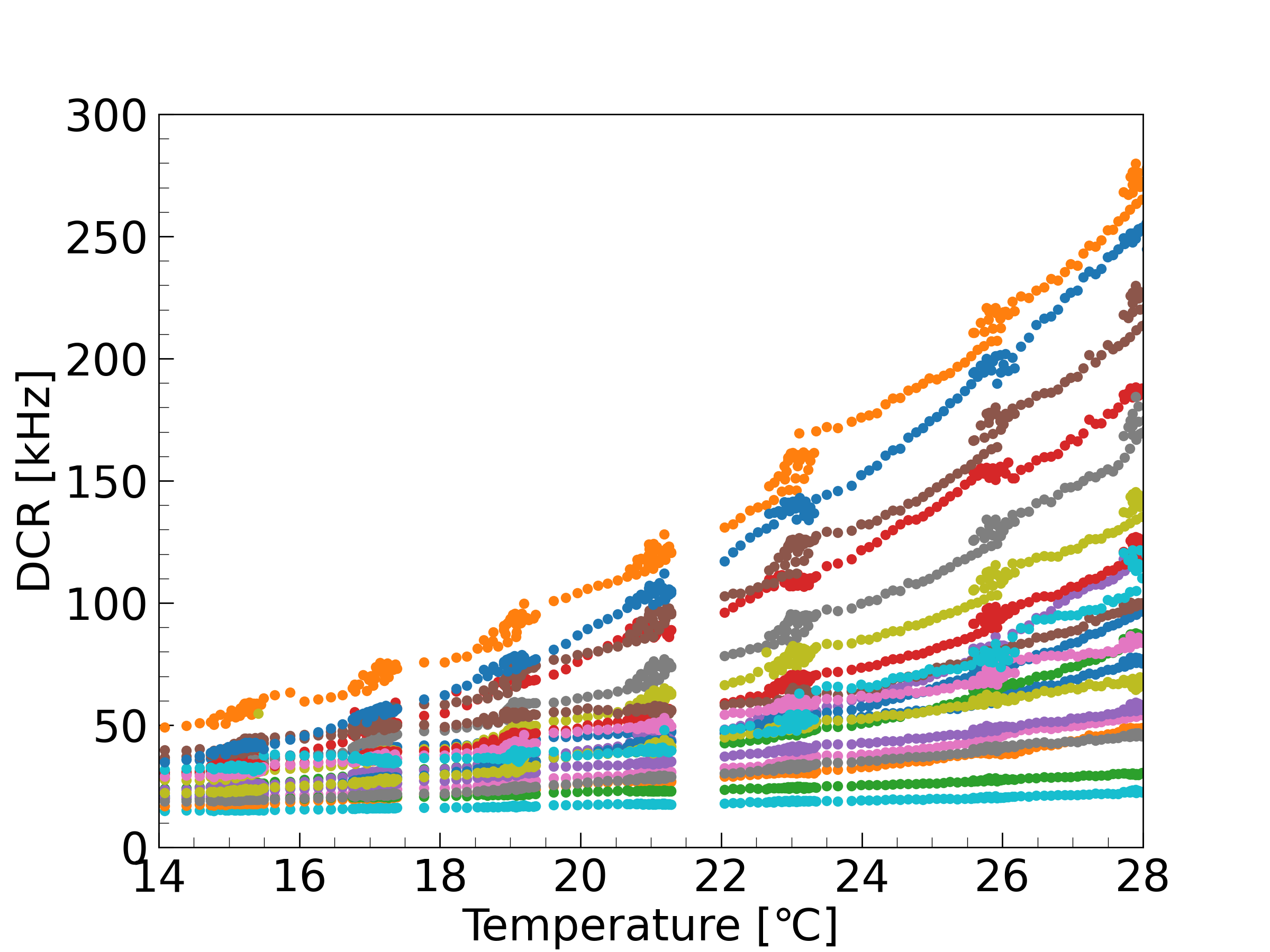}}
    \newline
     \subfloat[DCR and cooling]{\label{fig:dcrdistributionnnvtcooling}\includegraphics[width=0.495\textwidth,height=0.4\textwidth]{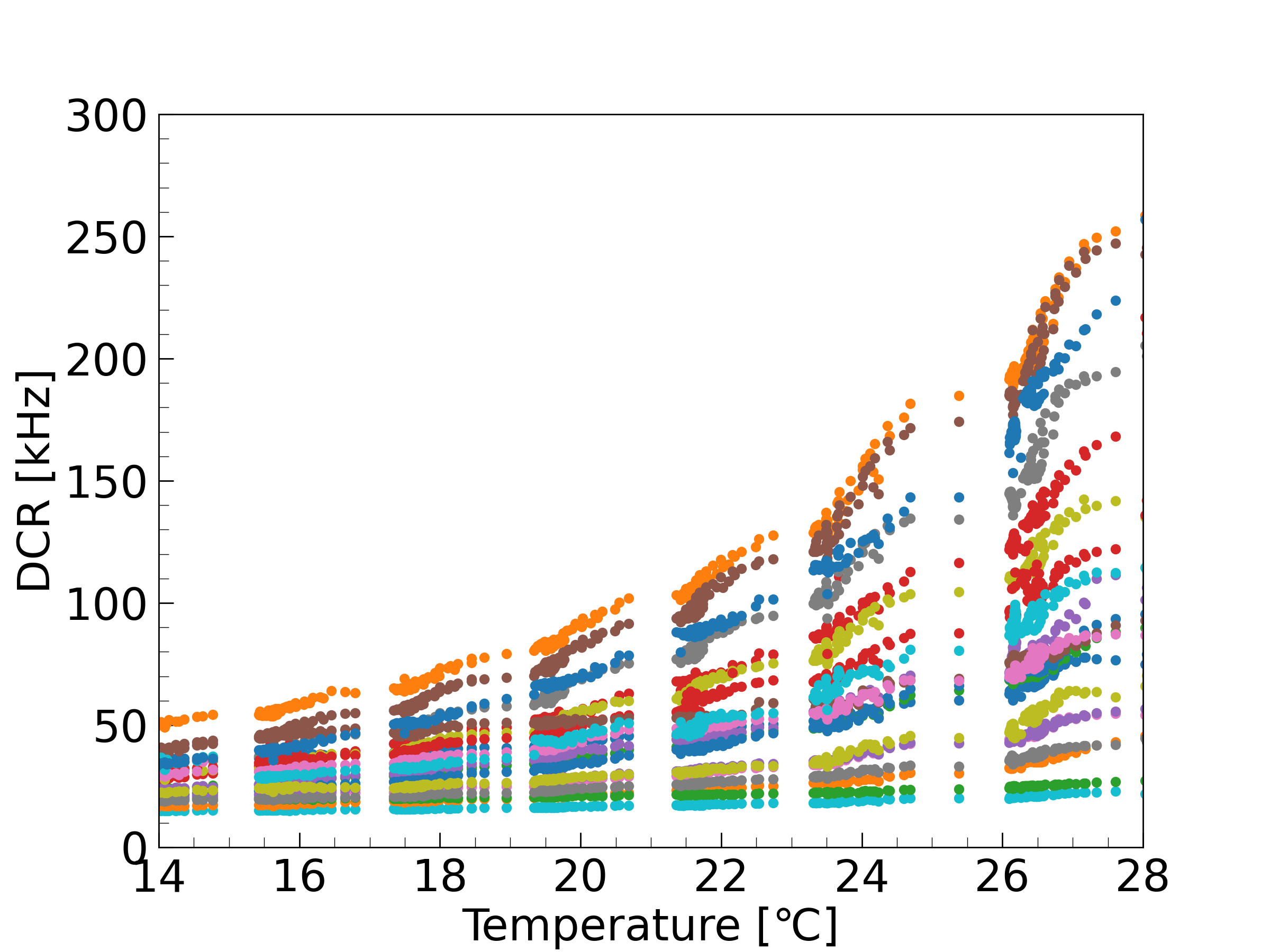}}
    \hfill 	
     \subfloat[DCR ratio relative to 21$^\circ$C~]{\label{fig:ratiobarennvt}\includegraphics[width=0.495\textwidth,height=0.4\textwidth]{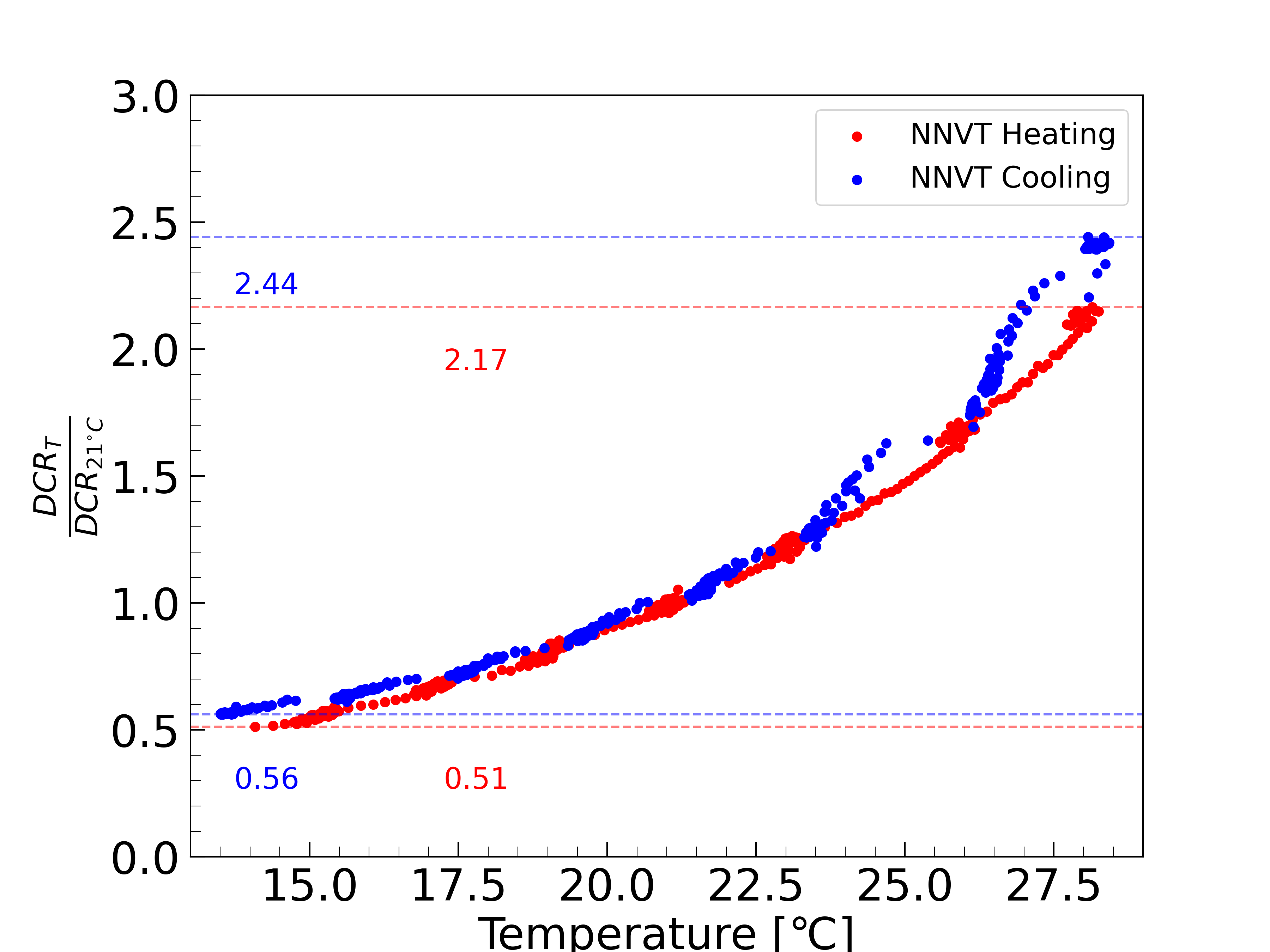}}
   \caption{DCR and heating and cooling of bare NNVT PMTs, where the temperature is measured inside the drawers aside the PMT as discussed in section.\ref{sec:setup}}
   \label{fig:barennvt}
 \end{figure}

%\subsubsection{HPK}
 \item{Heating and cooling of HPK PMTs}
\par A similar measurement was conducted with bare HPK PMTs following the same procedures used for bare NNVT PMTs. The results are presented in Figure \ref{fig:barehpk}. Figure \ref{fig:barehpk:T}, showing the monitored temperature during the cooling and heating phases, while Figure \ref{fig:dcrdistributionhpkheating} and Figure \ref{fig:dcrdistributionhpkcooling} display the monitored DCR as a function of temperature for the two processes, respectively.

In the measured temperature range, the variation of the DCR with respect to temperature is only 1.5\%/$^\circ$C and approximately 0.2\,kHz/$^\circ$C, respectively, as shown in Figure \ref{fig:ratiobarehpk}. In comparison to the bare NNVT PMTs, this variation is much smaller, indicating a more pronounced temperature effect on NNVT PMTs. The detailed results are also summarized in Table \ref{table:tempeffectparabare}.

 \begin{figure}[!ht]
     \subfloat[Monitored temperature]{\label{fig:barehpk:T}\includegraphics[width=0.495\textwidth,height=0.4\textwidth]{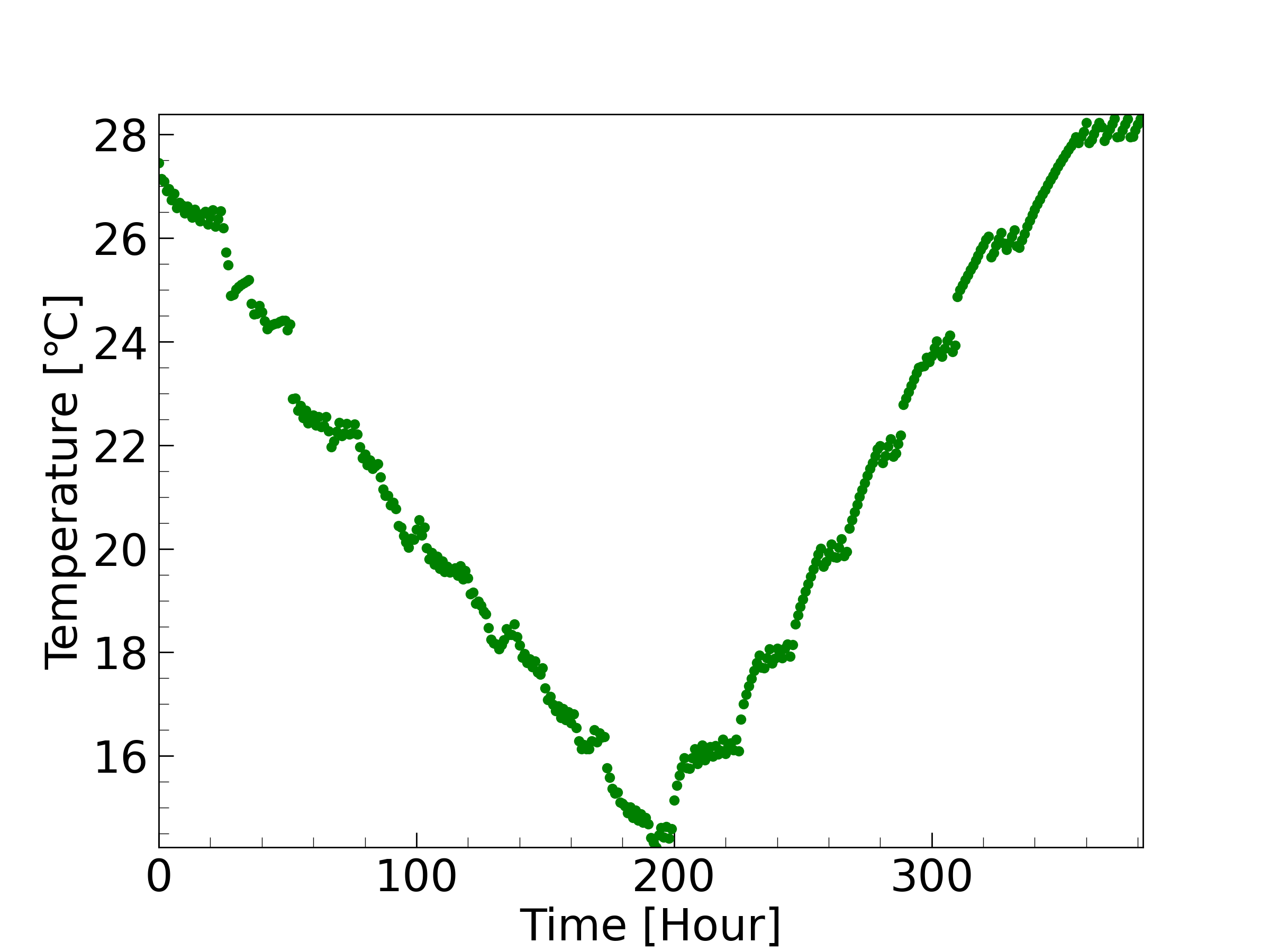}}
    \hfill 	
     \subfloat[DCR and heating]{\label{fig:dcrdistributionhpkheating}\includegraphics[width=0.495\textwidth,height=0.4\textwidth]{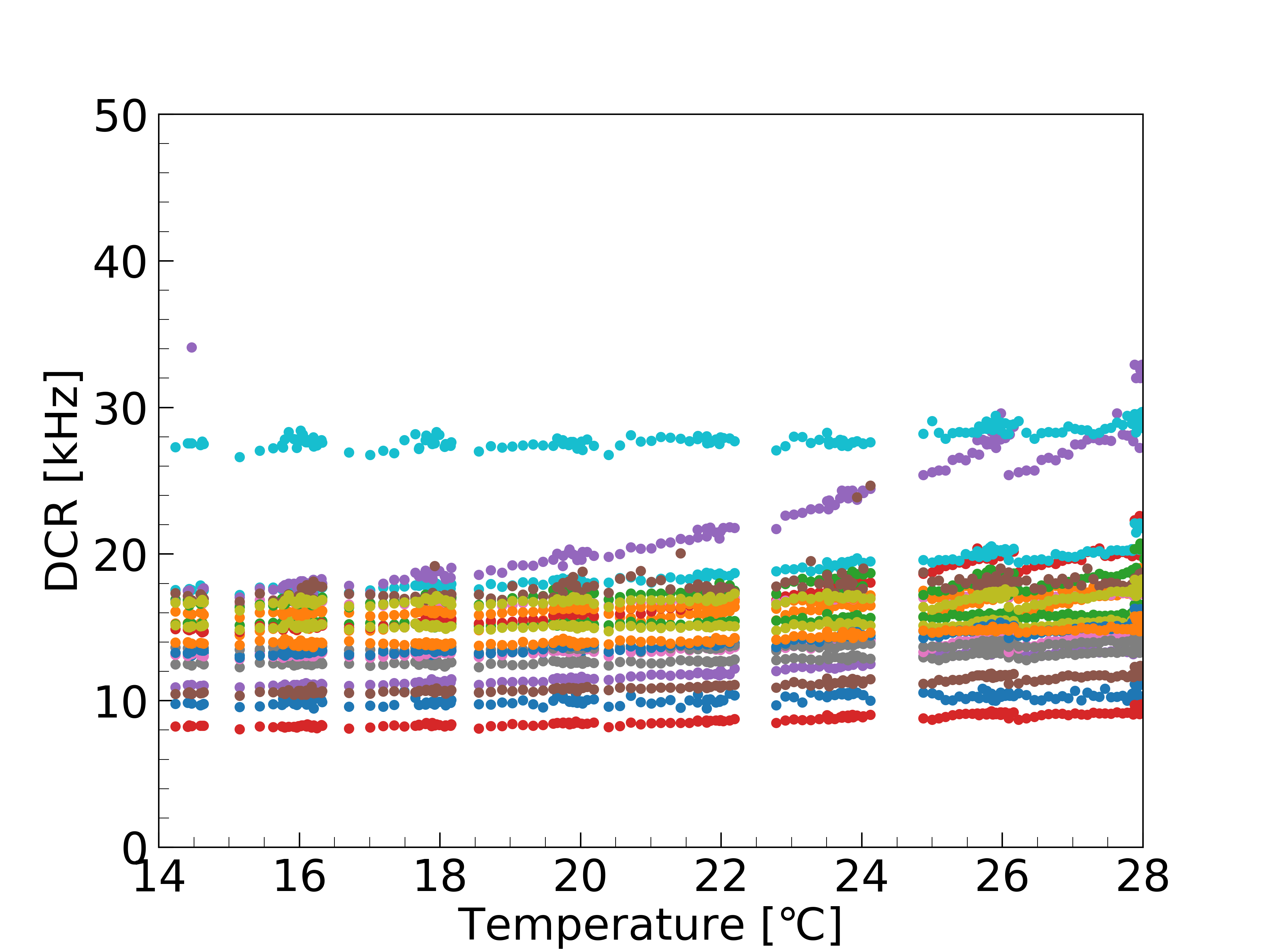}}
    \newline
     \subfloat[DCR and cooling]{\label{fig:dcrdistributionhpkcooling}\includegraphics[width=0.495\textwidth,height=0.4\textwidth]{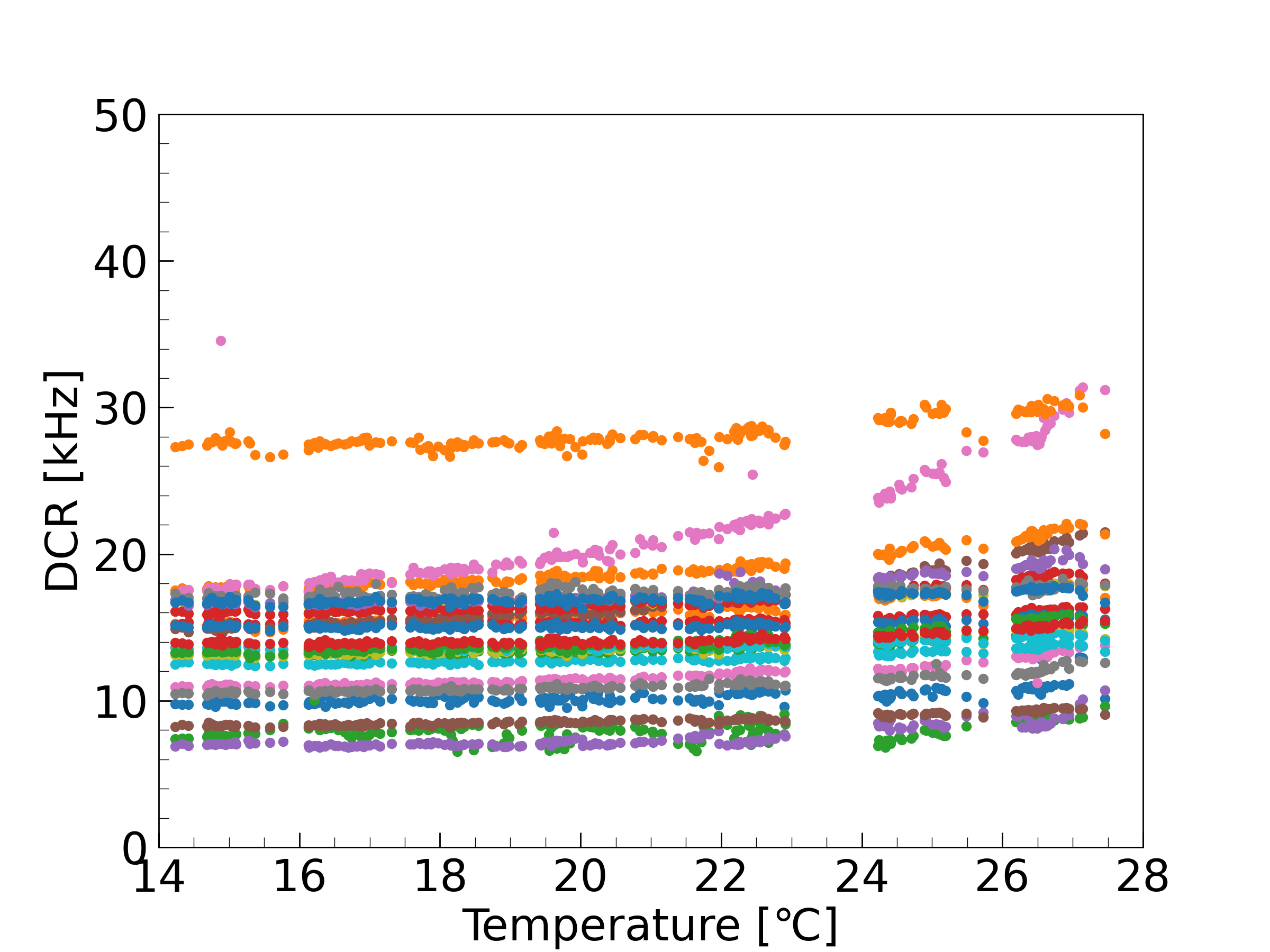}}
    \hfill 	
     \subfloat[DCR ratio relative to 21$^\circ$C~ ]{\label{fig:ratiobarehpk}\includegraphics[width=0.495\textwidth,height=0.4\textwidth]{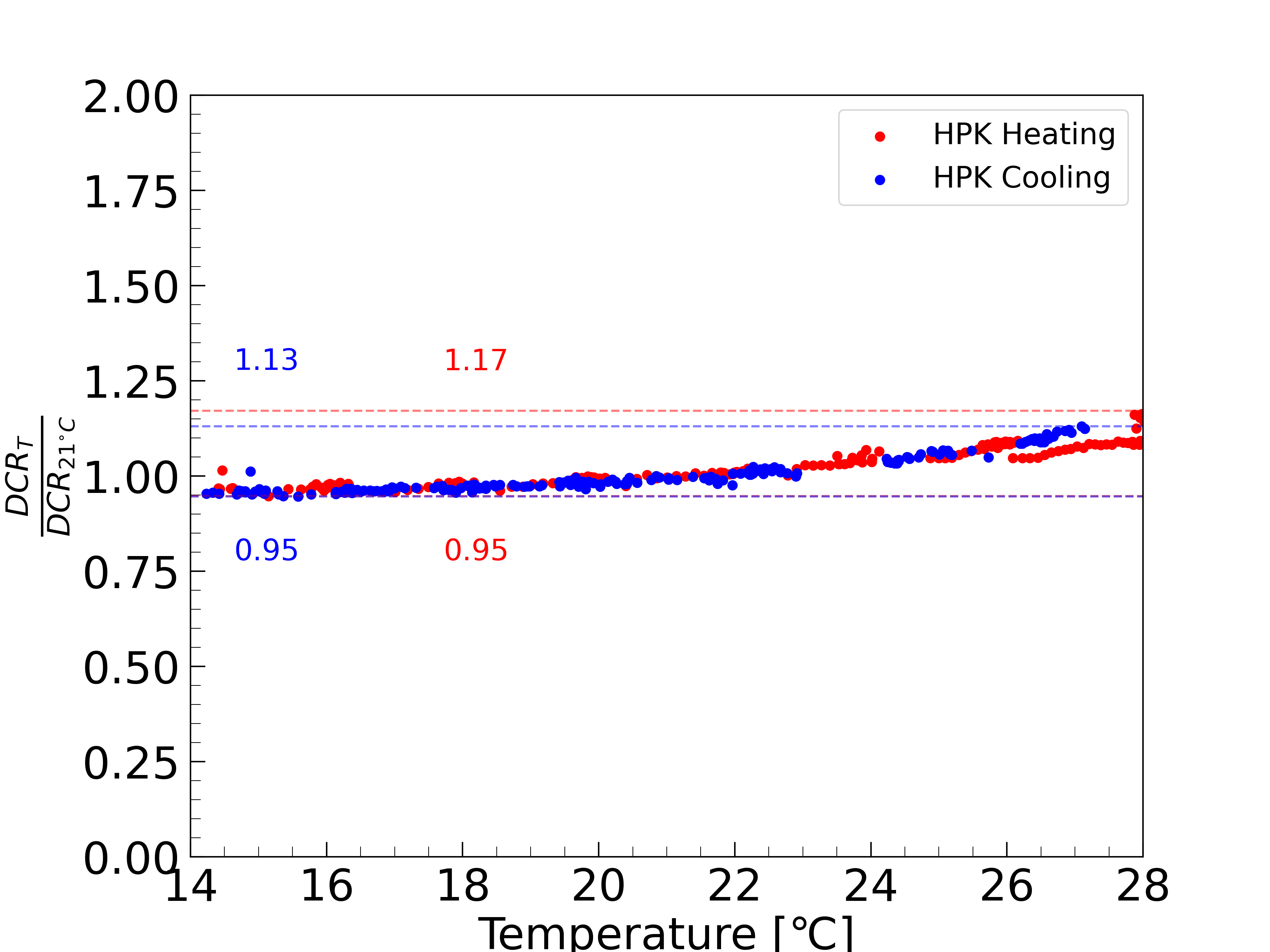}}
  \caption{DCR and heeting and cooling of bare HPK PMTs, where the monitored temperature is measured by sensors inside each drawers aside the PMT as discussed in section \ref{sec:setup}}
  \label{fig:barehpk}
 \end{figure}

 \begin{table}[H]
 \centering
 \caption{Parameters of temperature effect for bare PMTs}
 \label{table:tempeffectparabare}
 \begin{tabular}{cccccc}
 \hline
 PMT                   &         & \multicolumn{2}{c}{DCR/DCR\_21\,$^\circ$C} & Variation Ratio\,{[}\%/$^\circ$C{]} & DCR \,{[}kHz/$^\circ$C{]} \\
                       &         & max.            & min.            &                    &                          \\
                       \hline
 \multirow{2}{*}{HPK}  & heating & 1.17           & 0.95           & 1.60               & 0.24                    \\
                       & cooling & 1.13           & 0.95           & 1.43               & 0.22                    \\
 \multirow{2}{*}{NNVT} & heating & 2.17           & 0.51           & 11.67              & 6.23                    \\
                       & cooling & 2.44           & 0.56           & 12.60              & 6.06                   \\
 \hline
 \end{tabular}
 \end{table}
 
\end{itemize}
 
 \subsubsection{Potted PMT with 1F3 electronics}
\label{sec:temperaturepotted}

\iffalse
\fi

A series of tests on the potted NNVT and HPK PMTs have been conducted, using the 1F3 electronics housing within container D. These tests included a total of 11 HPK PMTs and 15 NNVT PMTs. The temperature was initially raised to 28\,$^\circ$C, which was the starting temperature of the monitoring of the DCR during the subsequent cooling process. To guarantee thermal equilibrium of the PMTs, a slow cooling rate of 1 $^\circ$C every 2 hours was chosen.
%To ensure taht the PMTs have reached thermal equilibrium, we configured the cooling rate to match the heating rate, decreasing the temperature by 1$^\circ$C every 2 hours.

\begin{figure}[!ht]
     \subfloat[Monitored temperature]{\label{fig:temppotted:T}\includegraphics[width=0.495\textwidth,height=0.4\textwidth]{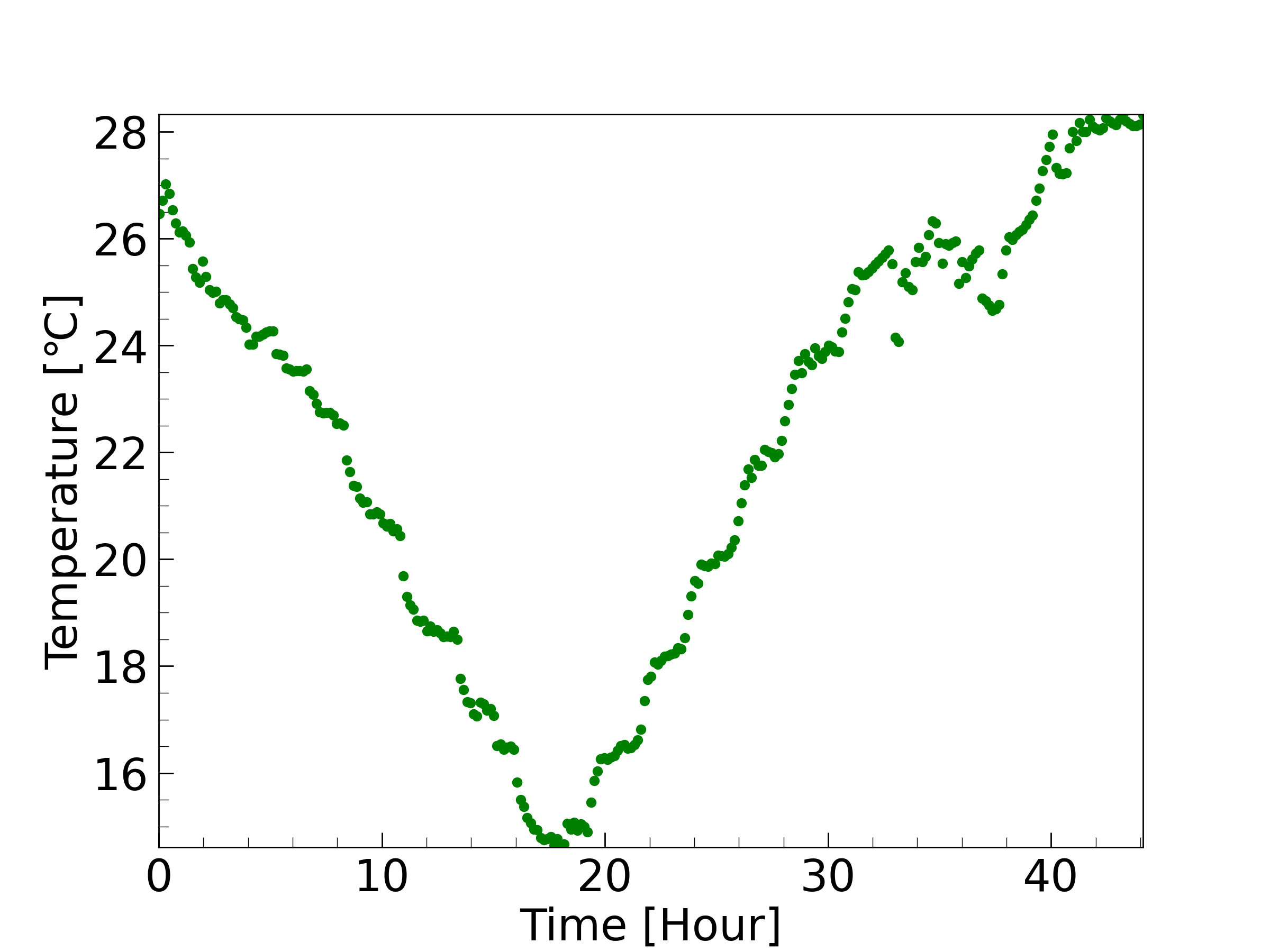}}
    \hfill 	
     \subfloat[DCR and heating]{\label{fig:potteddcr_heating}\includegraphics[width=0.495\textwidth,height=0.4\textwidth]{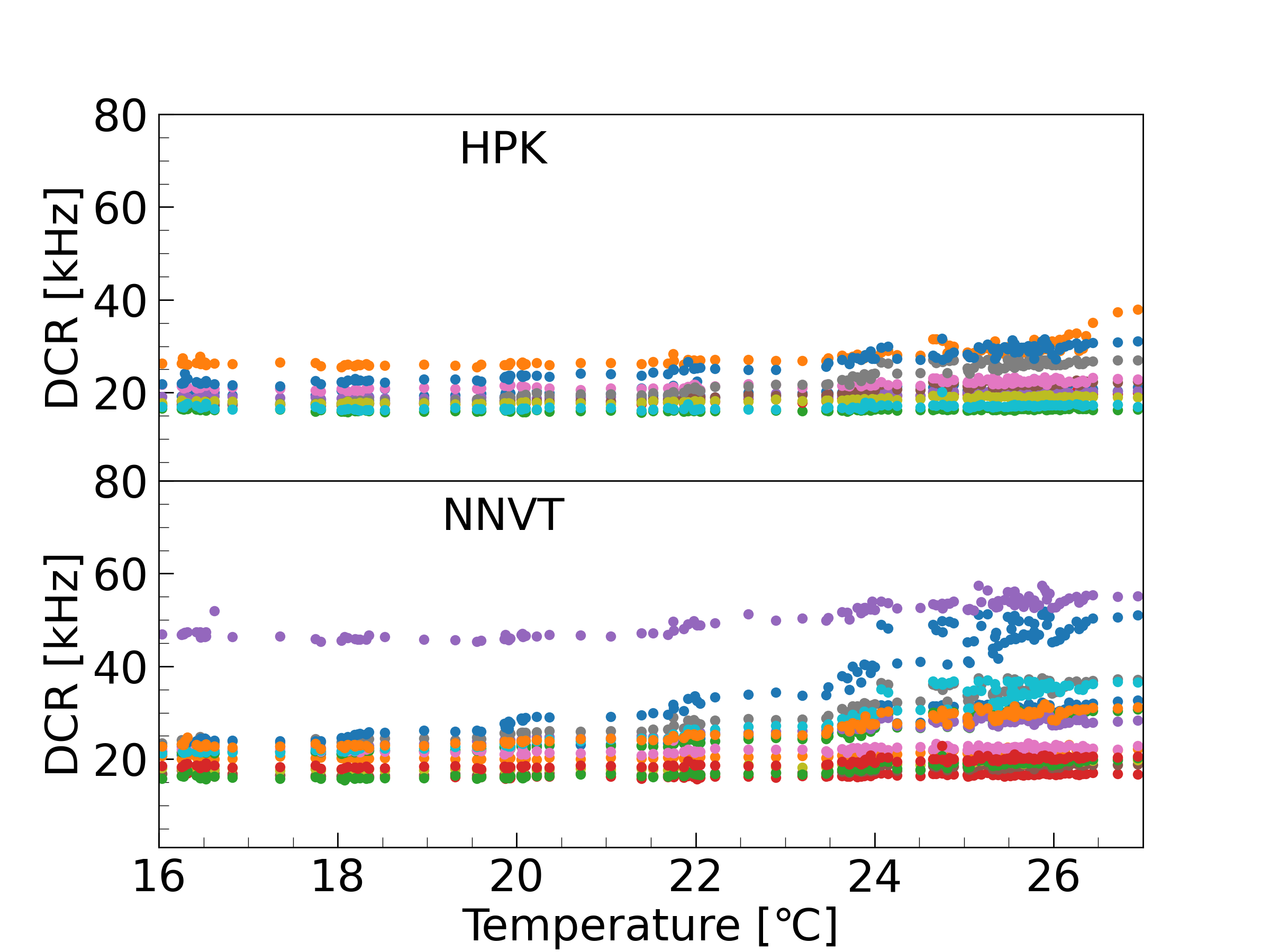}}
     \newline
         \subfloat[DCR and cooling]{\label{fig:potteddcr_cooling}\includegraphics[width=0.495\textwidth,height=0.4\textwidth]{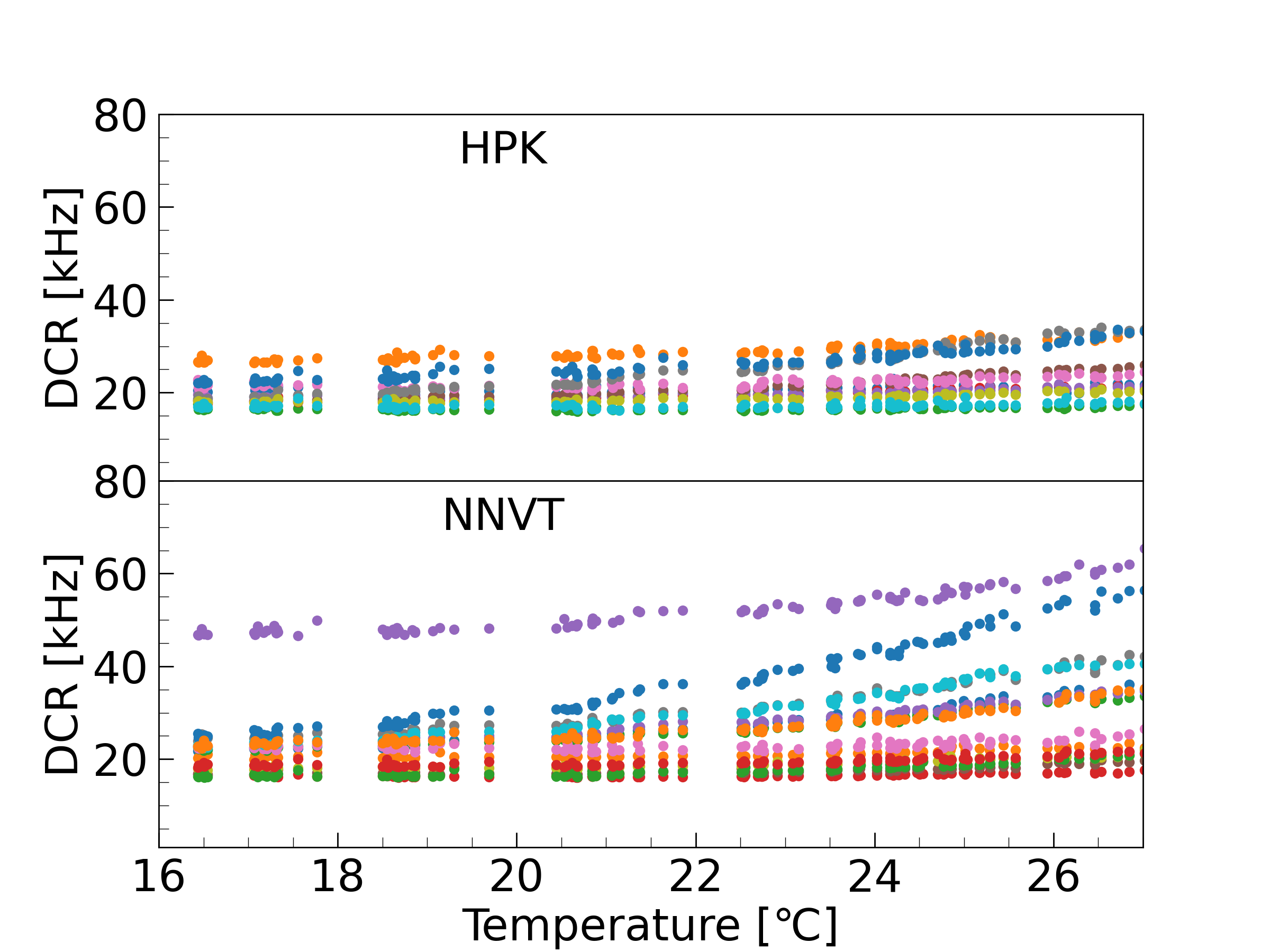}}
   \hfill 	
     \subfloat[DCR ratio relative to 21$^\circ$C~]{\label{fig:potteddcr_ratio_average}\includegraphics[width=0.495\textwidth,height=0.4\textwidth]{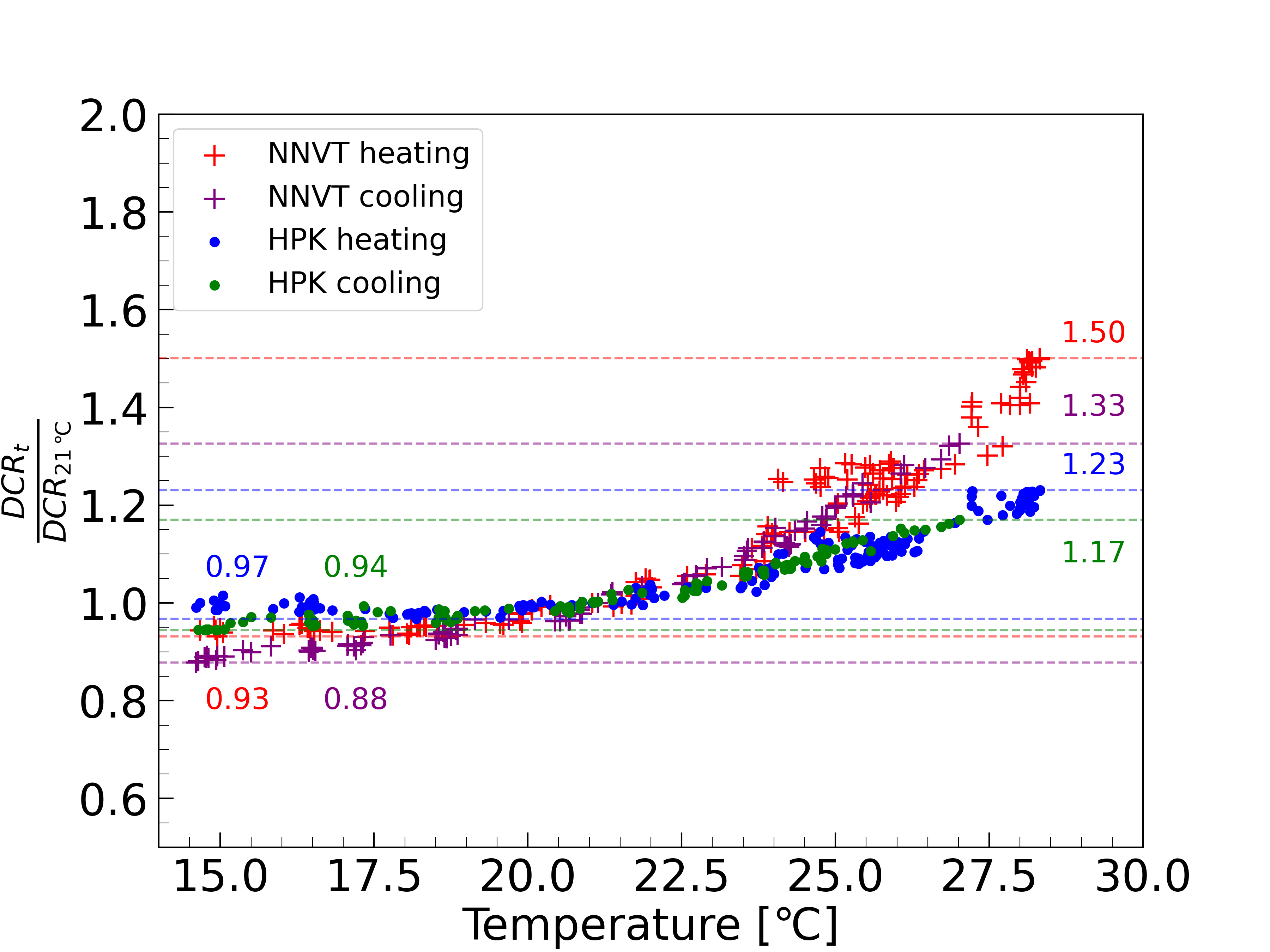}}
   \caption{DCR and heating and cooling process of potted HPK and NNVT PMTs}
   \label{fig:potteddcrtempeffect}
 \end{figure}

Figure \ref{fig:temppotted:T} illustrates the monitored temperature throughout the measurements, showing a cooling range from 28.3 $^\circ$C to 14.6 $^\circ$C, and a subsequent heating range from 14.6 $^\circ$C back to 28.3 $^\circ$C. %Notably, fluctuations were observed between 30 and 40 hours, attributed primarily to variations within the container system.

Figure \ref{fig:potteddcr_heating} and \ref{fig:potteddcr_cooling} display the absolute DCR measured during the heating and cooling processes, respectively. Additionally, Figure \ref{fig:potteddcr_ratio_average} presents the normalized DCR ratio of each PMT compared to their DCR at 21\,$^\circ$C, along with the average of all samples. The highest and lowest ratio are indicated by dashed lines.

In summary, as shown in Table \ref{table:tempeffectparapotted}, the behavior of the potted NNVT and HPK PMTs mirrors that of the bare PMTs during the heating and cooling processes, but with reduced factors. The potted HPK PMTs exhibited a temperature coefficient of approximately 2\%/$^\circ$C, which aligns with the behavior of bare HPK PMTs. In contrast, the potted NNVT PMTs displayed a coefficient of around 4\%/$^\circ$C within the temperature range of 14\,$^\circ$C to 28\,$^\circ$C, notably lower than the bare NNVT PMTs. This disparity might be due to two reasons: 1.) the trend in the temperature range of 14\,$^\circ$C to 21\,$^\circ$C is more stable. 2.) the waiting time for the PMTs to reach thermal equilibrium was shorter in the range of 14\,$^\circ$C to 21\,$^\circ$C. Additionally, the potting appears to contribute to a reduced DCR for the NNVT PMTs, as highlighted in Figure \ref{fig:dcrxompareresults}.
%This disparity may be due to a more stable trend observed in the temperature range of 14\,$^\circ$C to 21\,$^\circ$C, coupled with the shorter duration of the thermal equivalent test—where the factor reflects about 8\% if we consider only the range from 21$^\circ$C to 28$^\circ$C. Additionally, the potting appears to contribute to a reduced DCR for the NNVT PMTs, as highlighted in Figure \ref{fig:dcrxompareresults}.

The parameters gathered from the DCR and temperature tests for the two separate batches of PMTs are summarized in Table \ref{table:tempeffectparapotted}, demonstrating consistent results across both measurements.

 %The results are shown in Figure \,\ref{fig:cooling}. Similar to the heating process, it is about 3\%/$^\circ$C ($\sim$0.76\,kHz/$^\circ$C) for NNVT PMTs and about 2\%/$^\circ$C($\sim$0.36\,kHz/$^\circ$C) during the cooling process. Comparing the Figure \,\ref{fig:heating} and Figure \,\ref{fig:cooling}, the two processes have the same trend within error. And NNVT PMTs show an obvious temperature effect. fitting results also shown as Table.\ref{table:tempeffectpara}

 \begin{table}[!ht]
 \centering
 \caption{Parameters of DCR and temperature of potted PMTs}
 \label{table:tempeffectparapotted}
 \resizebox{0.98\textwidth}{!}{
 \begin{tabular}{cccccccccc}
 \hline
 PMT  & \multicolumn{5}{c}{run 2537}                                              & \multicolumn{4}{c}{run 2561}                                    \\
 \hline
      &         & max.  & min.  & variation {[}\%/$^\circ$C{]} & abs.change {[}kHz/$^\circ$C{]} & max.  & min.  & variation {[}\%/$^\circ$C{]} & abs.change {[}kHz/$^\circ$C{]} \\
 HPK  & heating & 1.23 & 0.93 & 1.92               & 0.37                     & 1.22 & 0.97 & 2.00               & 0.41                     \\
      & cooling & 1.17 & 0.88 & 1.82               & 0.36                     & 1.21 & 0.96 & 2.04               & 0.43                     \\
 NNVT & heating & 1.50 & 0.97 & 3.61               & 0.90                     & 1.42 & 0.90 & 3.61               & 0.76                     \\
      & cooling & 1.33 & 0.94 & 4.17               & 0.83                     & 1.39 & 0.94 & 4.12               & 0.90                    \\
     \hline
 \end{tabular}
 }
 \end{table}

 \subsubsection{DCR and Room Temperature}
 \label{sec:stability:DCRvariation}

\begin{figure}[!ht]
     \subfloat[Monitored DCR and temperature of NNVT PMTs]{\label{fig:NNVT temp effect}\includegraphics[width=0.495\textwidth,height=0.4\textwidth]{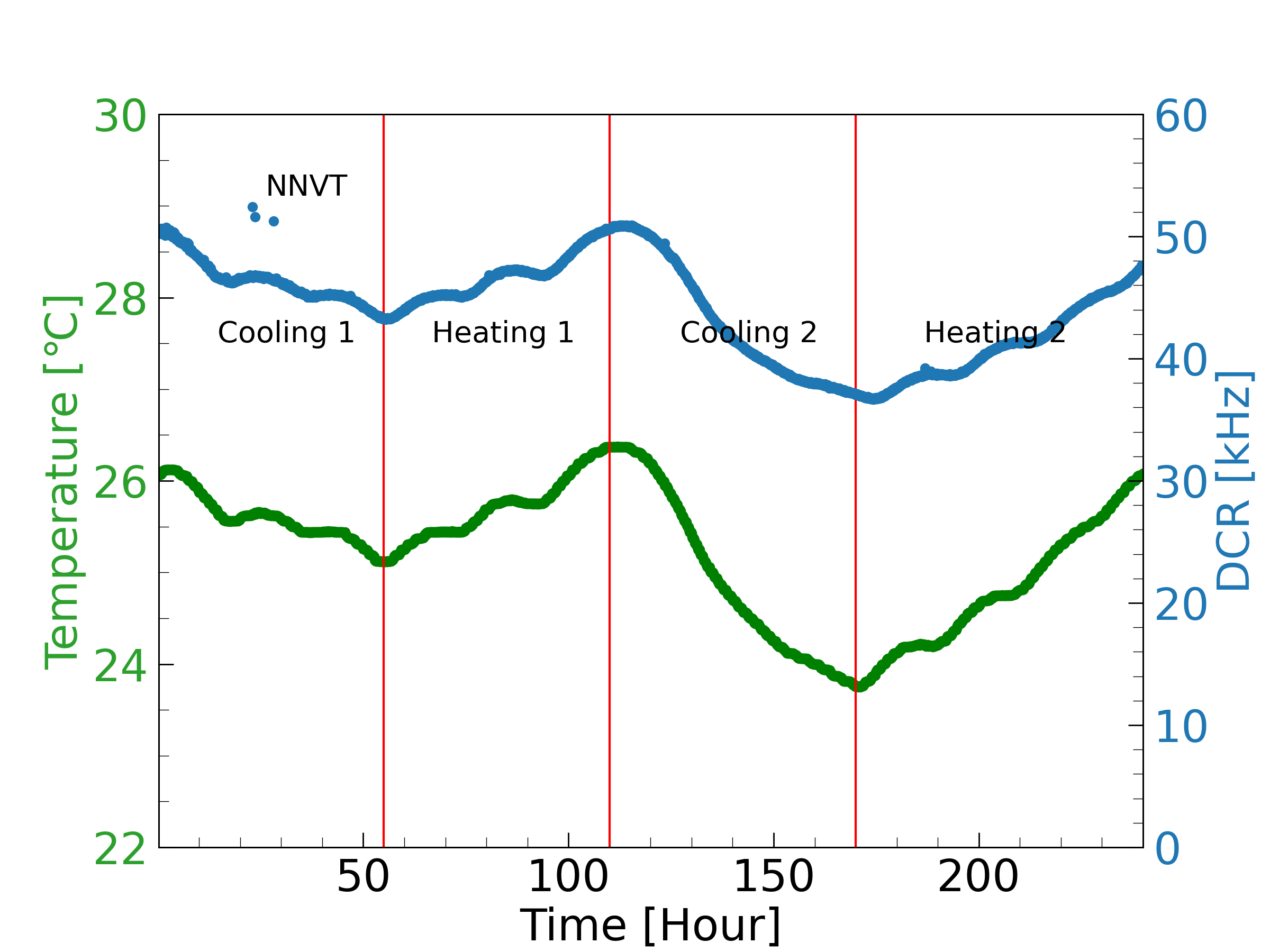}}
    \hfill 	
    \subfloat[Monitored DCR and temperature of HPK PMTs]{\label{fig:HPK temp effect}\includegraphics[width=0.495\textwidth,height=0.4\textwidth]{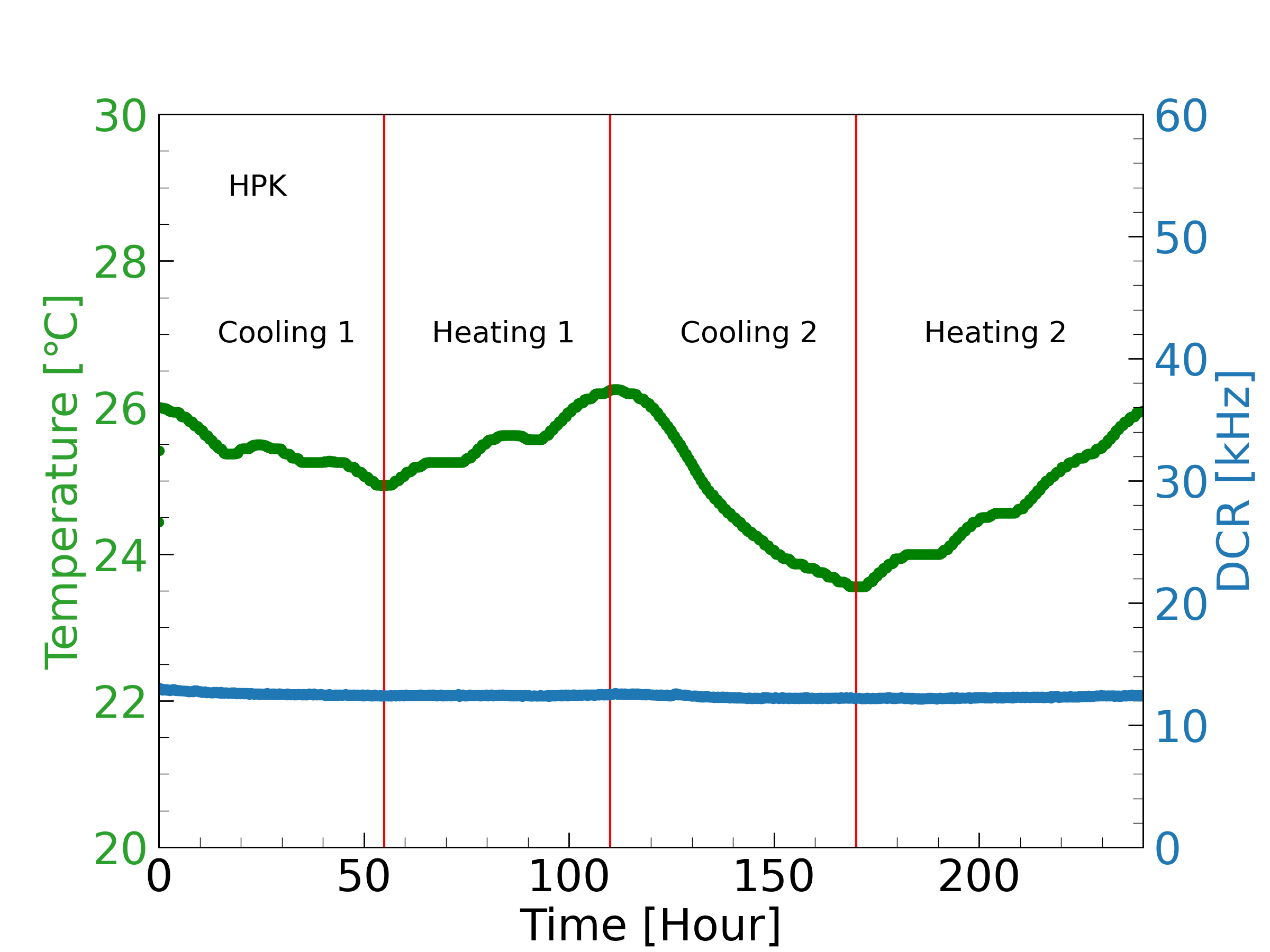}}
     \newline
    \subfloat[DCR and room temperature of NNVT PMTs]{\label{fig:nnvtvsroomtemp}\includegraphics[width=0.495\textwidth,height=0.4\textwidth]{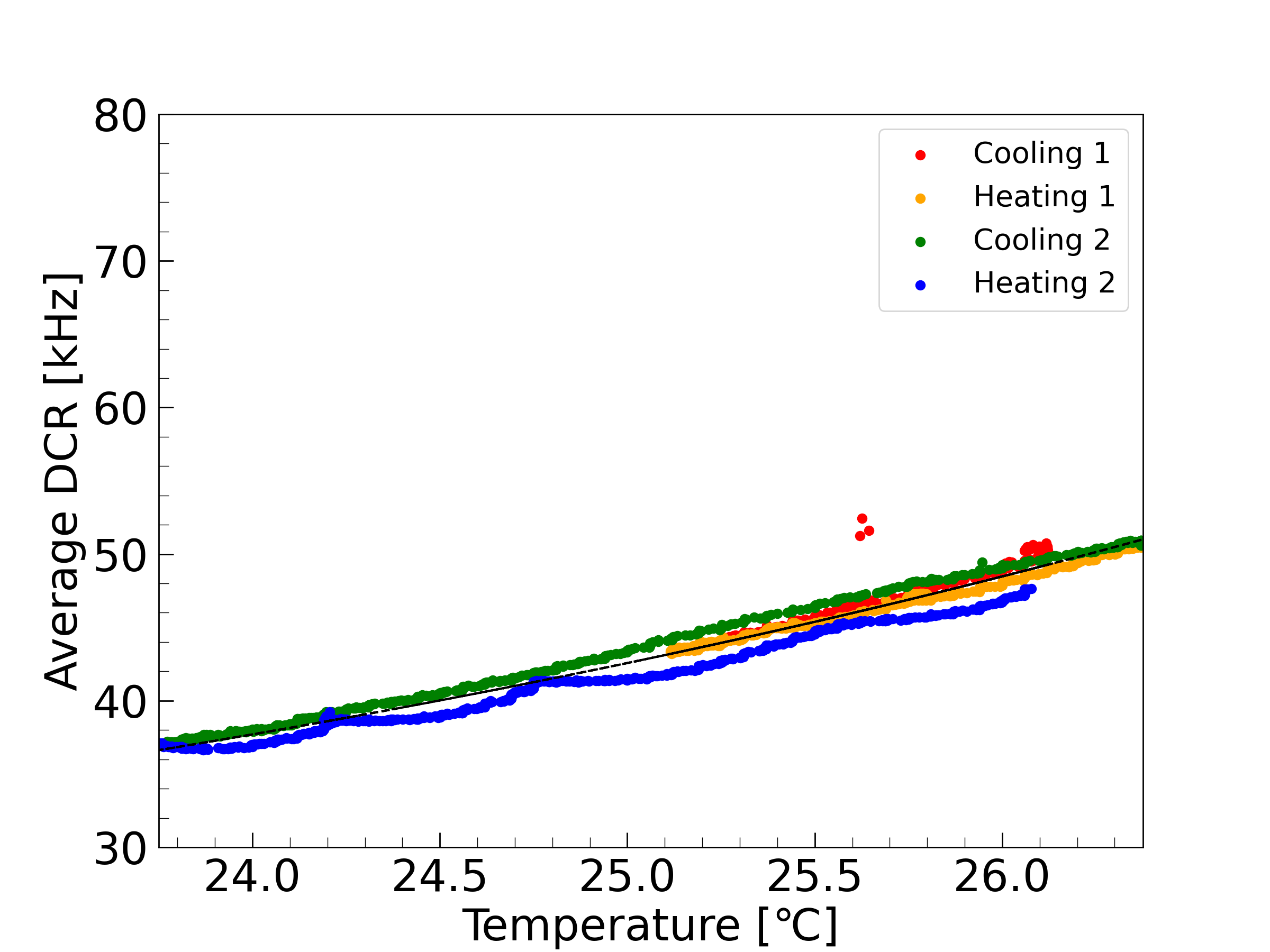}}
    \hfill 	
    \subfloat[DCR and room temperature of HPK PMTs]{\label{fig:hpkvsroomtemp}\includegraphics[width=0.495\textwidth,height=0.4\textwidth]{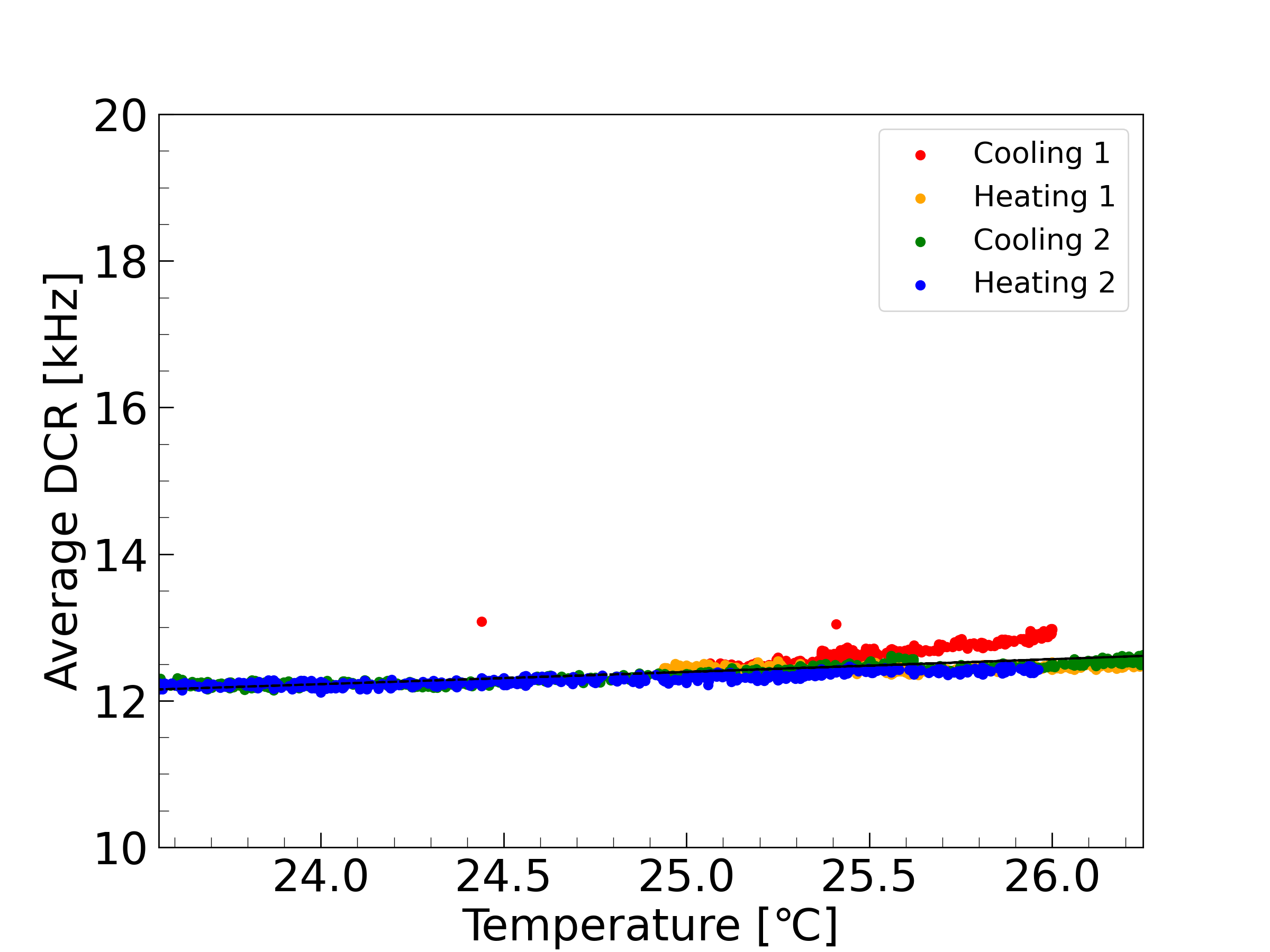}}
    \newline
\subfloat[Room temperature variation of NNVT]{\label{fig:nnvtroomtempvariation}\includegraphics[width=0.495\textwidth,height=0.4\textwidth]{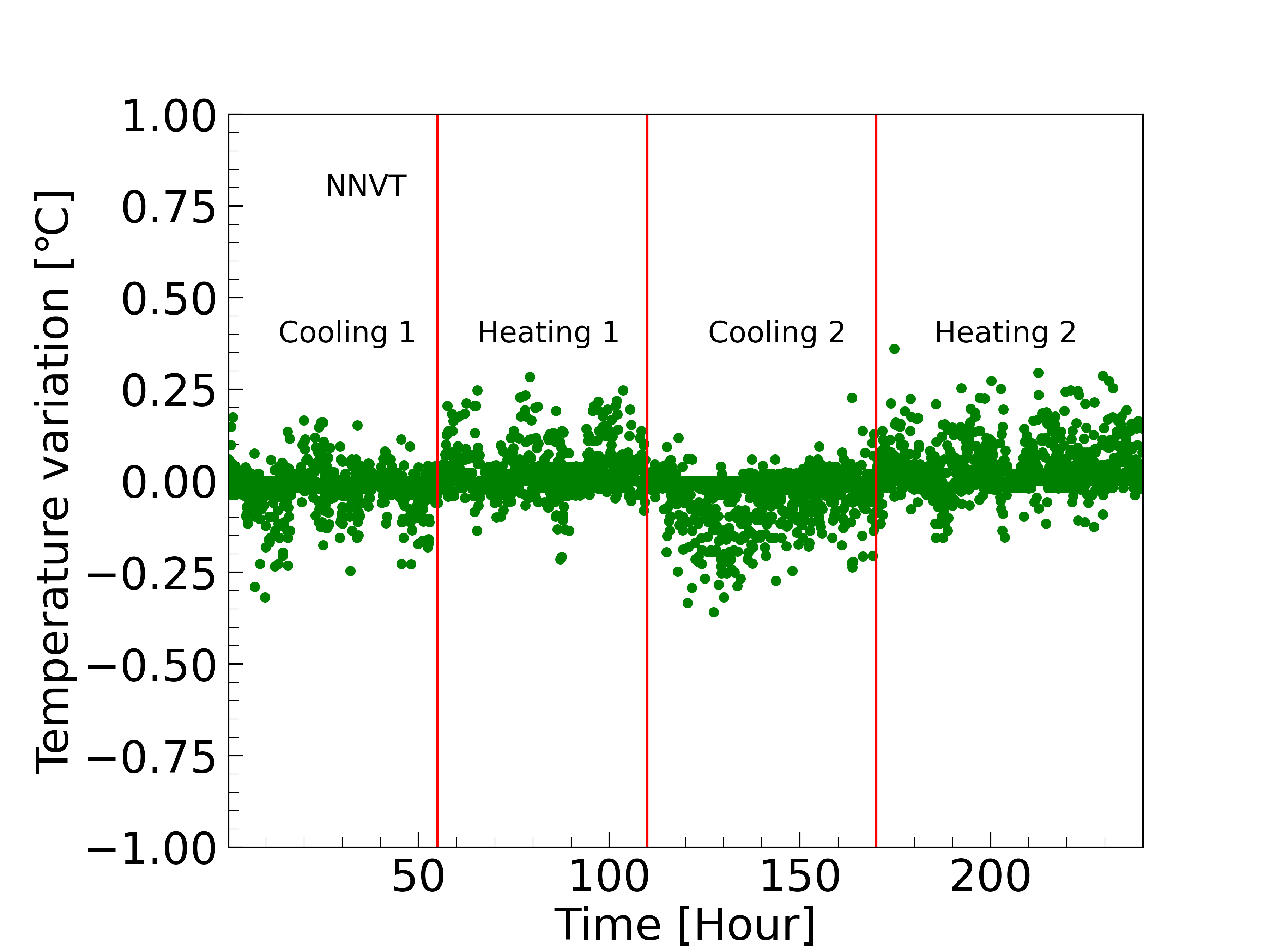}}
    \hfill 	
    \subfloat[Room temperature variation of HPK]{\label{fig:hpkroomtempvariation}\includegraphics[width=0.495\textwidth,height=0.4\textwidth]{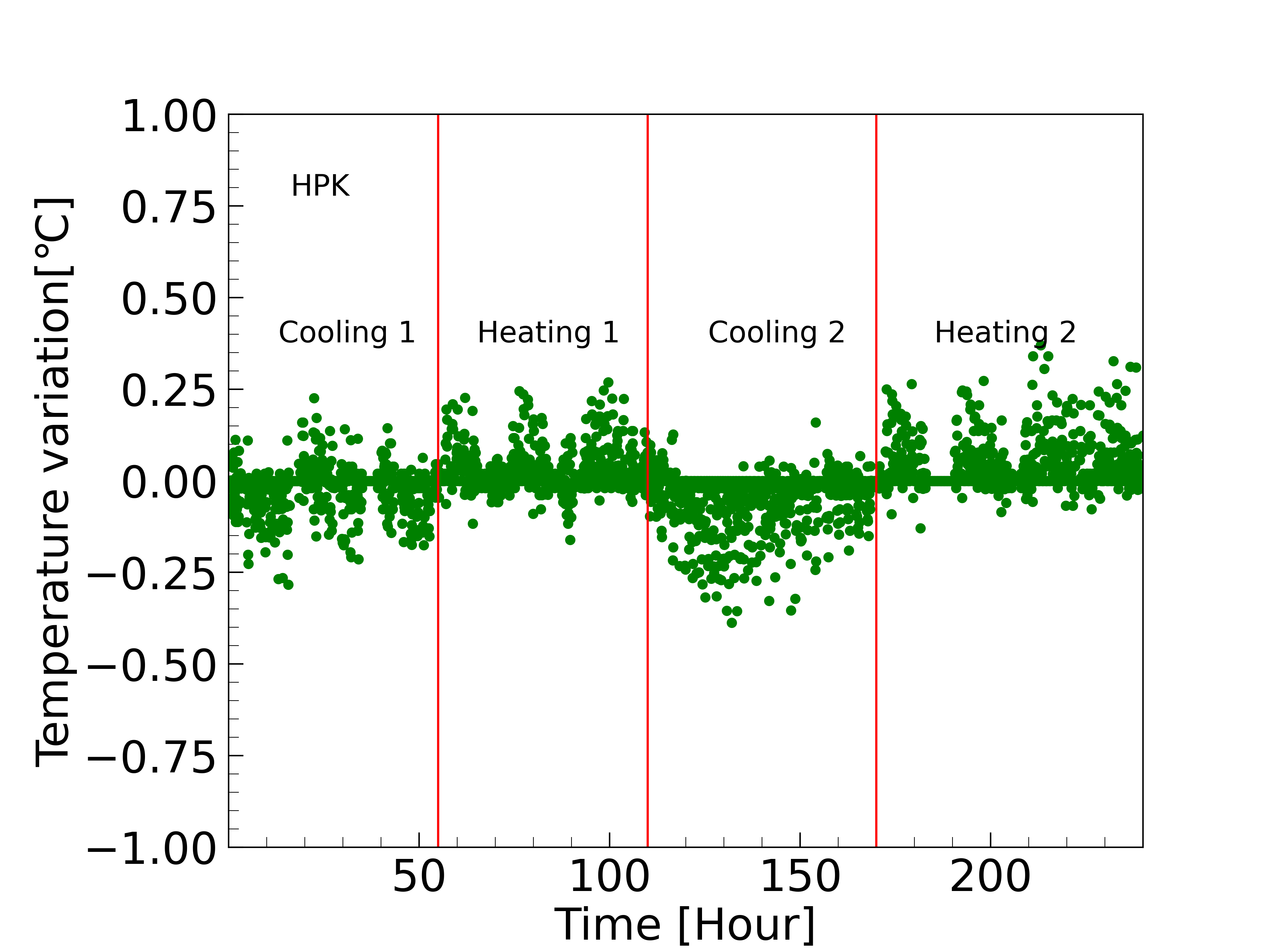}}
  \caption{DCR and room temperature variation}
   \label{fig:tempffect}
 \end{figure}

\iffalse
\fi
During PMT mass testing, the time duration of the PMTs occupied inside the container system was intentionally extended to assess how they responded across various temperature fluctuations around room temperature. In the analysis at hand, data set spanning 250 hours was examined. The monitored DCR during this test is illustrated in Figure \,\ref{fig:tempffect}. The average DCR measurements, aggregated from 6 potted NNVT PMTs and 12 potted HPK PMTs, are displayed alongside the temperature data in Figure \ref{fig:NNVT temp effect} (NNVT PMTs) and Figure \ref{fig:HPK temp effect} (HPK PMTs).

The observations can be categorized into four distinct stages: two heating phases (Heating 1 and Heating 2) and two cooling phases (Cooling 1 and Cooling 2), all occurring within a temperature range of 24\,$^\circ$C to 26\,$^\circ$C. The average absolute DCR across the tested samples in relation to temperature is presented in Figure \ref{fig:nnvtvsroomtemp} and Figure \ref{fig:hpkvsroomtemp} for each respective stage. The temperature variation inside during four stages are displayed in Figure \ref{fig:nnvtroomtempvariation} and \ref{fig:hpkroomtempvariation} for both PMT types.

The rate of temperature change was quite gradual, ranging from -0.25\,$^\circ$C/hour to 0.25\,$^\circ$C/hour. For the HPK PMTs, all stages exhibited a consistent trend, with an average variation factor of 0.35\,kHz/$^\circ$C. However, in the cooling 1 stage, a deviation was noted, primarily attributed to insufficient cooling time following the loading of PMTs into the containers, as discussed in Section \ref{sec:time}.

Similarly, for the NNVT PMTs, the trend remained consistent across all stages, with an average variation factor of 6.01\,kHz/$^\circ$C within the 24\,$^\circ$C to 26\,$^\circ$C range. Notably, the heating 2 stage displayed a sub-structure correlated to the temperature increase, particularly around 26\,$^\circ$C. This can likely be traced back to the rapid temperature change and the heightened sensitivity of NNVT PMTs to temperature variations.

Both short-term and long-term assessments confirm the consistency of performance among the potted PMTs throughout the heating and cooling processes within the specified temperature range.

%According to the measurement results of heating, thermal cooling, and room temperature variation, the temperature effect of PMTs are basically consistent as shown in Table XXXXXX.The variation is more obvious for NNVT PMTs. With the 2-D correlation of DCR and temperature, 

 \subsection{DCR Spikes and Time}
 \label{sec:stability:DCR}

 \subsubsection{Long Stability of DCR}
 \label{sec:2:longstability}
 
\iffalse
\fi

The long-term stability of the DCR is crucial for ensuring stable detector operation and accurate measurements. For JUNO experiment, fluctuations in the DCR during detector operation can lead to false triggers, thereby affecting the energy measurement accuracy \cite{Qian_2020-flasher,Yang_2020-base-flasher,dayabay-PMT-2008}.

\begin{figure}[!ht]
     \subfloat[NNVT PMTs in Container A without temperature control]{\label{fig:NNVT long effect1}\includegraphics[width=0.495\textwidth,height=0.4\textwidth]{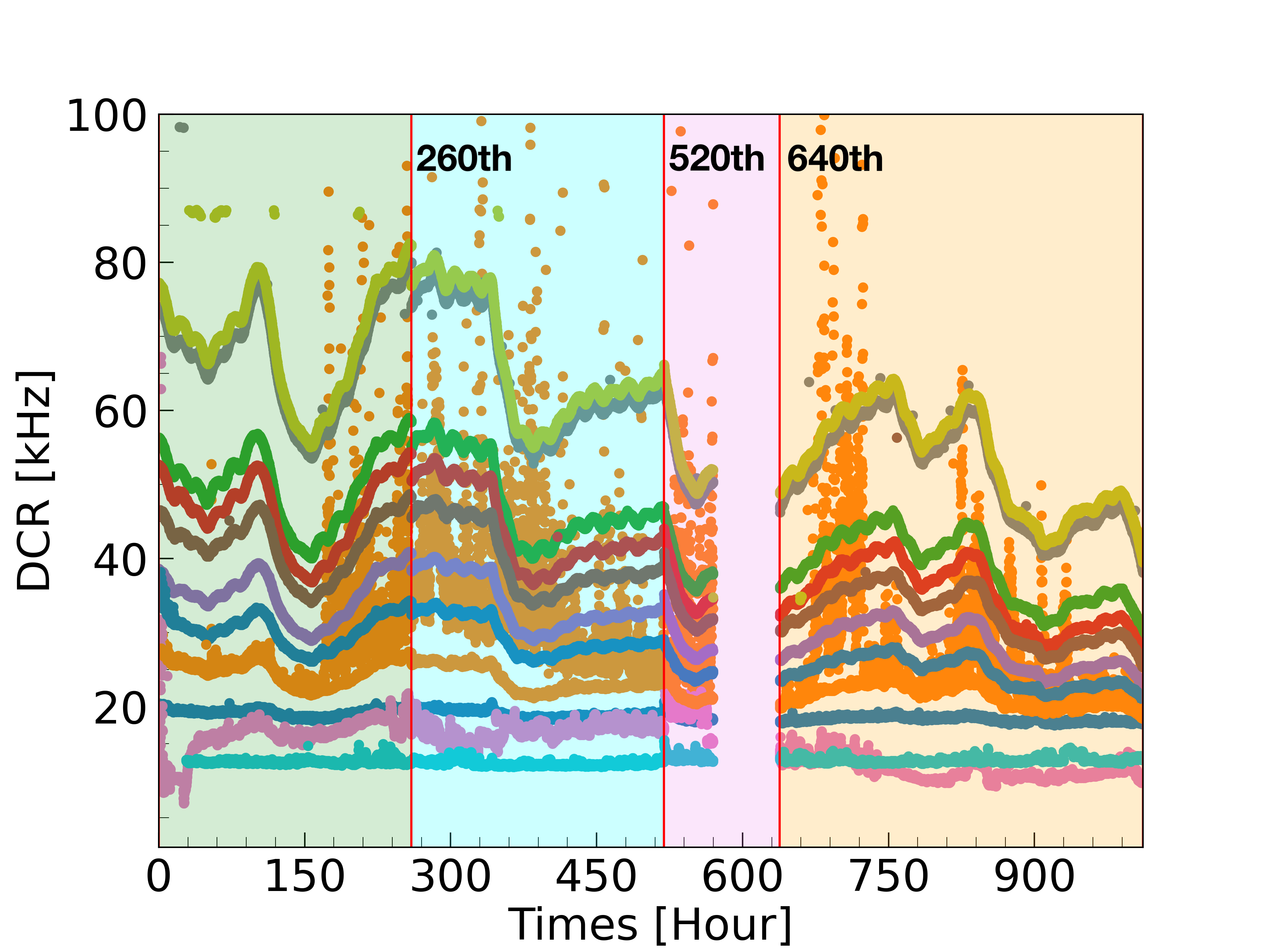}}
    \hfill 	
     \subfloat[HPK PMTs in Container A without temperature control]{\label{fig:HPK long effect1}\includegraphics[width=0.495\textwidth,height=0.4\textwidth]{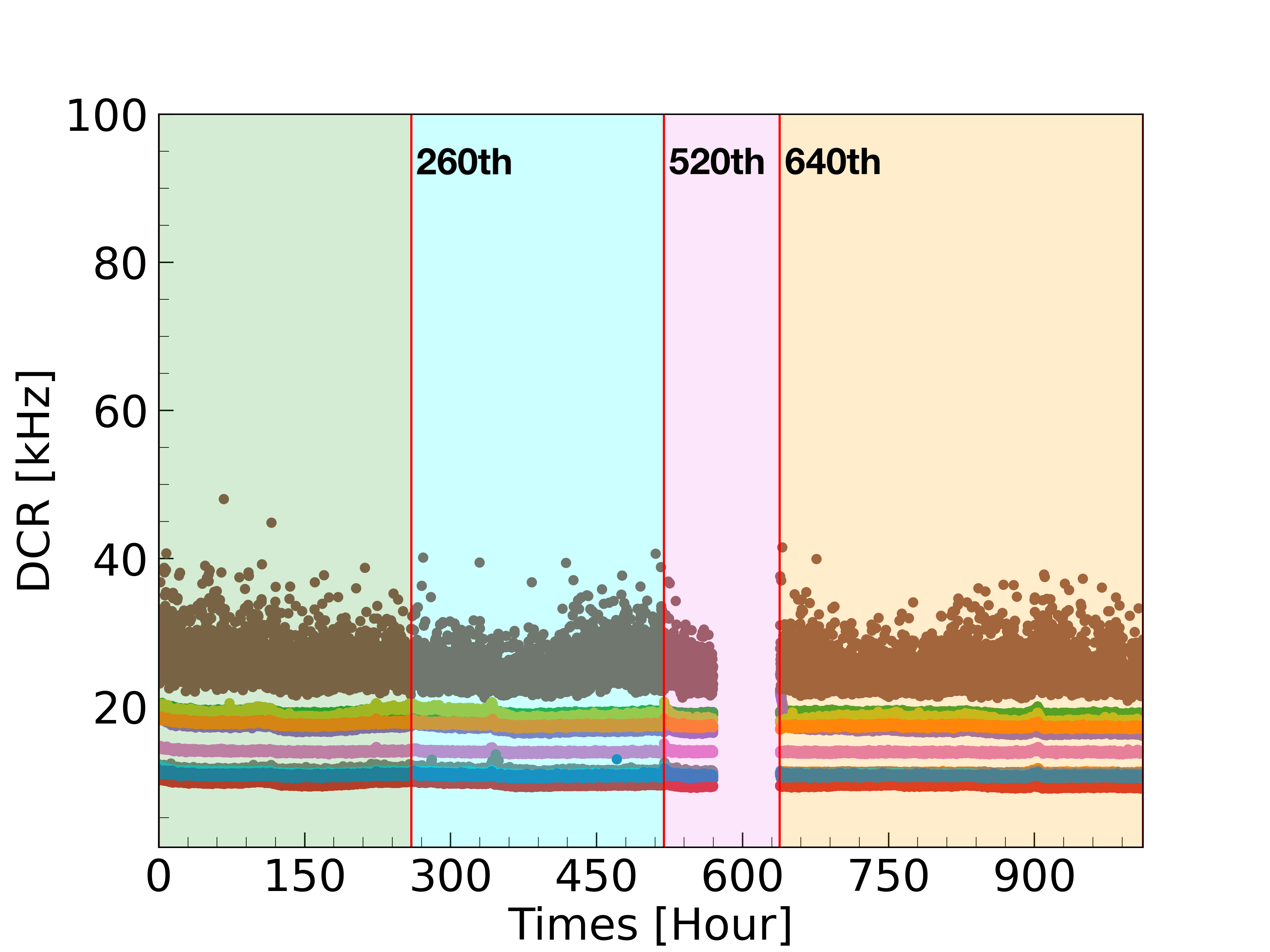}}
    \newline
     \subfloat[NNVT PMTs in Container D with temperature control]{\label{fig:NNVT long effect2}\includegraphics[width=0.495\textwidth,height=0.4\textwidth]{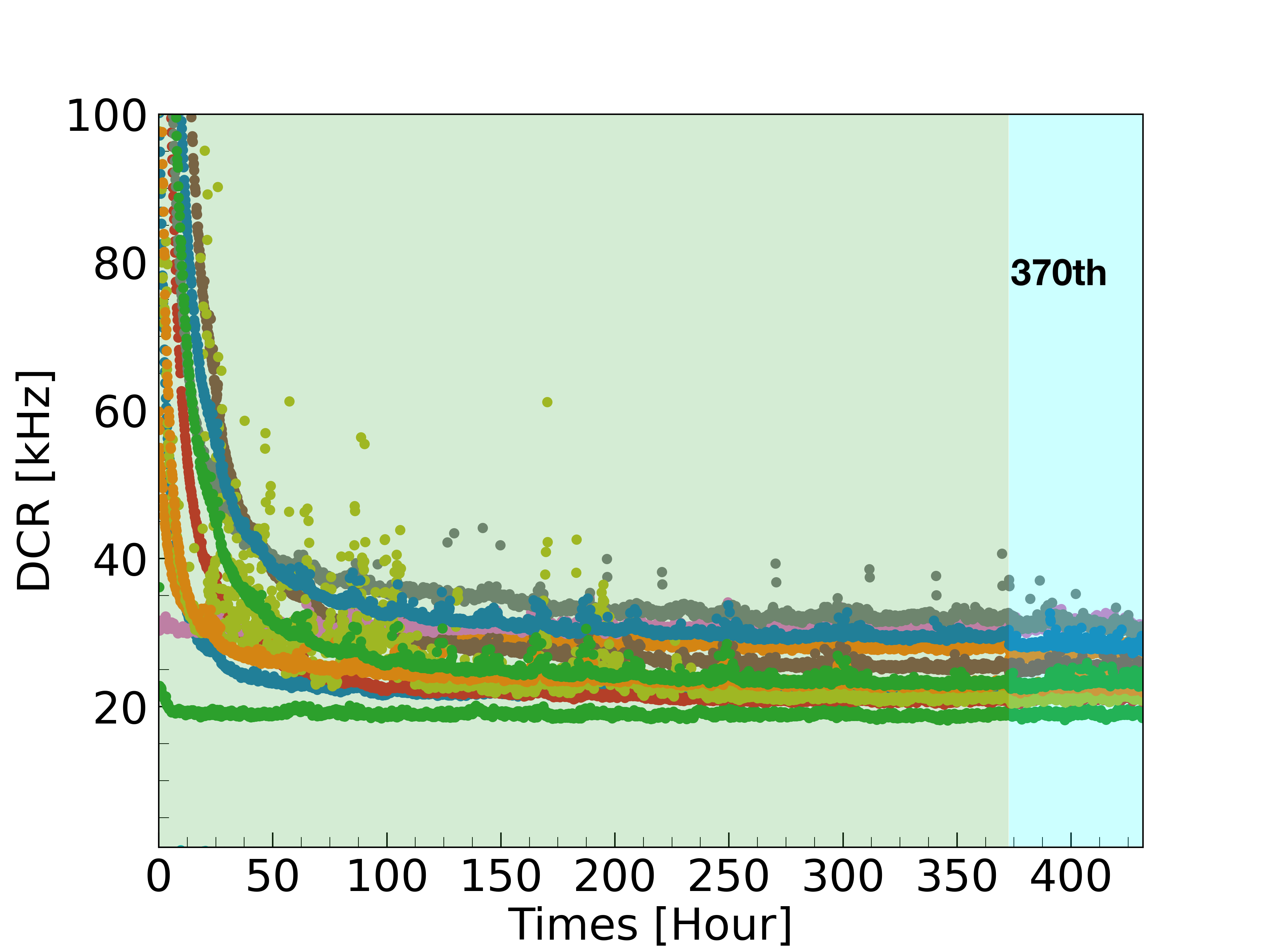}}
    \hfill 	
     \subfloat[HPK PMTs in Container D with temperature control]{\label{fig:HPK long effect2}\includegraphics[width=0.495\textwidth,height=0.4\textwidth]{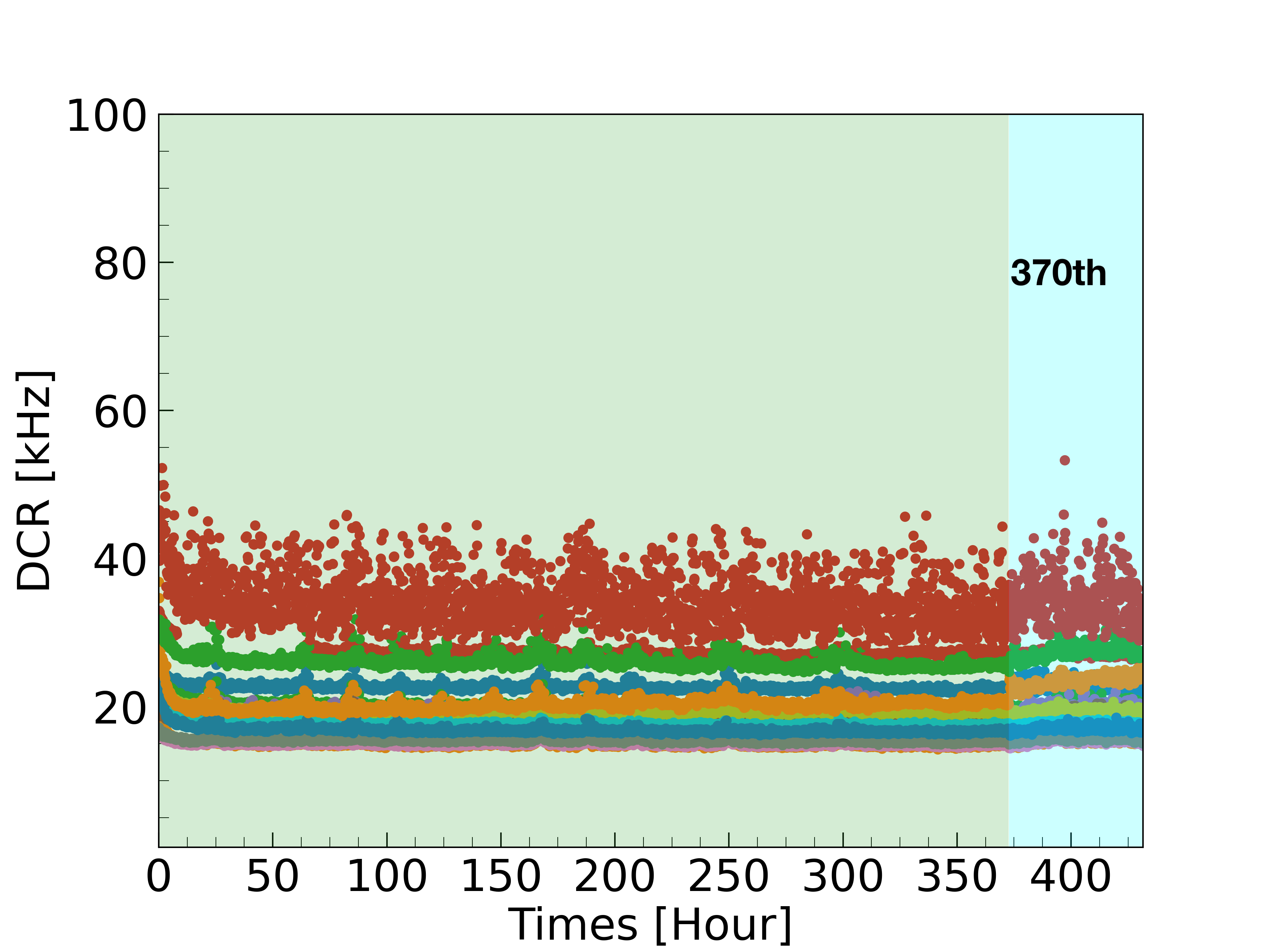}}
   \caption{DCR monitoring of bare NNVT and HPK PMTs for a long-term in container A without temperature control and potted PMTs in container D with temperature control.}
   \label{fig:longstablity}
 \end{figure}

To assess the stability of the DCR over time, various loads and runs of PMTs were conducted during the Pan-Asia testing, monitoring DCR stability for approximately 1000 hours. This is depicted in Figure \ref{fig:longstablity}, where the bare PMTs' performance is illustrated. The different colors in the background correspond to distinct periods during which the same PMTs were placed in various container channels.

Further analysis of the PMTs in container A, without temperature control, are shown in Figure \ref{fig:NNVT long effect1} and Figure \ref{fig:HPK long effect1}. In contrast, Figs. \ref{fig:NNVT long effect2} and \ref{fig:HPK long effect2} present results from PMTs in container D, which were tested under temperature control inside. The data indicates that NNVT PMTs exhibit more significant variations linked to temperature changes compared to HPK PMTs. This observation can primarily be attributed to the temperature effects, as discussed in Section \ref{sec:stability:DCRvariation}.

This study highlights the importance of maintaining stable DCR for optimal detector performance and suggests that temperature regulation can mitigate some of the variability in DCR readings.

\begin{figure}[!ht]
  \subfloat[DCR of normal potted PMTs over time]{\label{fig:normaldcr}\includegraphics[width=0.495\textwidth,height=0.4\textwidth]{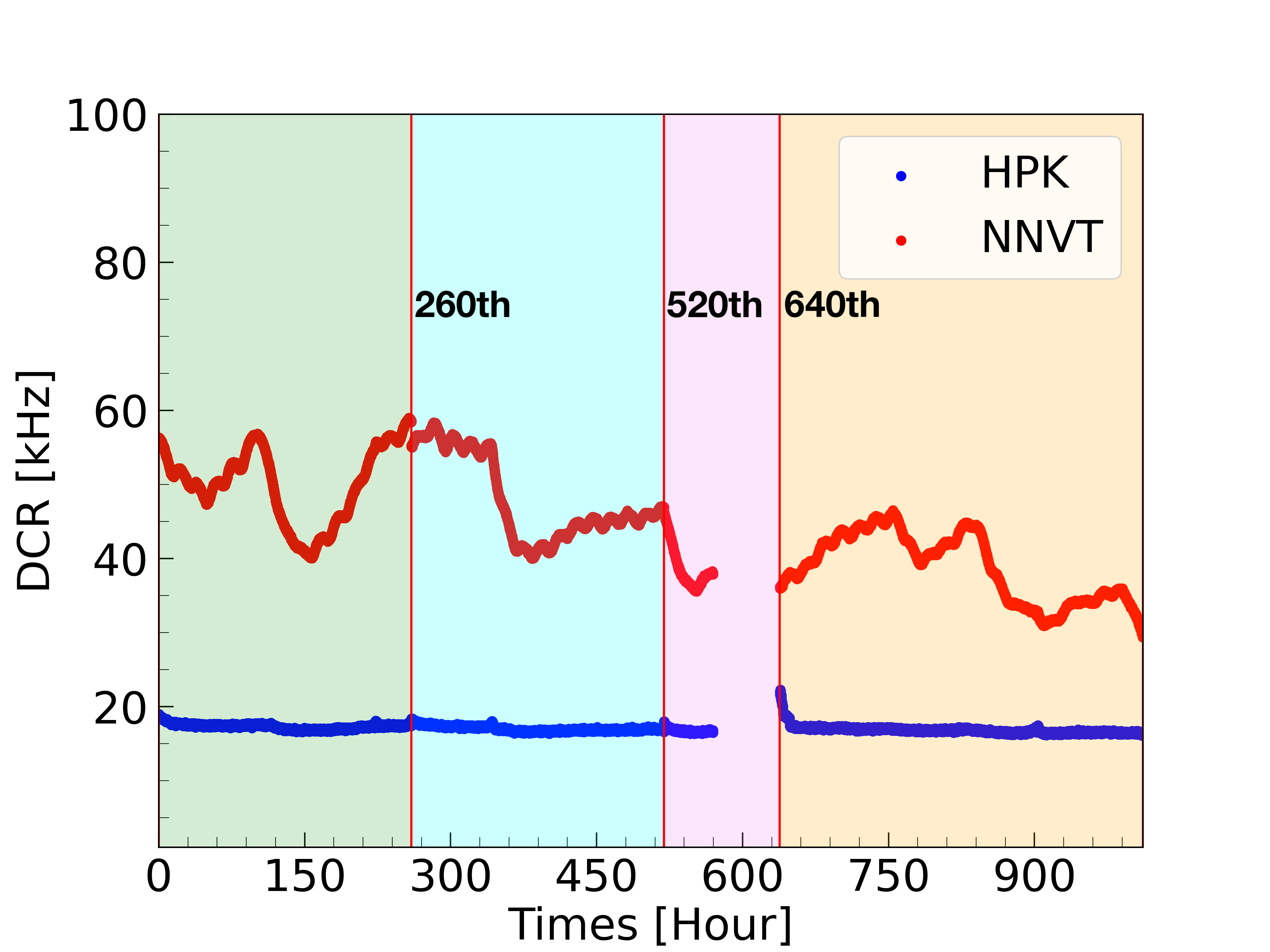}}
    \hfill 	
    \subfloat[DCR spikes of abormal potted PMTs over time]{\label{fig:potteddcrspe}\includegraphics[width=0.495\textwidth,height=0.4\textwidth]{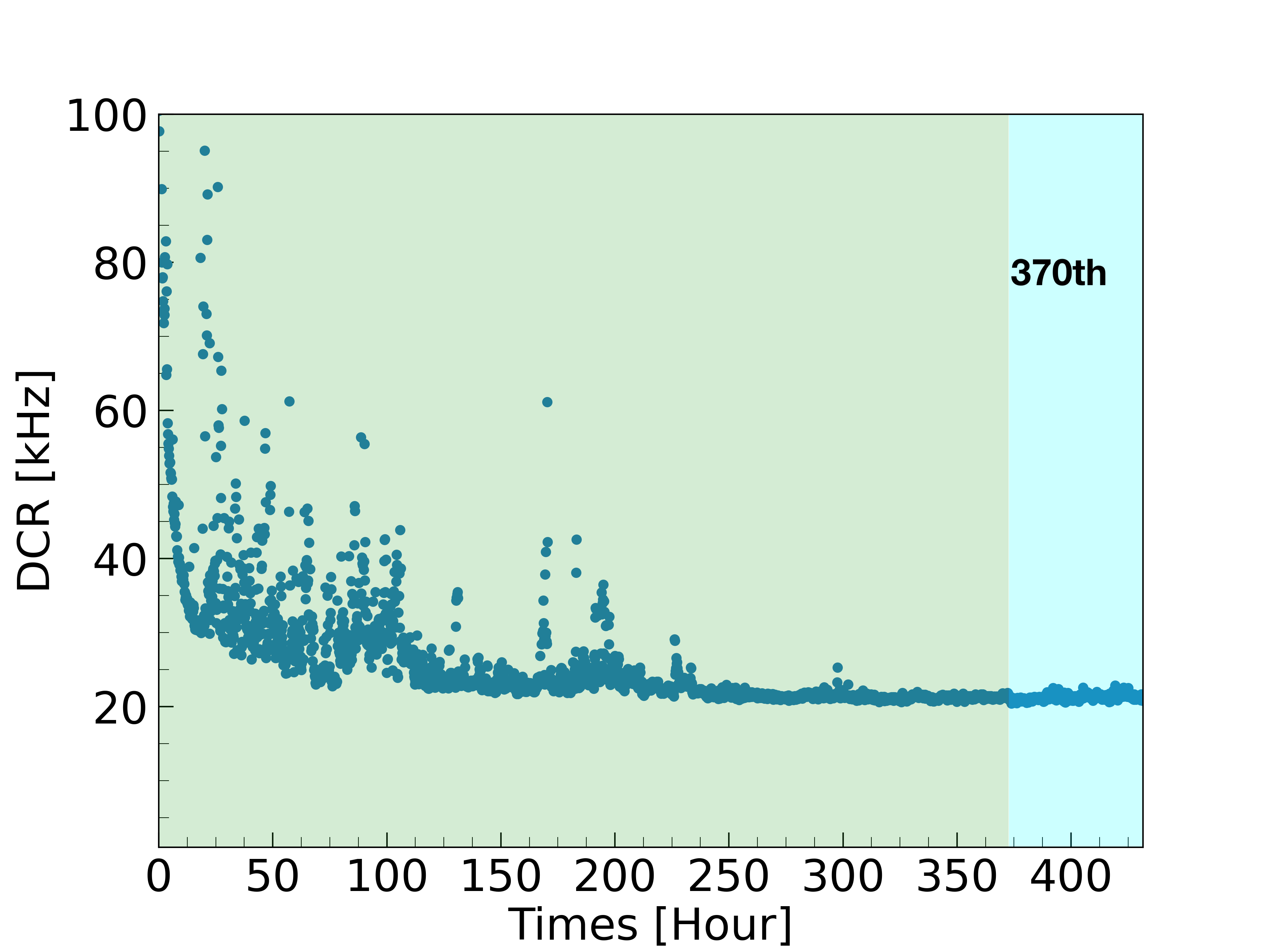}}
     \newline
     \centering
     \subfloat[DCR spikes of abormal potted PMTs with longer time duration]{\label{fig:potteddcrabnormal}\includegraphics[width=0.495\textwidth,height=0.4\textwidth]{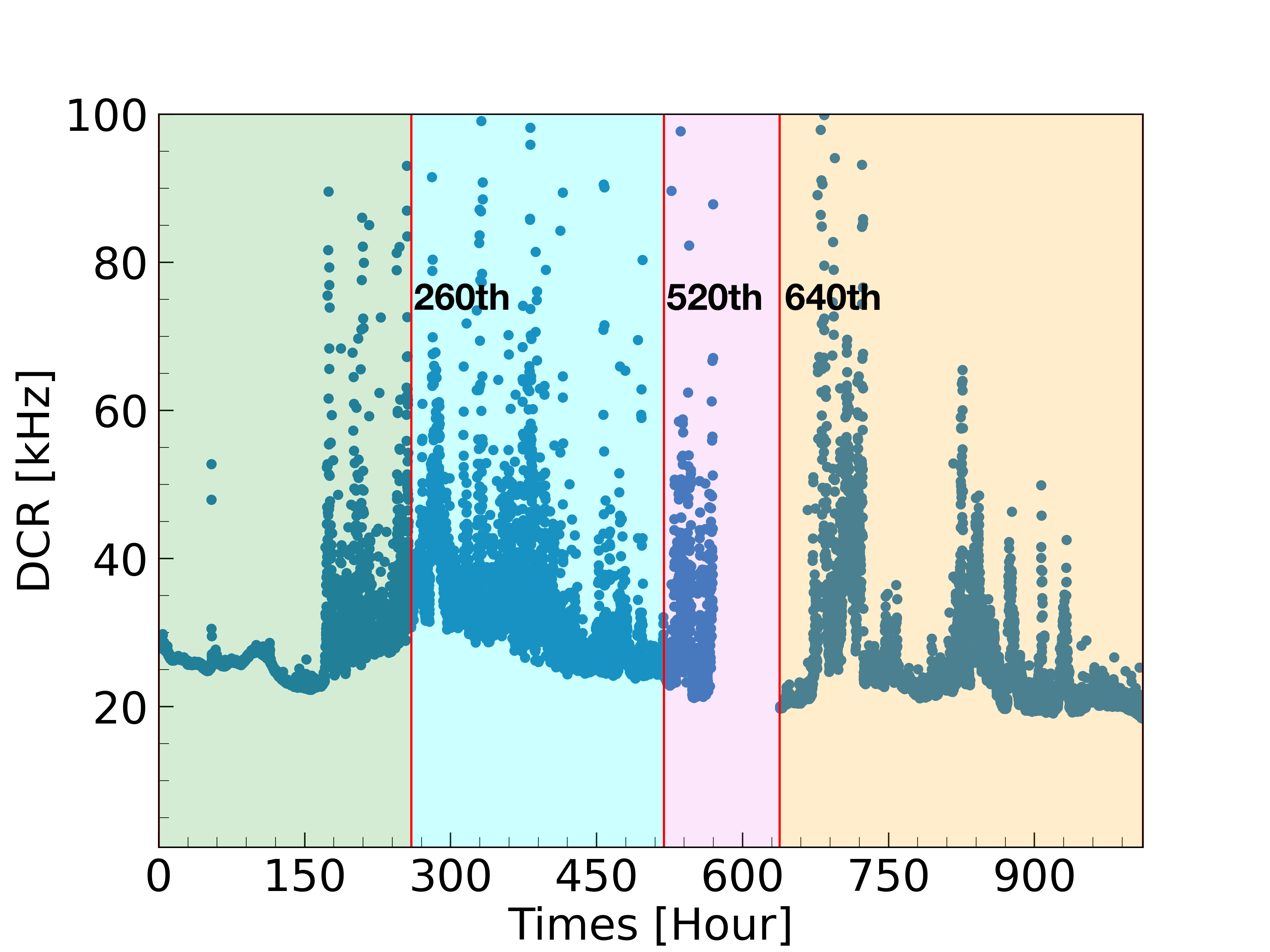 }}
   \caption{Long-term DCR monitoring: (a) DCR of normal potted PMTs over time (b) DCR spikes of abormal NNVT potted PMTs over time, but after testing drawer swapping recovered as tagged by the cyan blue background. (c) abnormal NNVT DCR with jumps, but after testing drawer swapping, it is not recovered as tagged by the different color.}
 \label{fig:specialpmt}
 \end{figure}

 A successful monitoring of the long-term stability of 117 PMTs under various load conditions was performed, comprising of 73 tubes from NNVT and 44 from HPK. The test results of selected samples are illustrated in Figure \ref{fig:specialpmt}. Among the monitored PMTs, the majority exhibited a stable and smooth distribution over time, as shown in Figure \ref{fig:normaldcr}. However, a subset of PMTs exhibited abnormal fluctuations during testing (see Figure \ref{fig:potteddcrspe}), where the DCR spiked to levels exceeding 100 kHz, surpassing our acceptance criteria. These anomalies could potentially be influenced by various hardware components, including electronics or the PMT units themselves, particularly during the occurrence of flashing PMTs.

To identify and diagnose potential issues stemming from connections or electronics, additional  measurements were conducted. The first step involved swapping testing drawers and replacing HV deviders, a straightforward approach to eliminate possible hardware problems. After these adjustments, some PMTs stabilized, as indicated by the cyan blue background in Figure \,\ref{fig:potteddcrspe}, and maintained this stability over an extended period. Conversely, several PMTs continued to exhibit instability, marked by a different background color in Figure \,\ref{fig:potteddcrabnormal}, despite the changes made. This persistent instability may indicate inherent issues with the PMTs themselves. Ultimately, approximately 7\% of the long-term monitored PMTs displayed unstable DCR characteristics.

In summary, the abnormal jumps in DCR are detrimental to detector performance, as they can lead to false triggering and adversely affect energy measurements. While the underlying causes warrant further investigation beyond the container system's capabilities, a standalone inquiry will be addressed in subsequent discussions.

 \subsubsection{Short-term DCR Monitoring}

 During the mass testing of the JUNO bare PMTs, the DCR of each tested PMT was monitored at 30-minute intervals throughout a cooling period lasting 12 hours or longer, depending on their respective configurations. To minimize the effects of DCR shifts during the initial cooling phase, the evaluation of the monitored DCR was started only after the third hour. A focus was set specifically for significant DCR fluctuations, defined as increases exceeding 50\,kHz or 50\% compared to the previously recorded DCR point, as detailed in Section \ref{sec:2:longstability}.

\begin{figure}[H]
 \subfloat[Normal and smooth curves of DCR over time]{\label{fig:potteddcrnormal2}\includegraphics[width=0.495\textwidth,height=0.4\textwidth]{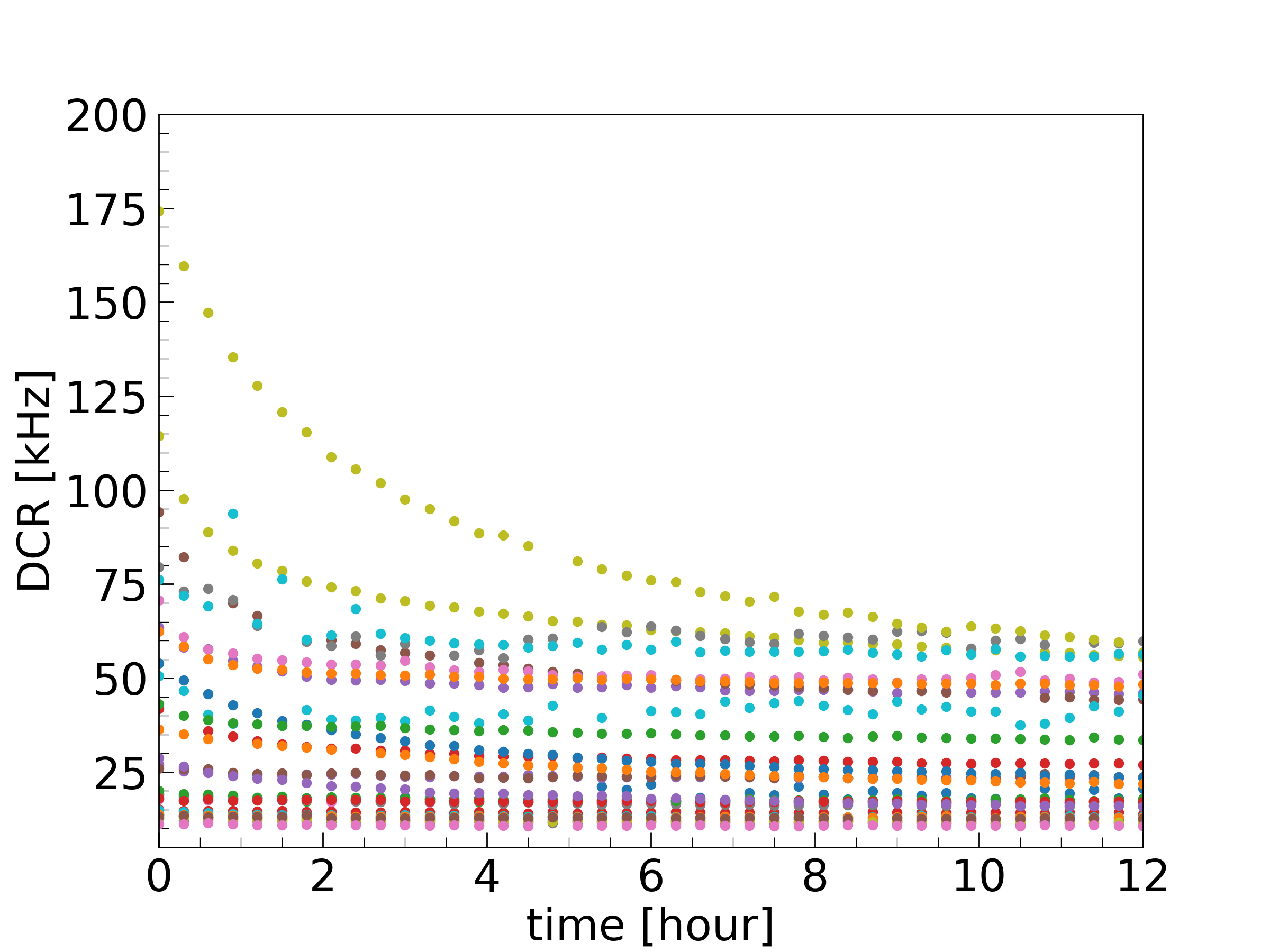}}
 \hfill 	
     \subfloat[DCR curve with identified spikes]{\label{fig:potteddcrabnormal2}\includegraphics[width=0.495\textwidth,height=0.4\textwidth]{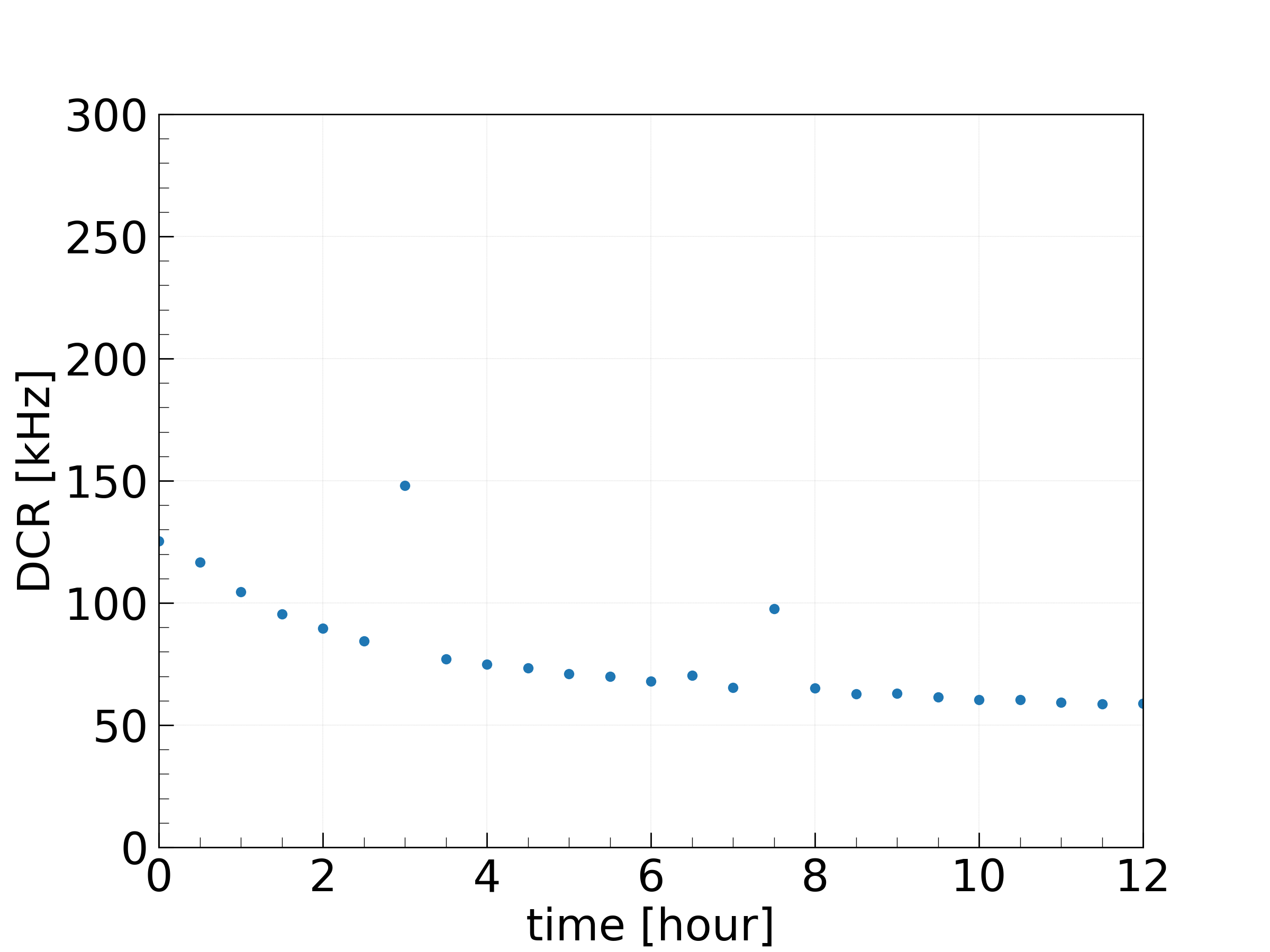}}
   \caption{Monitored DCR of PMTs during their cooling time for the acceptance test: (a) normal DCR with smooth curve and (b) abnormal DCR curve with some spikes}
 \label{fig:shortresult}
 \end{figure}

Figures \ref{fig:potteddcrnormal2} and \ref{fig:potteddcrabnormal2} illustrate examples of the DCR behavior. Figure \ref{fig:potteddcrnormal2} features curves exhibiting normal and stable DCR characteristics, while Figure \ref{fig:potteddcrabnormal2} shows a DCR curve with identified spikes.

The analysis found that approximately 20\% of the tested PMTs experienced at least one jump in DCR, with about 0.2\% exhibiting multiple jumps.

 \subsection{Discussion}
 \label{sec:discuss}
The cooling down time of the PMTs was found to be crucial for the DCR measurement. 
%Regarding the cooling down time of the DCR of PMTs, 
It is advisable to allow a longer cooling period of at least 50 hours for NNVT PMTs, while a minimum of 7 hours is sufficient for HPK PMTs. This extended cooling time ensures that DCR measurements achieve an uncertainty of less than 1\,kHz. Notably, the DCR values observed during the JUNO PMT mass testing are slightly elevated compared to the anticipated baseline.

For accurate DCR measurements, it is recommended to control the temperature within $\pm$1 $^{\circ}$C. This precision in temperature control is essential for achieving a DCR measurement accuracy of 1\,kHz in the temperature range of 2 $^{\circ}$C, which poses no challenge for the future JUNO detector operations.

Additionally, variations in DCR may be linked to several factors, including noise interference, cable and connector quality, electronic components, or even potential issues related to flashing PMTs. These aspects necessitate further investigation to understand their impact on measurement consistency.

\section{Possible Flasher Identification}
 \label{sec:flasher}

\begin{figure*}[!ht]
    \subfloat[System configuration]{\label{fig:longstabilitymoniter}\includegraphics[width=0.495\textwidth,height=0.3\textwidth]{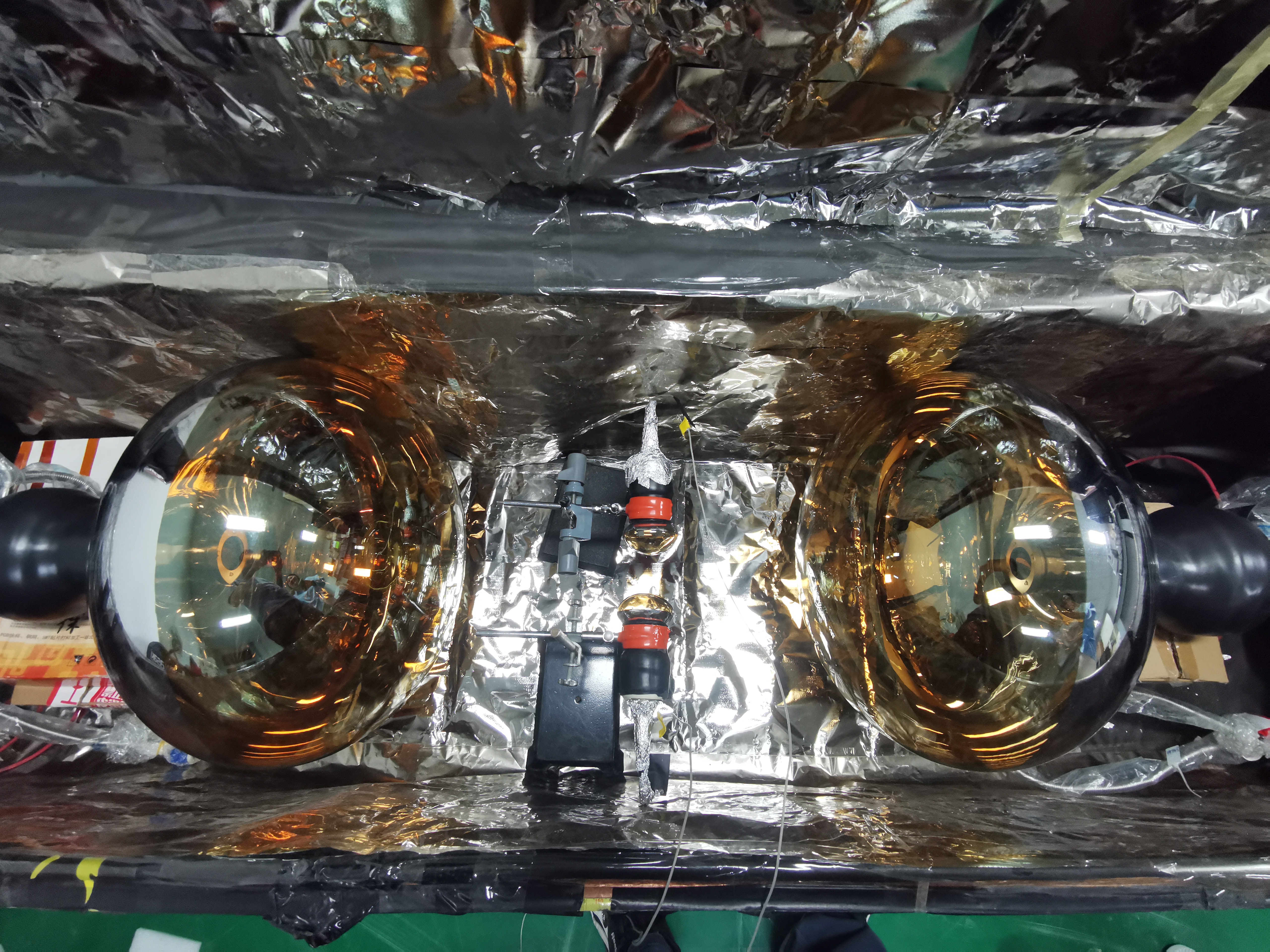}}
   \hfill 	
    \subfloat[System schematic]{\label{fig:systemsetup}\includegraphics[width=0.495\textwidth,height=0.3\textwidth]{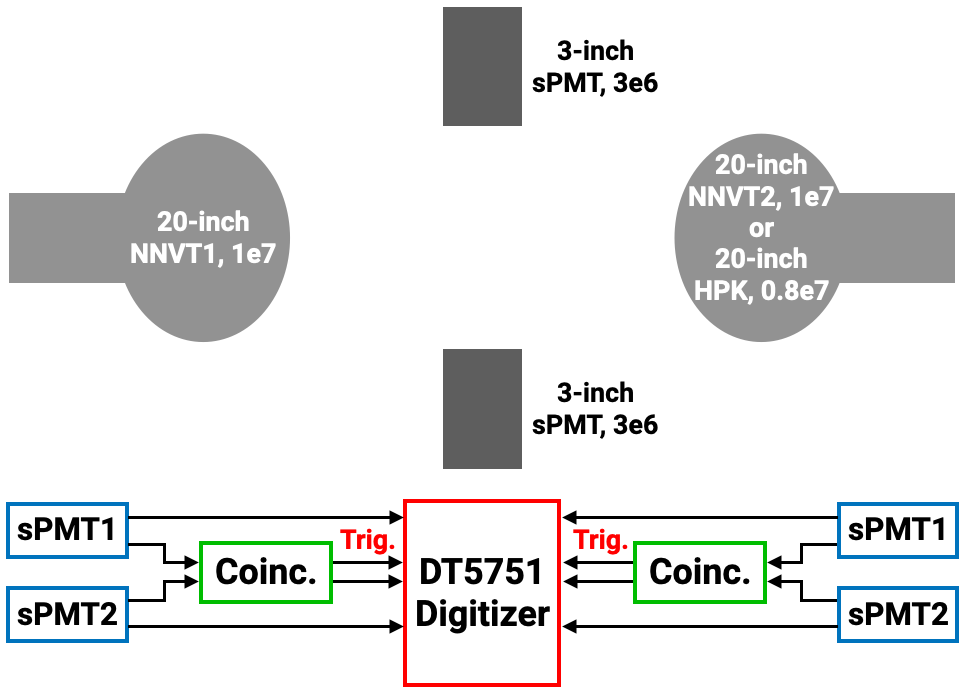}}
  \caption{Experimental Setup}
  \label{fig:setupsystem}
\end{figure*}

PMT flasher is an important background/noise for the rare events' experiment \cite{Yang_2020,Qian_2020}, which also can be identified by the monitored PMT DCR: the PMT itself and its neighbor PMTs. 
If the jumps originate from the PMTs or electronic sources, only the count rates for PMT random coincidences would significantly increase, without impacting other PMTs. Conversely, under sufficiently high emission intensity, all PMTs respond simultaneously, leading to concurrent increases in DCR across the board.
To gain deeper insights into DCR jump from possible PMT flasher, a series of experiments was conducted at the Institute of High Energy Physics (IHEP) laboratory, as illustrated in Figures \ref{fig:setupsystem} and \ref{fig:setup}. 
 
%\par The stability of the DCR is a critical parameter in the evaluation of PMT performance. Understanding this metric is essential for assessing the overall effectiveness of PMTs in various applications.

The setup configuration includes two 20-inch PMTs; NNVT and HPK tubes, and two 3-inch PMTs, produced by Hainan Zhanchuang Photonics Technology Co., Ltd (HZC), denoted as small PMT (sPMT). The 20-inch PMTs were positioned towards each other and two 3-inch PMTs were placed in between to enhance light collection and allow for potential flasher tagging. The two 20-inch PMTs operate at a gain of ($1 \times 10^7$) and exhibited abnormal DCR behavior during the Pan-Asia test. Two 3-inch PMTs function at a lower gain of ($3 \times 10^6$). Notably, the threshold for the 3-inch PMTs is set at 0.25\,photoelectrons (pe), while the DCR for the 20-inch PMTs is significantly higher, requiring a threshold of 10\,pe to maintain a lower event rate. The distribution of parameters for all PMTs is summarized in Table \ref{table:pmtparameter}.

%10 pe sounds very huge. Are sure about this? 10 pe is by far too high to be connected to DCR, which is single pe signal. 

\begin{table}[!ht]
\centering
\caption{The PMT configuration setup, where the amplitude of a single p.e.~for NNVT or HPK is around 8\,mV, while it is around 20\,mV for sPMT with an amplifier of 10 times}
\label{table:pmtparameter}
\begin{tabular}{ccc}
\hline
PMT   & HV {[}V{]} & Threshold  [pe] \\
\hline
NNVT1 & 1700       & 10            \\
NNVT2 & 1850       & 10            \\
HPK   & 2300      & 10            \\
SPMT1 & 1180       & 0.25          \\
SPMT2 & 1150       & 0.25     \\
\hline
\end{tabular}
\end{table}

\subsection{DCR Spike}

%\textcolor{red}{Monitoring rewrite?}

\begin{figure*}[!ht]
    \subfloat[Two NNVT PMTs]{\label{fig:longstabilitymoniternnvtnnvt}\includegraphics[width=0.495\textwidth,height=0.3\textwidth]{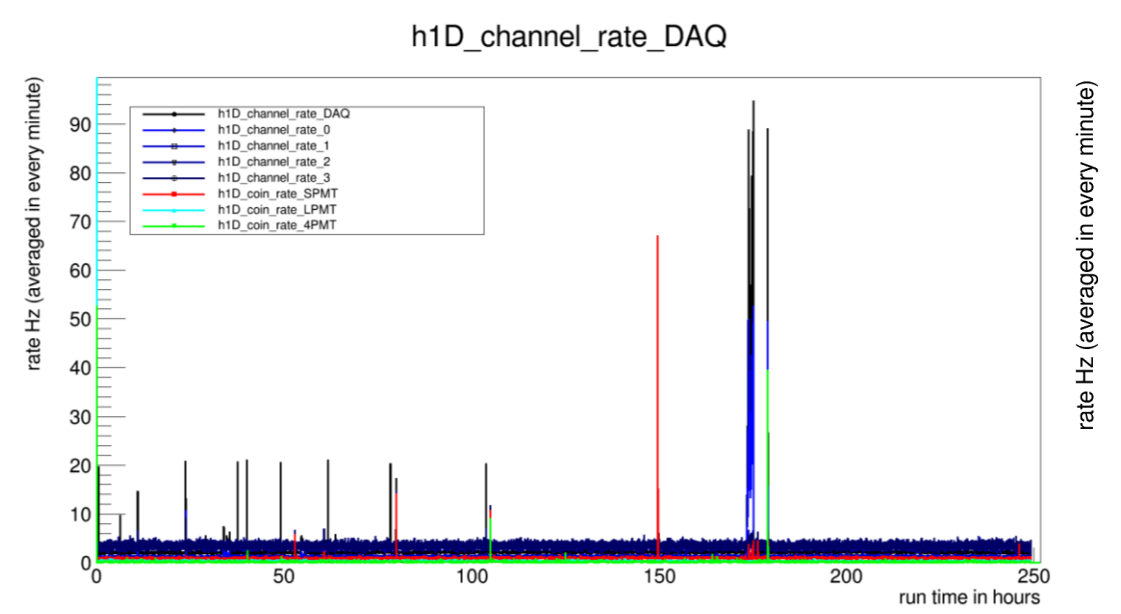}}
   \hfill 	
    \subfloat[One NNVT and one HPK PMTs]{\label{fig:longstabilitymoniternnvthpk}\includegraphics[width=0.495\textwidth,height=0.3\textwidth]{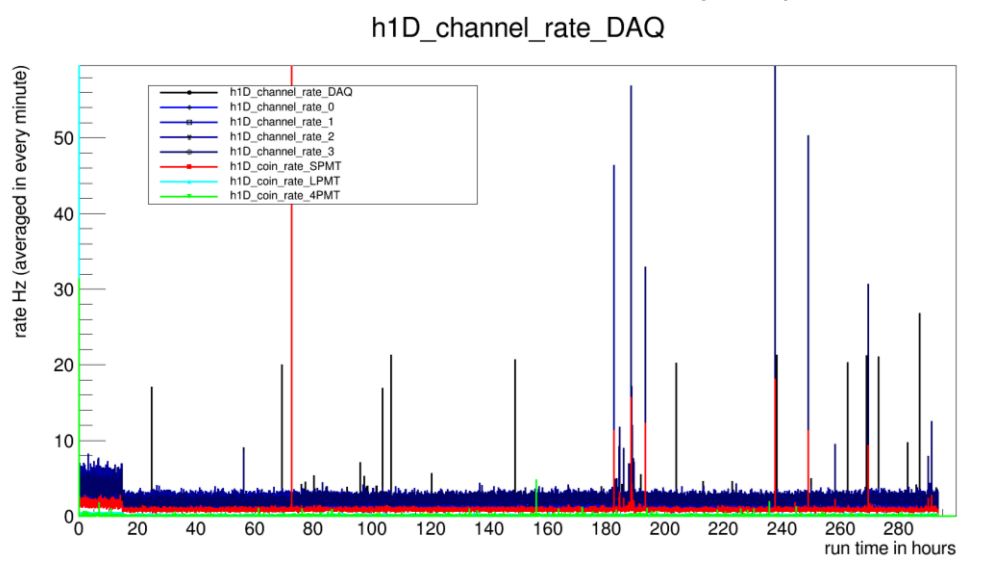}}
  \caption{DCR monitoring in long term}
  \label{fig:longterm}
\end{figure*}

\par With the setup shown in Figure \ref{fig:setupsystem}, several notable DCR spikes can be identified in both 3-inch and 20-inch PMTs during the long-term monitoring as shown in Figure \ref{fig:longstabilitymoniternnvtnnvt} (Two NNVT PMTs, NNVT+NNVT) and Figure \ref{fig:longstabilitymoniternnvthpk} (One NNVT and one HPK PMTs, NNVT+HPK).
In both cases, the instances of all four PMTs simultaneously experiencing DCR spikes within a monitoring duration of less than 300 hours are rare. An analysis of the monitoring results yields a random coincidence rate of 0.4$\times$10$^{-5}$\,Hz for NNVT/NNVT and 3.6 $\times$10$^{-5}$\,Hz for NNVT/HPK, whereas the measured coincidence rates exceed 0.5\,Hz. This disparity primarily arises from the characteristics of the 20-inch PMT itself.

%: a, b, c, and d signify PMT random accidental coincidences. Meanwhile, the e and f colors represent SPMTs and LPMTs coincidences, respectively. 

\subsection{Investigation on DCR Spikes}

\begin{figure*}[!ht]
    \subfloat[System configuration]{\label{fig:longstabilitymoniter2}\includegraphics[width=0.495\textwidth,height=0.3\textwidth]{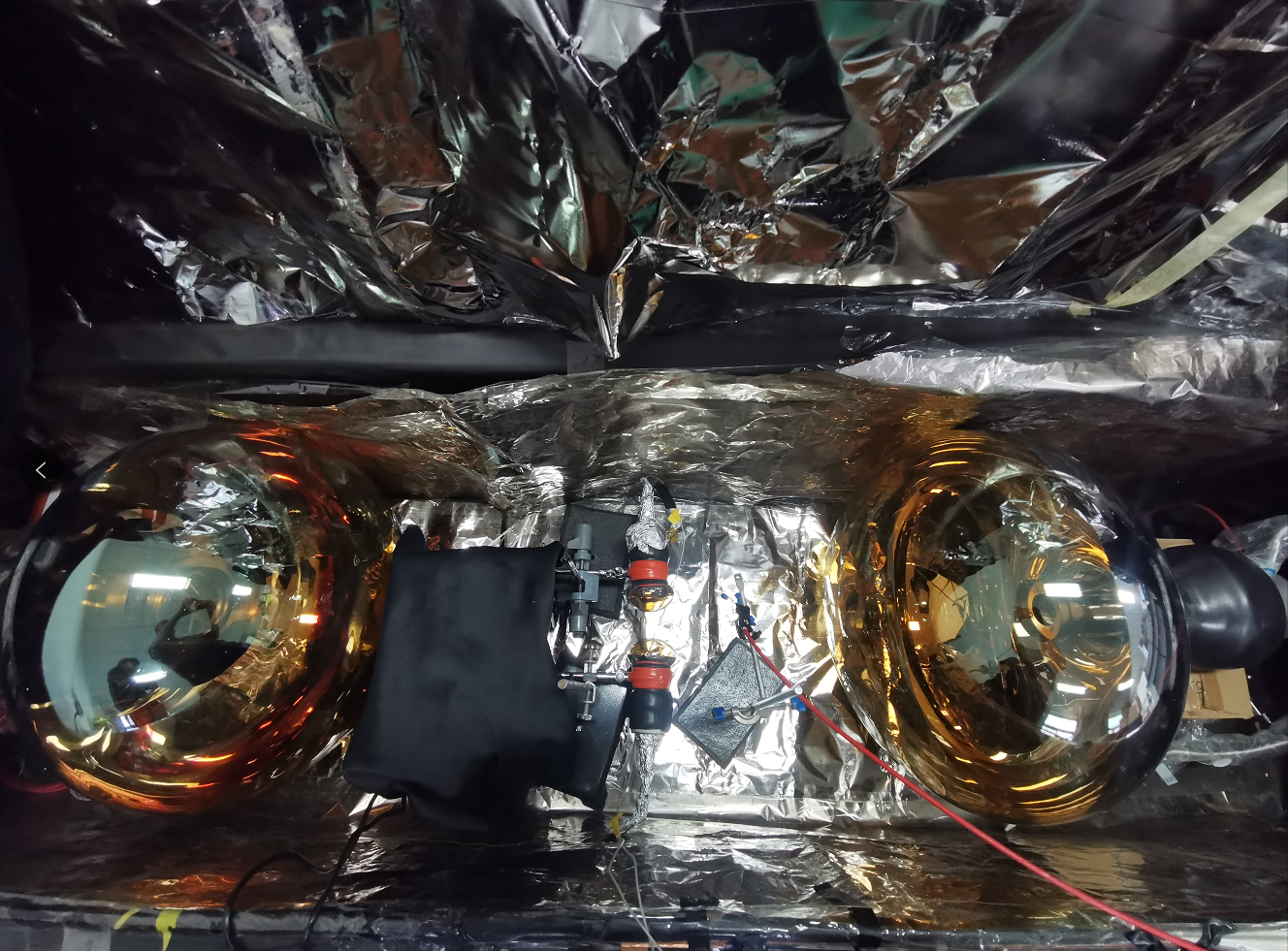}}
   \hfill 	
    \subfloat[System schematic]{\label{fig:systemsetup2}\includegraphics[width=0.495\textwidth,height=0.3\textwidth]{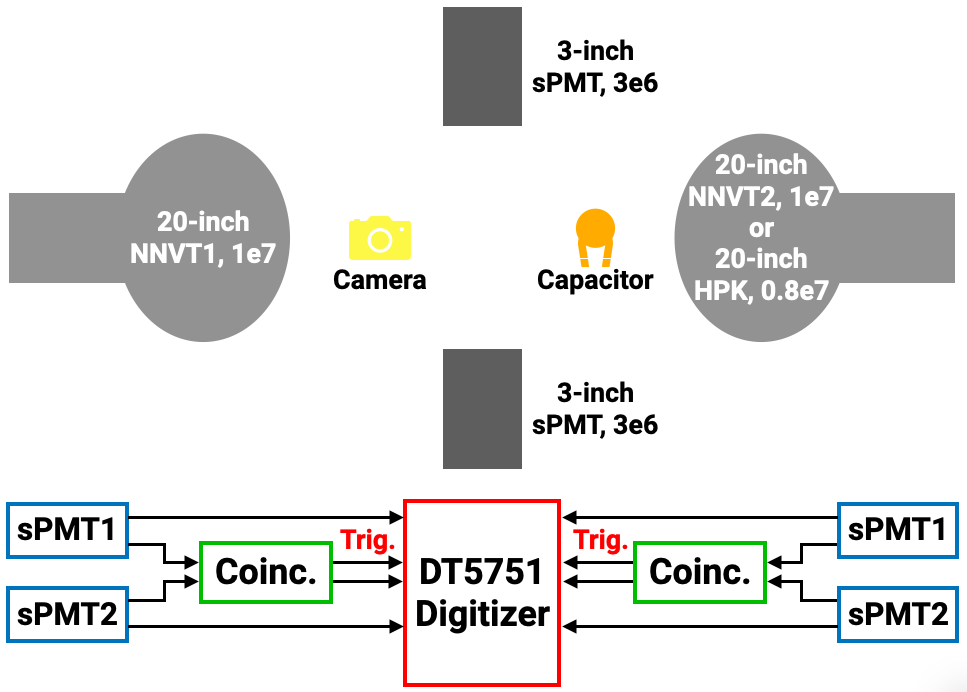}}
  \caption{Experimental Setup for DCR spike with camera}
  \label{fig:setup}
\end{figure*}

%Tobis new Point of reading

\par A single photon sensitive camera (ORCA-Quest qCMOS C15550-20UP from Hamamatsu Photonics \cite{Wang_2023-track,Wang_2023}) is a useful tool to detect a weak light sources to investigate DCR spikes, where a voltage test and a capacitor test were designed in conjunction with the camera, as shown in Figure ~\ref{fig:setup}. In these tests, the camera was oriented towards a 20-inch PMT. In the voltage test, we systematically increased the working voltage of PMT in increments of 100\,V from a start of 1500\,V and monitored the PMT's coincidence rates by a counter. Conversely, in the capacitor test, as illustrated in the diagram (Figure \ref{fig:systemsetup2}), we introduced a capacitor chain and applied high voltage to make it spark. The breakdown of the capacitor generated flashes (artificial sparks), enabling the observation of the PMT's response through this test.

%it is not clear to my, how the light of the sparking capacitor was brought to/is coupled with the PMTs. Additonally, providing the camera model would be a nice addition to the experimental setup.

\subsubsection{Tagging with Artificial Spark}

\begin{figure}[!ht]
    \subfloat[Capacitor w/ visual light]{\label{fig:capacitance}\includegraphics[width=0.495\textwidth,height=0.45\textwidth]{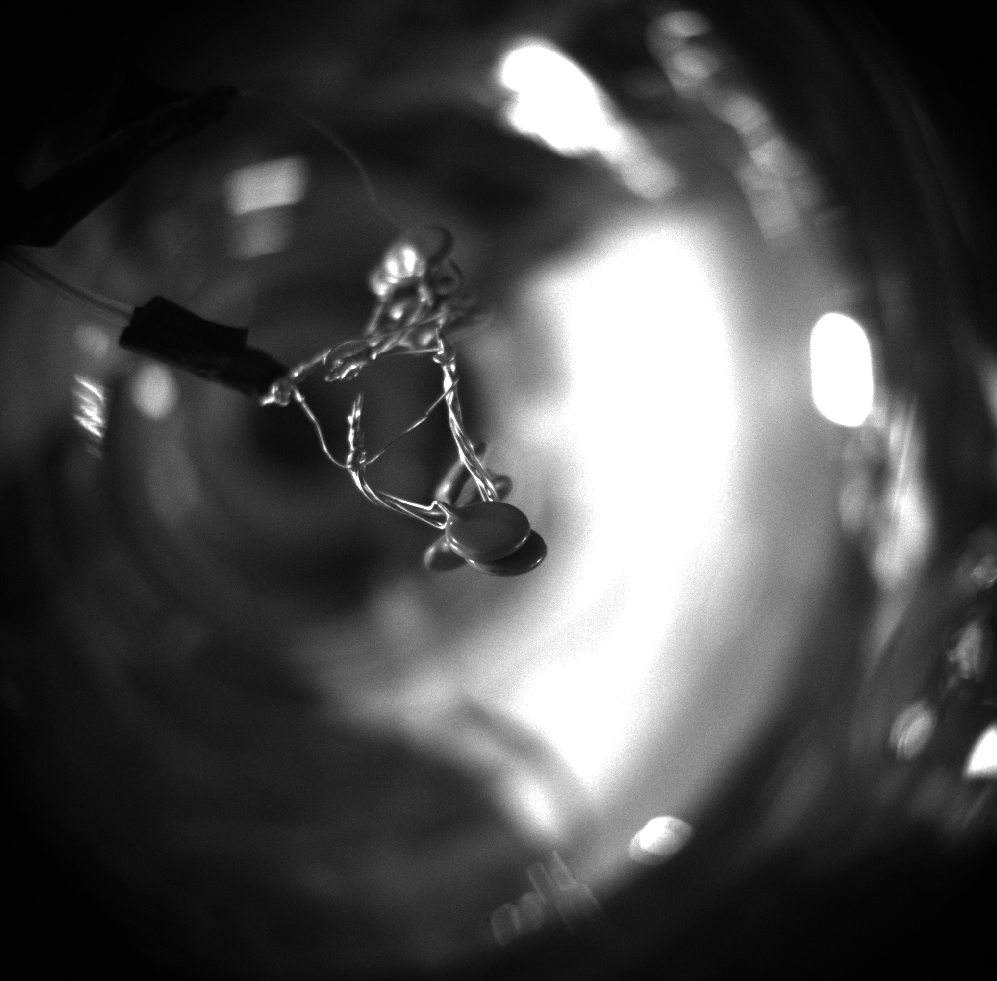}}
   \hfill 	
    \subfloat[Capture photo before the flasher]{\label{fig:flashercapture1}\includegraphics[width=0.495\textwidth,height=0.45\textwidth]{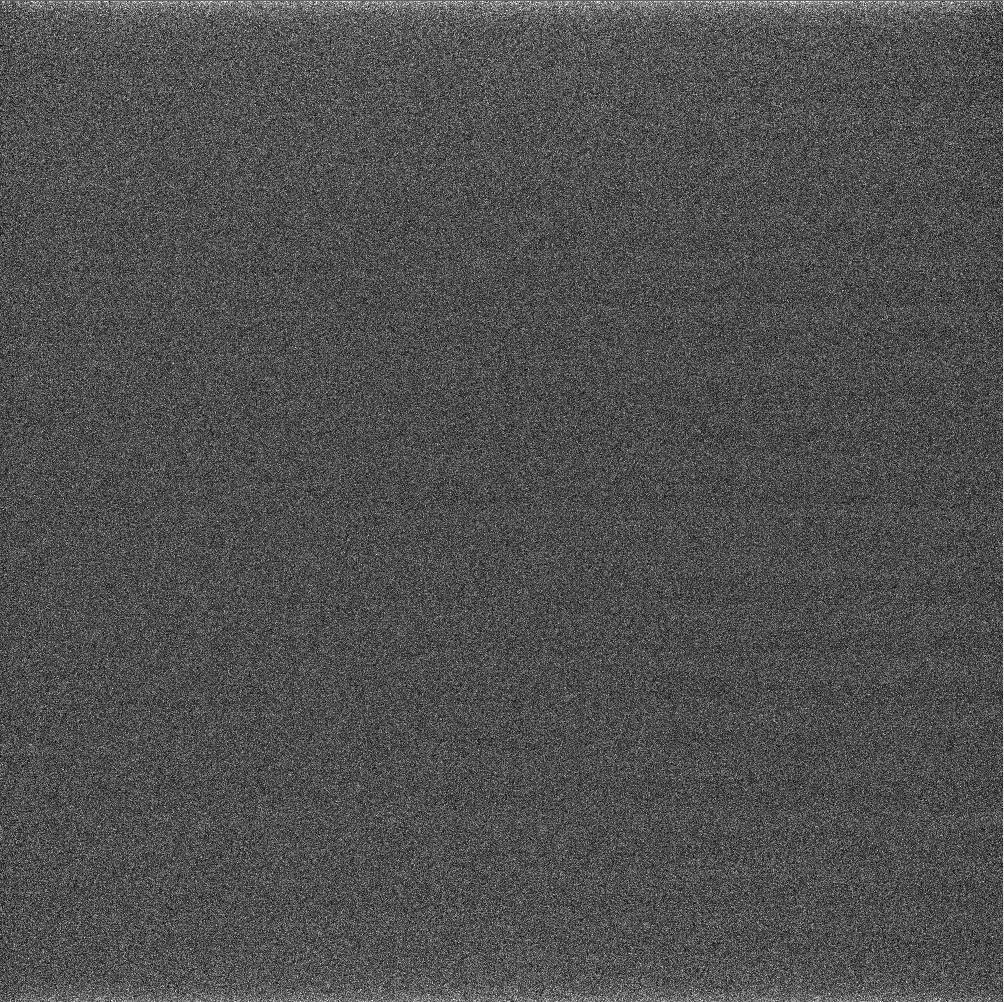}}
    \hfill 	
    \subfloat[Capture of the spark]{\label{fig:flashercapture2}\includegraphics[width=0.495\textwidth,height=0.45\textwidth]{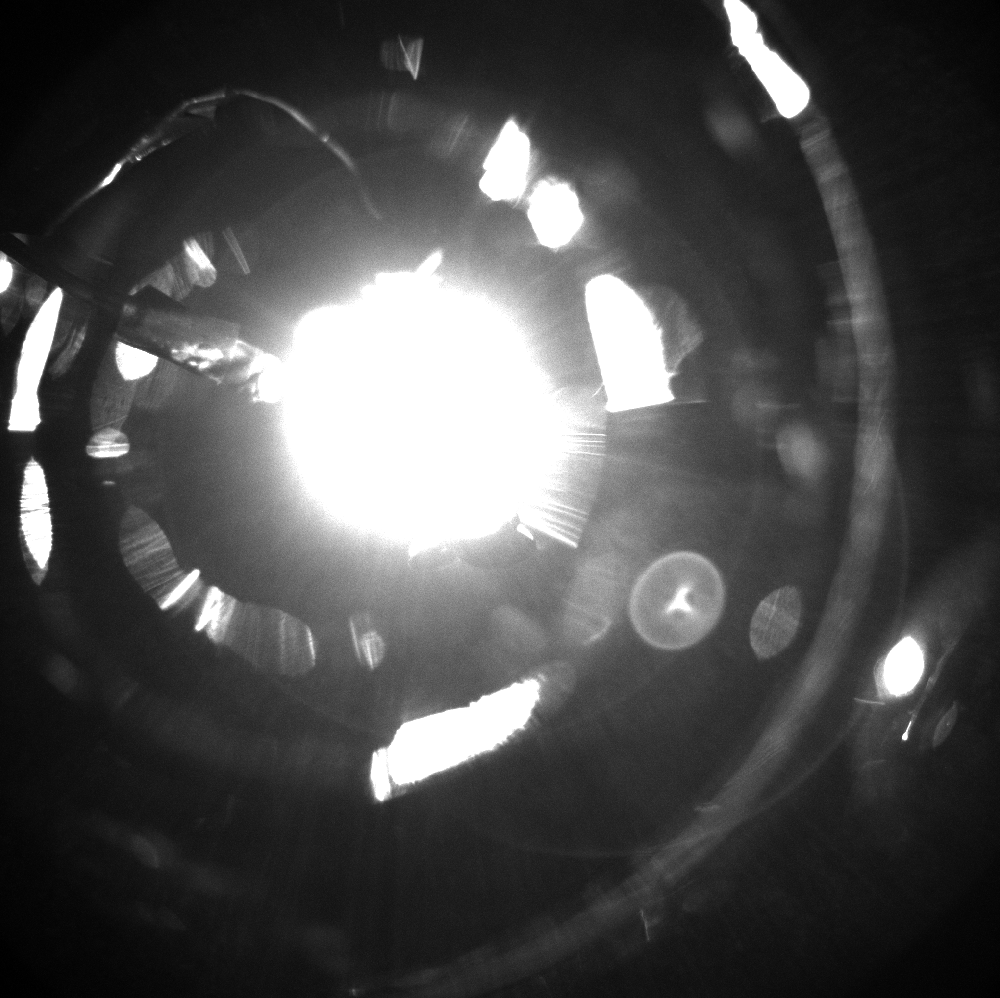}}
    \hfill 	
    \subfloat[Capture when the flasher decaying]{\label{fig:flashercapture3}\includegraphics[width=0.495\textwidth,height=0.45\textwidth]{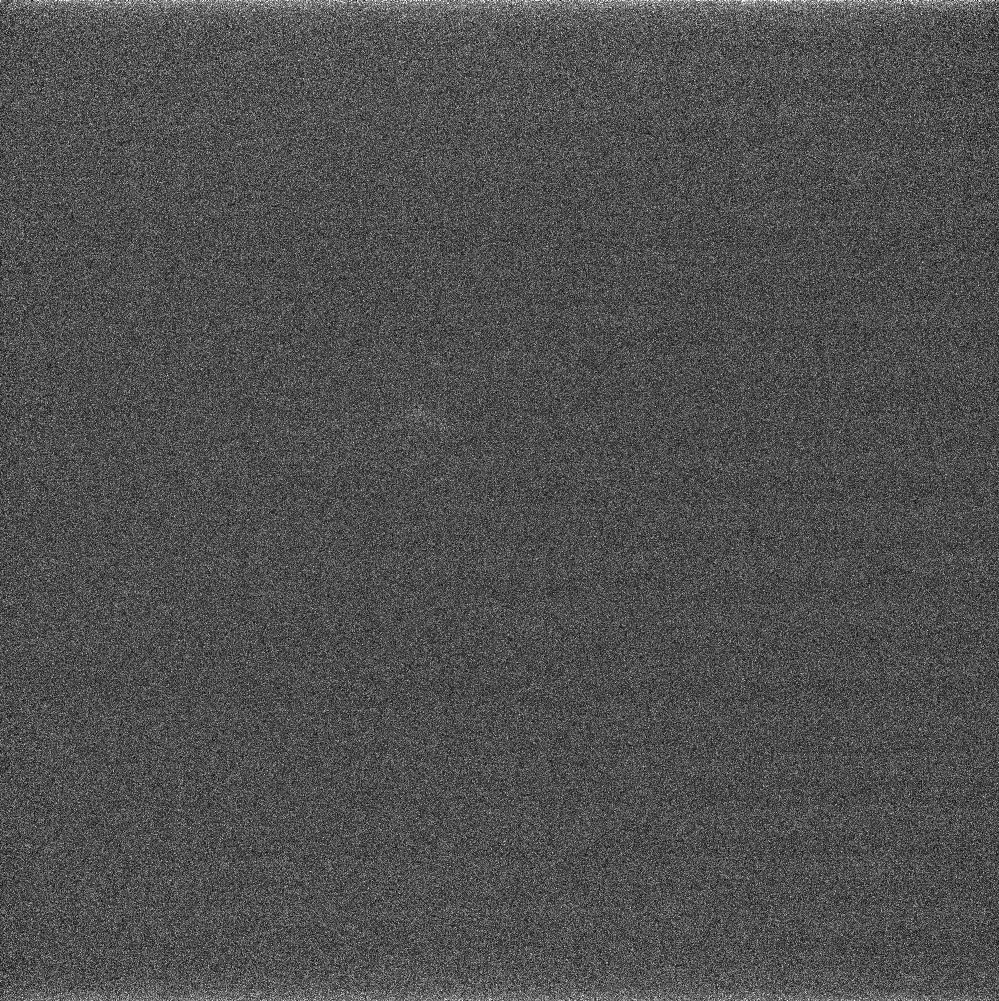}}
  \caption{Visual capture of the artificial spark by capacitor}
  \label{fig:capacitanceflasher}
\end{figure}

The camera effectively captured the distinctive characteristics of the sparks produced by the capacitor at 2500\,V. Figure \ref{fig:capacitanceflasher} elucidates the specific events observed within the dark box. Additionally, Figure \ref{fig:capacitance} visually represents the positioning of the capacitor under visible light. The subsequent three images document the sparking process as captured by the camera with an exposure time of 500\,ms. Figure \ref{fig:flashercapture1} depicts the state prior to the activation of the artificial spark. Following the breakdown of the capacitor, which resulted in the generation of sparks, the camera recorded the transient conditions within the dark box, as illustrated in Figure \ref{fig:flashercapture2}. The photograph demonstrates a gradual weakening of the flash over the subsequent 500\,ms, as evidenced in Figure \ref{fig:flashercapture3}.

\begin{figure}[!ht]
\centering
  \includegraphics[width=0.6\textwidth,height=0.5\textwidth]{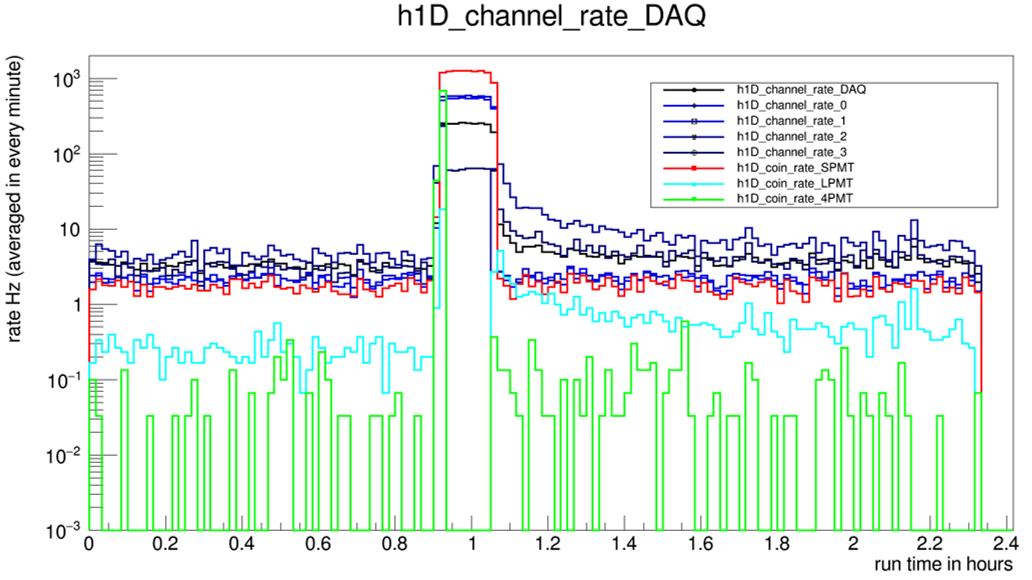}
 \caption{Flasher monitoring by PMT signal coincident rate}
  \label{fig:flahser}
 \end{figure}

Simultaneously, the coincidence rate of the PMTs recorded by the counter is shown in Figure \ref{fig:flahser}. During sparking, the rates of the PMTs exhibited distinct patterns, as indicated in the graph. It is clear that the rate increases rapidly when the spark occurs, demonstrating that the flash at this intensity can be simultaneously monitored by all PMTs. Notably, the rates of two 3-inch PMTs, along with their coincidence rate, showed a significant increasing. The rate of the 20-inch PMT also experienced a notable surge at the moment of ignition, although it remained static due to saturation effects. During the artificial spark, the system’s coincidence rate reached saturation (approximately 1\,kHz) and persisted for around 10 minutes. Following the spark, a gradual decrease in the coincidence rate was observed, in contrast to the rapid diminishing of the spark as documented by the camera. However, after the saturation effects subsided, the DCR of the 20-inch PMT declined. Concurrently, the accidental coincidence rate among all four PMTs exhibited a decreasing until its back to normal levels. These characteristic fluctuations in PMT coincidence counts are typical outcomes triggered by the externally applied ignition of the capacitor.

\subsubsection{Check with Higher PMT HV}

\begin{figure}[!ht]
\centering
  \includegraphics[width=0.6\textwidth,height=0.5\textwidth]{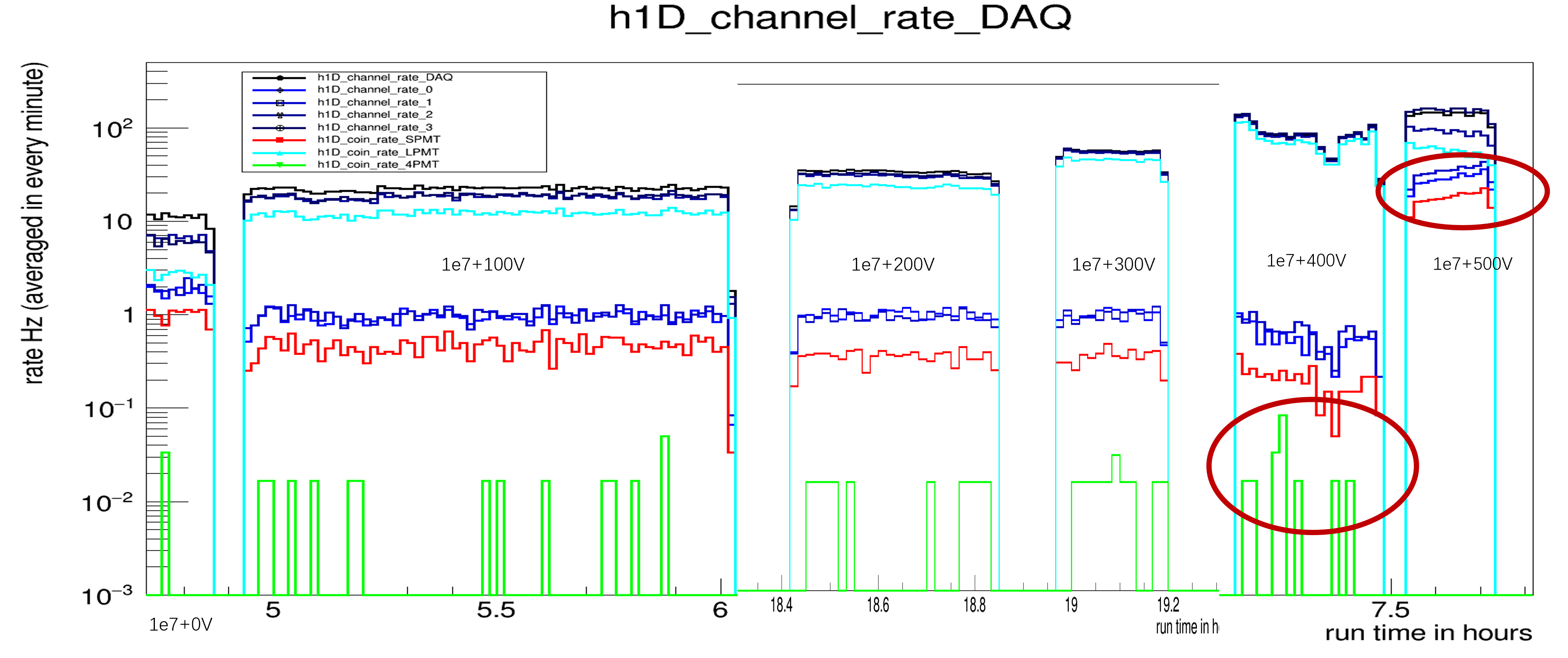}
 \caption{Monitored DCR during 20-inch PMT HV increasing in steps}
  \label{fig:hvadd}
 \end{figure}

\begin{figure*}[!ht]
    \subfloat[20-inch NNVT PMT by the camera]{\label{fig:photocathodennvt}\includegraphics[width=0.495\textwidth,height=0.495\textwidth]{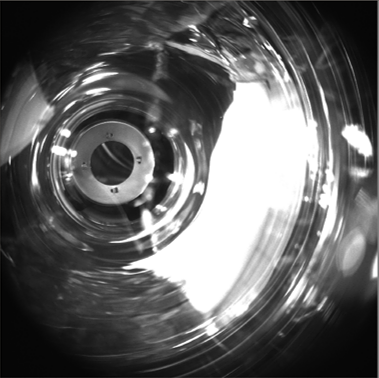}}
   \hfill 	
    \subfloat[Possible candidate of flasher captured by the camera]{\label{fig:pictureofnnvtphotocathode}\includegraphics[width=0.495\textwidth,height=0.495\textwidth]{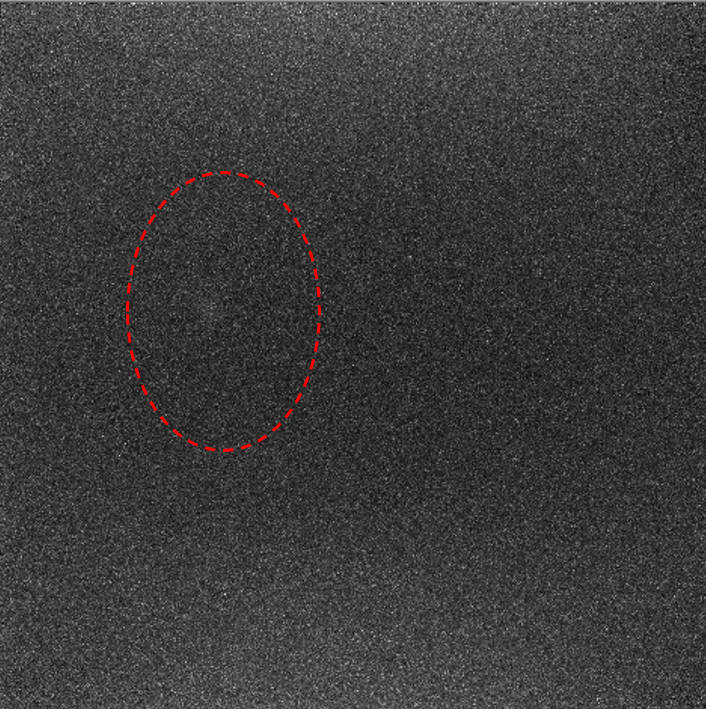}}
  \caption{Possible flasher monitoring by camera when the 20-inch PMT, applied +500\,V additional to its gain  1$\times$10$^{7}$. These two photos are taken under the same position under the visible light and in dark.}
  \label{fig:cameramonitor}
\end{figure*}

%The gain of large PMT operating voltage for Juno is at 1$\times$10$^{7}$. 
In this test, we investigate the intrinsic characteristics of the PMTs by increasing their working high voltage (HV) to enhance potential flasher. The study of flashers involves analyzing the coincidence rates at various operating voltages, as illustrated in Figure \ref{fig:hvadd}. The coincidence rate of the 20-inch PMTs increases with the rise in working voltage with a same threshold. Consequently, monitoring the coincidence rate of the smaller PMT can provide insights into the behavior of the larger PMT. The red line indicates the coincidence outcome for the 3-inch PMTs. Notably, when the working voltage of PMTs is raised by 500\,V higher than the nominal HV of a gain of 1$\times$10$^{7}$, the event rate experiences a significant increase, while the event rates at other voltages remain stable. This observation suggests that under these conditions, the 3-inch PMT detects the PMT flashers emitted by the 20-inch PMT. In each configuration, the camera was used with a same configuration. However, it captured (one second exposure time) a possible flasher from the microchannel plate only at +500\,V higher than the norminal HV, as shown in Figure ~\ref{fig:cameramonitor}.

% I think it's hard to mention "several factors" without providing them. This will for sure been on a reviewers' list. I'd recommend to either provide some of these factors, or to dismiss the last sentence.

%It's plausible to discern the PMTs' flashers by examining waveform characteristics. 
%Hence, this study analyzed waveform traits at the +500\,V operating voltage, as depicted in the figure. However, further analysis is necessary to elucidate the distinct features of the flashers.

\subsubsection{Discussion}

With the artificial spark, the performance of the setup is checked: camera, PMT DCR spikes, PMT recovery, the coincidence among PMTs, which is valuable for better understanding the features of possible sparks.
With the higher PMT working HV test to enhance the possible flashing of PMT inside, a possible flasher of NNVT PMT from the MCP is identified, and its image, recovery is monitored by the PMTs and the camera, which is valuable for further investigation.

\section{Summary}
\label{sec:conclu}

This paper investigates the dark count rate (DCR) of the JUNO 20-inch PMTs occupied in a container system, focusing on the effects of cooling time, temperature, and long-term stability. It was found that the DCR reduces during the cooling process of the bare PMTs. In addition an influence of both thermal effects and (previous) light exposure was found. 
%The findings indicate that the DCR reduces during the cooling process following high voltage (HV) activation of the JUNO bare PMTs; this reduction is influenced by both exposure and thermal effects.
Specifically, the NNVT PMTs exhibit a DCR decrease to approximately 56\% of the measured value after a sufficient cooling time of 12 hours. In contrast, the HPK PMTs show a reduction to about 83\% under the same conditions.
%Even though this is not wrong, I think we might provide a shifting picture with this last sentence with only providing the reduction rate. If we do so, we also have to point out that the median DCR of the NNVT is much higher than the HPK ones. 

In terms of temperature effects on DCR variation, the HPK PMTs demonstrate a more consistent response to cooling and heating (within the temperature range of 14\,$^\circ$C to 28\,$^\circ$C) as well as at room temperature (ranging from 24\,$^\circ$C to 26\,$^\circ$C). For the HPK PMTs, DCR varies by about 2.0\%/$^\circ$C (or 0.55\,kHz/$^\circ$C) relative to the measurement at 21\,$^\circ$C.

Furthermore, the bare NNVT PMTs display a more pronounced temperature dependency, with a DCR variation of approximately 12.0\%/$^\circ$C (6\,kHz/$^\circ$C) against the value at 21\,$^\circ$C, when observed between 14\,$^\circ$C and 28\,$^\circ$C. For the potted NNVT PMTs, the DCR temperature coefficient is around 4.0\%/$^\circ$C (2\,kHz/$^\circ$C) in the same range. This behavior suggests that the DCR becomes flatter with respect to temperature after the potting process for NNVT PMTs, particularly between 14\,$^\circ$C and 21\,$^\circ$C. Notably, the trend remains consistent with room temperature monitoring, showing similar stability for both bare and potted PMTs in the range of 24\,$^\circ$C to 26\,$^\circ$C.

During mass testing, approximately 20\% of the PMTs exhibited abnormal DCR jumps at least once while monitoring cooling time, while long-term monitoring (involving around 110 PMTs) indicated abnormalities in about 7\% of cases. These DCR jumps might be associated with noise, cables, connectors, and electronics components, as well as potential flashers of the PMTs, warranting further investigation.
%Should we provide information about what "long-term" means in this case?

Laboratory measurements on the NNVT PMTs with abnormal DCR revealed characteristics suggestive of artificial sparks with significant intensity, along with indications related to the microchannel Plate (MCP) when supplied with higher voltage. Photographic evidence captured instances of both potential flashers. Additionally, the time-dependence of DCR was monitored through the coincidence events of four PMTs.

\acknowledgments
\label{sec:1:acknow}
This work was supported by the National Natural Science Foundation (NSFC) of China No. 11875282, the Strategic Priority Research Program of the Chinese Academy of Sciences, Grant No.\,XDA10011100, the CAS Center for Excellence in Particle Physics, the Joint Institute of Nuclear Research (JINR), Russia and Lomonosov Moscow State University in Russia, the joint Russian Science Foundation (RSF), DFG (Deutsche Forschungsgemeinschaft).
The authors acknowledge all colleagues from JUNO collaboration for operating the 20-inch PMT testing system.

\bibliographystyle{JHEP}
\bibliography{biblio.bib}
\end{document}